\documentclass{SCGE}
\usepackage[colorlinks,linkcolor=blue,citecolor=blue,urlcolor=blue]{hyperref}
\usepackage{graphicx}

\usepackage{color}

\let\citedash\relax
\makeatletter \providecommand{\citedash}{\hbox{-}\penalty\@m}
\makeatother

\usepackage{amsfonts}
\usepackage{bm}
\usepackage{bbold}
\usepackage{psfrag}

\begin{document}
\begin{picture}(0,0){\rm
\put(0,-20){\makebox[160truemm][l]{\bf {\sanhao\raisebox{2pt}{.}}
Invited Review {\sanhao\raisebox{1.5pt}{.}}}}}
\put(0,-34){\jiuwuhao {\textcolor[rgb]{0.5,0.5,0.5}{\sf 
}}}
\end{picture}

\Year{}
\Month{}
\Vol{}
\No{}
\BeginPage{1}
\AuthorMark{{\rm M. Modugno \textit{et al.}}}
\AuthorMarkCite{M. Modugno \textit{et al.}}
\DOI{} 
\ArtNo{}

\title[Tight-binding models for ultracold atoms in optical lattices]{Tight-binding models for ultracold atoms in optical lattices: \\general formulation and applications}
\author[1,2]{Michele Modugno}{}
\author[3]{Julen Iba\~nez-Azpiroz}{}
\author[4]{Giulio Pettini}{}

\address[{\rm1}]{Departamento de Fisica Te\'orica e Historia de la Ciencia, Universidad del Pais Vasco UPV/EHU, 48080 Bilbao, Spain;}
\address[{\rm2}]{IKERBASQUE, Basque Foundation for Science, 48011 Bilbao, Spain;}
\address[{\rm3}]{\mbox{Peter Gr\"unberg Institute and Institute for Advanced Simulation, Forschungszentrum J\"ulich \& JARA, D-52425 J\"ulich, Germany}}
\address[{\rm4}]{Dipartimento di Fisica e Astronomia, Universit\`a di Firenze,
and INFN, 50019 Sesto Fiorentino, Italy}

\maketitle \vspace{-3.5mm}{\footnotesize\begin{center} 
\end{center}}\vspace*{-5mm}

\begin{center}
\rule{16.5cm}{0.4pt}
\parbox{16.5cm}
{\begin{abstract}
Tight-binding models for ultracold atoms in optical lattices can be
properly defined by using the concept of maximally localized Wannier
functions for composite bands. The basic principles of this approach are
reviewed here, along with different applications to lattice potentials
with two minima per unit cell, in one and two spatial dimensions.
Two independent methods for computing  the tight-binding coefficients
- one \textit{ab initio}, based on the maximally localized Wannier functions, the other through analytic expressions in terms of the energy spectrum - are considered.
In the one dimensional case, where the tight-binding coefficients can be
obtained by designing a specific gauge transformation, we consider both the 
case of quasi resonance between the two lowest bands, and that between $s$
and $p$ orbitals. In the latter case, the role of the
Wannier functions in the derivation of an effective Dirac equation is also reviewed.
Then, we consider the case of a two dimensional honeycomb potential,
with particular emphasis on the Haldane model, its phase diagram, and
the breakdown of the Peierls substitution. Tunable honeycomb lattices,
characterized by movable Dirac points, are also considered.
Finally, general considerations for dealing with the interaction terms are presented.
\end{abstract}}
\end{center}\vspace*{-0.6cm}

\begin{center}
\parbox{16.5cm}
{\bf\jiuhao ultracold atoms, optical lattices, tight-binding models, Wannier functions, effective Dirac equation, honeycomb lattices}
\end{center}

\begin{center}
{}
\CITA
\end{center}

\textwidth=178truemm 
\textheight=236truemm

\wuhao\vspace*{1.5mm}

\tableofcontents

\begin{multicols}{2}

\renewcommand{\baselinestretch}{1.08} \baselineskip 12.2pt\parindent=10.8pt

\section{Introduction}

Experiments with ultracold atoms in optical lattices have undergone an enormous development in recent years to the point that nowadays they represent a solid platform for the quantum simulation of condensed matter physics\cite{bloch2008,lewenstein2012}. These experiments, where atoms are trapped in crystal-like structures made by laser light, offer the possibility to tune most of the relevant parameters with great flexibility and precision, and even to control the dimensionality of the system. Depending on the beam geometry, one can realize one-, two-, or three-dimensional periodic lattices, with one or more wells per unit cell \cite{yukalov2008,yukalov2009}, as well as quasiperiodic structures \cite{lewenstein2007,sanchez-palencia2005,fallani2008,roati2008,modugno2009}. Among the various possibilities, honeycomb lattices are attracting an increasing interest owing to the presence of topological defects in their spectrum, the so-called Dirac points, which leads to remarkable relativistic effects \cite{zhu2007,wu2007,wu2008,wunsch2008,lee2009,soltan-panahi2011,soltan-panahi2012,gail2012,tarruell2012,lim2012,hasegawa2012,sun2012,fuchs2012}, in analogy to the case of graphene \cite{wallace1947,cloizeaux1963,cloizeaux1964a,reich2002}.

Though continuous potentials describing optical lattices can be expressed in simple analytic forms as the combination of a number of sinusoidal potentials, from the theoretical point of view it is often convenient to employ a description in terms of tight-binding models defined on a discrete lattice, as for electrons in a crystal lattice.
Paradigmatic models are the Hubbard model for fermions \cite{hubbard1963}, the Bose-Hubbard model for bosons \cite{fisher1989}, and the Haldane model \cite{haldane1988} in the presence of an external vector gauge field.
The motivation for using a tight-binding model is twofold. First of all, it allows to reduce the complexity of the continuous description to a limited set of parameters, each one playing a specific role (for example, the hopping between different states). In addition, it permits a sort of pictorial description in terms of particles sitting at specific lattice sites, that can tunnel to other sites, or interact between each other. The first aspect is more general, as in principle one can use a projection over any complete basis set, not necessarily of localized functions. In that case the particles would occupy specific basis states, not necessarily associated to a precise position in space. However, having a basis of functions that are localized in real space not only permits to play with a pictorial description, but also reduces the number of parameters needed for an accurate tight-binding description (e.g. tunneling amplitudes between distant sites are suppressed).

The tight-binding regime is easily accessible with ultracold atoms in optical lattices, as the lattice intensity can be tuned to sufficiently high values so that the atoms are deeply localized in the lowest vibrational states of the potential wells. Therefore, each well can be associated to a site of a discrete lattice, making the tight-binding description the natural choice for theoretical calculations. Usually, these models are restricted to a few coefficients associated to the hopping between neighboring sites, and to the onsite interaction among the atoms \cite{bloch2008}. Obviously, additional terms are also possible (for example, next-to-nearest or density-assisted tunneling terms), depending on the order of the tight-binding expansion.
In all cases, the existence of a basis of functions localized around the potential minima is not only important conceptually - in order to justify the tight-binding expansion - but also from the practical point of view, as a precise knowledge of the basis functions is needed to connect the tight-binding coefficients with the actual parameters that can be accessed experimentally.

In the case of optical lattices with a cubic-like arrangement -- with a single well per unit cell -- the natural basis is provided by the exponentially decaying Wannier functions discussed by Kohn \cite{wannier1937,kohn1959}. Notably, in this case the expressions for the tunneling coefficients depend just on the Bloch spectrum \cite{he2001}, being therefore independent on the basis choice. Instead, the interaction coupling still depends on the specific basis, by construction. Analytic expressions for both coefficients can be obtained by means of different approximations \cite{zwerger2003,gerbier2005}. Nevertheless, in general this approach is not suitable when the potential has more than one well per unit cell, because the Kohn-Wannier functions display the same symmetry as the local potential structure, and may not be maximally localized \cite{kohn1959}. For example, for a potential with two degenerate minima in the unit cell, these functions occupy both wells and cannot be associated to a single lattice site \cite{cloizeaux1963,cloizeaux1964a}. Then, in order to deal with non-trivial cell structures, one has to resort to different strategies. A common approach that is found in the literature is that of the so-called atomic orbitals \cite{ashcroft1976,wallace1947,reich2002}, that has been recently employed e.g. for the case of a symmetric double-well unit cell of two-dimensional graphene-like optical lattices \cite{lee2009}. This method is based on a specific ansatz, according to which tight-binding Wannier functions are constructed from linear combinations of wave functions deeply localized in the two potential wells of the unit cell. 

A more general and powerful approach, that has been successfully employed for describing real 
material structures \cite{marzari2012}, is represented by the maximally localized Wannier functions (MLWFs) introduced in a seminal paper by Marzari and Vanderbilt \cite{marzari1997}. The MLWFs are obtained by minimizing the spread of a set of generalized Wannier functions by means of a suitable gauge transformation of the Bloch eigenfunctions for composite bands, and they usually present an exponential decay \cite{brouder2007,panati2013}. This approach reproduces the results discussed by Kohn for the single band case, and it can be extended to more complex situations when \textit{generalized} MLWFs for composite bands are needed. This method is currently implemented by means of a software package, and is largely employed for computing MLWFs of real condensed matter systems \cite{mostofi2008}.

Here we shall review these concepts, following the lines of Refs. \cite{modugno2012,ibanez-azpiroz2013,ibanez-azpiroz2013a,lopez-gonzalez2014,ganczarek2014,ibanez-azpiroz2014,ibanez-azpiroz2015}. First, in sect. \ref{sec:mlwf} we introduce the tight-binding expansion from general principles, by considering the specific implementation for periodic structures with two lattice sites per unit cell. Here we also discuss different strategies for the numerical implementation. Various tight-binding models in one and two spatial dimensions are then considered in the rest of the paper. Sect. \ref{sec:oned} is devoted to the one dimensional case, where it is possible to write down a set of differential equations for the gauge mixing transformation that allow to efficiently compute both the MLWFs and the tunneling coefficients. Explicit results for the case of quasi resonance between the two lowest bands, and that of $s$ and $p$ orbitals, are discussed. Here we also review the role of the Wannier functions in the derivation of an effective Dirac equation. In sect. \ref{sec:twod} we consider the case of two dimensional honeycomb lattices, which exhibit Dirac points in their energy spectrum and are therefore closely connected to the physics of graphene. In particular, there we discuss in detail the \textit{ab initio} derivation of the celebrated Haldane model, that is characterized by the presence of a periodic magnetic field, with vanishing flux through the unit cell. We analyze the corresponding  topological phase diagram, as well as the breakdown of the Peierls substitution. In addition, we also review the case of stretched honeycomb lattices, and derive a low-energy expansion around the merging points of the Dirac points, that can be moved and merged by tuning the lattice parameters.
Then, in sect. \ref {sec:interactions} various possible forms of interaction terms, and general considerations for dealing with them, are reviewed. Conclusions and perspectives are drawn in sect. \ref{sec:conclusions}.

\section{Tight-binding expansion and MLWFs}
\label{sec:mlwf}

Let us consider a system of non interacting bosonic or fermionic particles in the presence of a $D$-dimensional periodic potential.
The tight-binding expansion can be fully carried out both in a first or second-quantized formalism; here we adopt 
the latter and write the non-interacting many-body Hamiltonian as
\begin{equation}
\hat{\cal{H}}_{0}=\int d^{D}\bm{r}~{\hat{\psi}}^\dagger(\bm{r})\hat{H}_{0}{\hat{\psi}}(\bm{r}),
\label{eq:genham}
\end{equation}
where $\hat{\psi}(\bm{r})$ is the field operator, ${\bm{r}}$
is the position vector in $D-$dimensions, ${\hat{H}}_{0}=-(\hbar^{2}/2m)\nabla^{2} + V(\bm{r})$ the one-particle Hamiltonian, and $V(\bm{r})$ the optical potential describing the lattice, generated by laser beams of wavevectors with amplitude $k_{L}$\footnote{In the rest of the paper we shall fix $k_{L}=1$, $\hbar=1$, $m=1/2$ without loss of generality. This corresponds to measuring lengths in units of $1/k_{L}$ and energies in units of the recoil energy $E_{R}=\hbar^{2}k_{L}^{2}/2m$.
}. 
The periodicity of the potential implies that
$V(\bm{r})= V(\bm{r}+\bm{R})$, where $\bm{R}$ belongs to the associated Bravais lattice ${\cal{B}}=\{{\bm{R}}~;~
{\bm{R}}=n_1{\bm{a}}_{1}+\dots+n_{D}{\bm{a}}_{D}~;~n_{1},\dots,n_{D}=0,\pm 1,\pm 2,\dots\}$.
The corresponding reciprocal space is generated by the vectors $\bm{b}_{j}$ that satisfy $\bm{a}_{i}\cdot\bm{b}_{j}=2\pi\delta_{ij}$.

The Hamiltonian (\ref{eq:genham}) can be conveniently mapped onto a discrete lattice corresponding to the minima of the potential $V(\bm{r})$ by expanding the field operator in terms of a complete set of functions $\{w_{\bm{j}\nu}(\bm{r})\}$
localized at each minimum,
\begin{equation}
\hat{\psi}(\bm{r})\equiv \sum_{\bm{j}\nu}{\hat{a}}_{\bm{j}\nu}w_{\bm{j}\nu}(\bm{r}),
\label{eq:psiexpf}
\end{equation}
where $\nu$ is a bandlike index and 
$\hat{a}_{\bm{j}\nu}^{\dagger}$ ($\hat{a}_{\bm{j}\nu}$)
 the creation (destruction) operator of a single particle in the $\bm{j}$-th cell.
These operators satisfy the usual commutation (or anti commutation) rules 
following from those of the field $\hat{\psi}(\bm{r})$.
In the following, we shall consider generalized Wannier functions obtained by performing a unitary mixing of $N$ Bloch eigenstates
\begin{align}
w_{\bm{j}\nu}({\bm{r}})&=\frac{1}{\sqrt{V_{\cal B}}}
\int_{\cal B} \!\!d\bm{k} ~e^{-i\bm{k}\bm{R}_{\bm{j}}}\sum_{m=1}^{N}U_{\nu m}(\bm{k})\psi_{m\bm{k}}({\bm{r}}),
\label{eq:mlwfs}
\end{align}
with $V_{\cal B}$ being the volume of the first Brillouin zone and $U_{\nu m}({\bm{k}})\in U(N)$ a unitary matrix
obeying periodicity conditions in order to preserve the Bloch theorem. In general, $N$ corresponds to the number of minima in the unit cell.
Then, the Hamiltonian (\ref{eq:genham}) can be written in terms of Wannier states $|w_{\bm{j}\nu}\rangle$ as
\begin{equation}
{\hat{\cal{H}}}_0 = \sum_{\nu,\nu'}\sum_{\bm{j,j'}}{\hat{a}}^{\dagger}_{\bm{j}\nu}{\hat{a}}_{\bm{j}'\nu'}
\langle w_{\bm{j}\nu}|{\hat{H}}_0|w_{\bm{j'}\nu '}\rangle,
\label{singparth0}
\end{equation}
where the matrix elements $\langle w_{\bm{j}\nu}|{\hat{H}}_0|w_{\bm{j'}\nu '}\rangle$ depend 
only on $\bm{i}=\bm{j'-j}$ due to the translational invariance of the lattice. 
These matrix elements correspond to tunneling amplitudes between different lattice sites, 
except for the special case $\bm{i=0}$, $\nu=\nu'$, that corresponds to onsite energies\footnote{
The terms with the same index $\bm{j}$ and different $\nu$ refer to different sites inside the same cell.}.
Then, by defining
\begin{equation}
\hat{d}_{\nu{\bm{k}}}=\frac{1}{\sqrt{V_{B}}}
\sum_{\bm{j}} ~e^{-i{\bm{k}}\cdot{\bm{R}}_{\bm{j}}}\hat{a}_{\bm{j}{\nu}},
\label{eq:opk}
\end{equation}
${\hat{\cal{H}}}_0$ is transformed as
\begin{equation}
\hat{\cal{H}}_{0}=\sum_{\nu,\nu'}\int_{\cal B} {d}^{2}{\bm{k}}~h_{\nu\nu'}({\bm{k}})
\hat{d}_{\nu{\bm{k}}}^{\dagger}\hat{d}_{\nu'\bm{k}},
\label{eq:intermh0}
\end{equation}
with
\begin{align}
&h_{\nu\nu'}(\bm{k})=\sum_{\bm{i}}e^{i{\bm{k}}\cdot{\bm{R}}_{\bm{i}}}
\langle {w}_{\bm{0}\nu}|\hat{H}_{0}|w_{\bm{i}\nu'}\rangle\nonumber\\
&=\frac{1}{V_{B}}\sum_{\bm{i}}\int_{\cal{B}}d\bm{q}~e^{i(\bm{k-q})\cdot\bm{R}_{\bm{i}}}
\sum_{n}U^{*}_{\nu n}(\bm{q})U_{\nu'n}(\bm{q})\varepsilon_{n}(\bm{q})
\label{eq:hamnu}
\end{align}
being the Hamiltonian density in quasimomentum space. 
Notice that the above expression is exact -- we have restricted the analysis to a specific subset composed by $N$ Bloch bands, but made no further approximations. Then, by using the following summation rule (valid for an infinite lattice)
\begin{equation}
\frac{1}{V_{B}}\sum_{\bm{i}}e^{i{\bm{R}}_{{\bm{i}}}\cdot (\bm{k}'-\bm{k})}=\delta(\bm{k}'-\bm{k}),
\label{eq:deltak}
\end{equation}
eq. (\ref{eq:hamnu}) can be rewritten as
\begin{equation}
 h_{\nu\nu'}(\bm{k})=\sum_{n}U^{*}_{\nu n}(\bm{k})U_{\nu'n}(\bm{k})\varepsilon_{n}(\bm{k}),
\end{equation}
whose eigenvalues coincide with the exact bands $\varepsilon_{\nu}(\bm{k})$ by construction, and are therefore independent on the specific choice of MLWFs. This is an obvious result owing to the completeness of any Wannier basis.

However, for practical purposes the summation over $i$ must be truncated by retaining only a finite number of matrix elements. This, in a nutshell, is the essence of the tight-binding expansion. Notice that the actual number of terms needed to reproduce the properties of the system within a certain degree of accuracy crucially depends on the properties of the basis functions $w_{\bm{j}\nu}(\bm{r})$.
Given a continuous Hamiltonian, the optimal choice of the Wannier basis requires to fix both $N$, 
which is the number of bands to be mixed, and the specific form for the matrix $U({\bm{k}})$. 
In the following, it is convenient to distinguish between the case $N=1$ and $N>1$.

\textit{Single band case.}
When there is just one well per unit cell the band mixing is not necessary, so that $N=1$. In this case the procedure greatly simplifies as the gauge group is $U(1)$ and the matrix $U({\bm{k}})$ takes the form of a diagonal phase transformation 
\begin{equation}
U_{\nu m}(\bm{k})=e^{i\phi_{\nu}(\bm{k})}\delta_{\nu m},
\end{equation}
with the phases $\phi_{\nu}(\bm{k})$ being periodic over the first Brillouin zone.
Then, from eq. (\ref{eq:hamnu}) one has \cite{he2001}
\begin{equation}
\langle {w}_{\bm{0}\nu}|\hat{H}_{0}|w_{\bm{i}\nu'}\rangle
=\frac{\delta_{\nu \nu'}}{V_{B}}\int_{\cal{B}}d\bm{q}~e^{-i\bm{q}\cdot\bm{R}_{\bm{i}}}
\varepsilon_{\nu}(\bm{q}),
\label{eq:singletunnel}
\end{equation}
so that both the onsite energies and the tunneling coefficients are independent of
the phases $\phi_{\nu}(\bm{k})$. Therefore, in the single band case the 
tight-binding expansion is \textit{gauge independent}, namely it does not depend on the choice of the Wannier functions.

\textit{Composite band case.}
Let us now consider a lattice potential with $N$ minima per unit cell, with $N>1$.  In this case, it is easy to verify that the tight-binding parameters depend explicitly on the specific choice of the matrices $U_{nm}(\bm{k})$, so that these parameters -- and any other physical quantity calculated at a finite order of the tight-binding expansion -- are \textit{gauge-dependent}.

Let us now turn to the specific set of MLWFs introduced by Marzari and Vanderbilt in Ref. \cite{marzari1997}.
They are defined via a transformation $U_{nm}(\bm{k})$ that minimizes the total spread 
$\Omega\equiv\sum_{\nu}[\langle \bm{r}^2\rangle_{\nu}-\langle \bm{r}\rangle_{\nu}^{2}]$. This quantity can be decomposed as $\Omega=\Omega_I+\tilde{\Omega}$, the first term being 
gauge invariant. In turn, the gauge dependent part, $\tilde{\Omega}$, can be written as the sum of a diagonal and an off-diagonal component\footnote{Both componets are non negative. The off-diagonal term is absent in case of a single band, $N=1$. }, $\tilde{\Omega}=\Omega_D+\Omega_{OD}$.
Both $\Omega_{D}$ and $\Omega_{OD}$ can be expressed in terms of the generalized Berry vector 
potentials $\bm{A}_{\nu\nu'}(\bm{k})$ defined as 
\begin{equation}
\bm{A}_{\nu\nu'}(\bm{k})=iV_{B}\langle u_{\nu\bm{k}}|\bm{\nabla}_{\bm{k}}|u_{\nu'\bm{k}}\rangle,
\label{eq:connections}
\end{equation}
which is an hermitian matrix.
In particular, one has
\begin{align}
\label{eq:omegad}
\Omega_D&=\sum_{\nu}\langle \left(\bm{A}_{\nu\nu}(\bm{k})-\langle \bm{A}_{\nu\nu}\rangle_{\cal{B}}\right)^{2}\rangle_{\cal{B}}=\sum_{\nu}\Omega_{D\nu}\,,
\\
\Omega_{OD}&=\sum_{\nu\neq \nu'}\langle |\bm{A}_{\nu\nu'}|^{2}\rangle_{\cal{B}}\,,
\label{eq:omegaod}
\end{align}
with $\langle \dots\rangle_{\cal{B}}$ representing the integral over the first Brillouin zone.
In one-dimension (1D), both $\Omega_D$ and $\Omega_{OD}$ can be made strictly vanishing, and the corresponding gauge is called \textit{parallel transport gauge}, as the off-diagonal Berry connections $\bm{A}_{\nu\nu'}$ (with $\nu\neq\nu'$) vanish. This may not be the case in higher dimensions, where usually the minimal value of the spread is finite. 
We also remark that in general (though in the absence of a formal proof) one may assume that the gauge where the spread of Wannier functions is minimal corresponds to the one that provides the best tight-binding approximation of individual Bloch bands.

As anticipated, the use of \textit{composite} instead of \textit{single band} transformations 
is required in case of a set of almost degenerate bands (well separated from the others), that usually 
corresponds to having more that one minimum per unit cell.
Here we focus our attention on systems whose Wigner-Seitz cell contains two basis points, say $A$ and $B$. 
When two Bloch bands are sufficiently separated from the others, the optimal tight-binding expansion for the corresponding sector of the spectrum is achieved by means of the MLWFs for those two Bloch bands, via the gauge transformation in eq. (\ref{eq:mlwfs}). Then, for a two-level system it is customary to write the Hamiltonian density in eq. (\ref{eq:hamnu}) as
\begin{equation}
h(\bm{k})=\left(\begin{array}{cc}
 \epsilon_{A}(\bm{k}) & z(\bm{k}) \\
 z^{*}(\bm{k}) & \epsilon_{B}(\bm{k})
\end{array}\right),
\label{eq:hmatrix}
\end{equation}
where (see eq. (\ref{eq:hamnu})) 
\begin{align}
\label{eq:epsnu}
\epsilon_{\nu}({\bm{k}})&=\sum_{{\bm{i}}}\langle w_{\nu}^{\bm{0}}|\hat{H}_{0}|w_{\nu}^{\bm{R}_{\bm{i}}}\rangle e^{i{\bm{k}}\cdot{\bm{R}}_{\bm{i}}}\equiv
E_{\nu} + \sum_{{\bm{i\neq 0}}}J_{{\bm{i}}}^{\nu}e^{i{\bm{k}}\cdot{\bm{R}}_{\bm{i}}}
\equiv
E_{\nu} + f^{\nu}({\bm{k}}),
\\
z({\bm{k}})&=\sum_{{\bm{i}}}\langle w_{A}^{\bm{0}}|\hat{H}_{0}|w_{B}^{\bm{R}_{\bm{i}}}\rangle e^{i{\bm{k}}\cdot{\bm{R}}_{\bm{i}}}\equiv
-\sum_{{\bm{i}}}T_{{\bm{i}}}e^{i{\bm{k}}\cdot{\bm{R}}_{\bm{i}}}.
\label{eq:zetanu}
\end{align}
Above, $E_{\nu}$ are the onsite energies, whereas $J_{{\bm{i}}}^{\nu}$ and $T_{{\bm{i}}}$ represents the tunneling amplitudes between sites of the same type ($A$ or $B$), and between sites of type $A$ and $B$. 
The sign convention is chosen in such a way that when the tunneling coefficients are real, they are all positive defined. 
Above, the index $\nu=1,2$ (see eq. (\ref{eq:psiexpf})) has been traded to $\nu=A,B$ since 
the associated MLWFs are located around the minima $A$ and $B$.
In the following, it is also convenient to fix the arbitrary energy offset such that $E_{A}=\epsilon$ and $E_{B}=-\epsilon$. 

The matrix $h_{\nu\nu'}(\bm{k})$ in eq. (\ref{eq:hmatrix}) can also be rewritten in a compact form, by using the basis formed by the $2\times 2$ identity matrix, $I$, and of the three Pauli matrices, $\sigma_{i}$. One has \cite{shao2008}
\begin{equation}
 h(\bm{k})=h_{0}(\bm{k})I+\bm{h}(\bm{k})\cdot\bm{\sigma},
\end{equation}
with ${\bm{h}}\equiv(h_{1},h_{2},h_{3})$ and 
\begin{align}
h_{0}(\bm{k})&=\frac{f_{A}(\bm{k})+f_{B}(\bm{k})}{2}\equiv f_{+}(\bm{k}),
\label{generalhmatrixh0}\\
h_{1}(\bm{k})&= {\rm{Re}}[z(\bm{k})]
,\label{generalhmatrixh1}\\
h_{2}(\bm{k})&= -{\rm{Im}}[z(\bm{k})]
,
\label{generalhmatrixh2}\\
h_{3}(\bm{k})&=\epsilon+\frac{f_{A}(\bm{k})-f_{B}(\bm{k})}{2}\equiv \epsilon + f_{-}(\bm{k}).
\label{generalhmatrixh3}
\end{align}

Finally, the tight-binding expansion for the two energy bands under consideration are obtained as the eigenvalues of the matrix (\ref{eq:hmatrix}), namely
\begin{align}
\epsilon_{\pm}(\bm{k})&= h_{0}(\bm{k})\pm|{\bm{h}}(\bm{k})|
\nonumber\\
&= f_{+}(\bm{k})\pm\sqrt{|\epsilon+f_{-}(\bm{k})|^2+|z(\bm{k})|^2}.
\label{eq:tbenergies}
\end{align}
This expression can be further simplified when the two minima of type $A$ and $B$ are degenerate ($\epsilon=0$) so that $f^{A}({\bm{k}})=f^{B}({\bm{k}})\equiv f({\bm{k}})$ and
\begin{equation}
\epsilon_{\pm}(\bm{k})= f({\bm{k}})\pm |z({\bm{k}})|.
\label{eq:degeneratecase}
\end{equation}

\textit{Numerical implementation.} The final step for completing the tight-binding expansion corresponds to determining the specific values of the tight-binding coefficients for a given configuration of the underlying continuous potential. Here we review two independent complementary approaches. 

\begin{figure}[H]
\centerline{\includegraphics[width=\columnwidth]{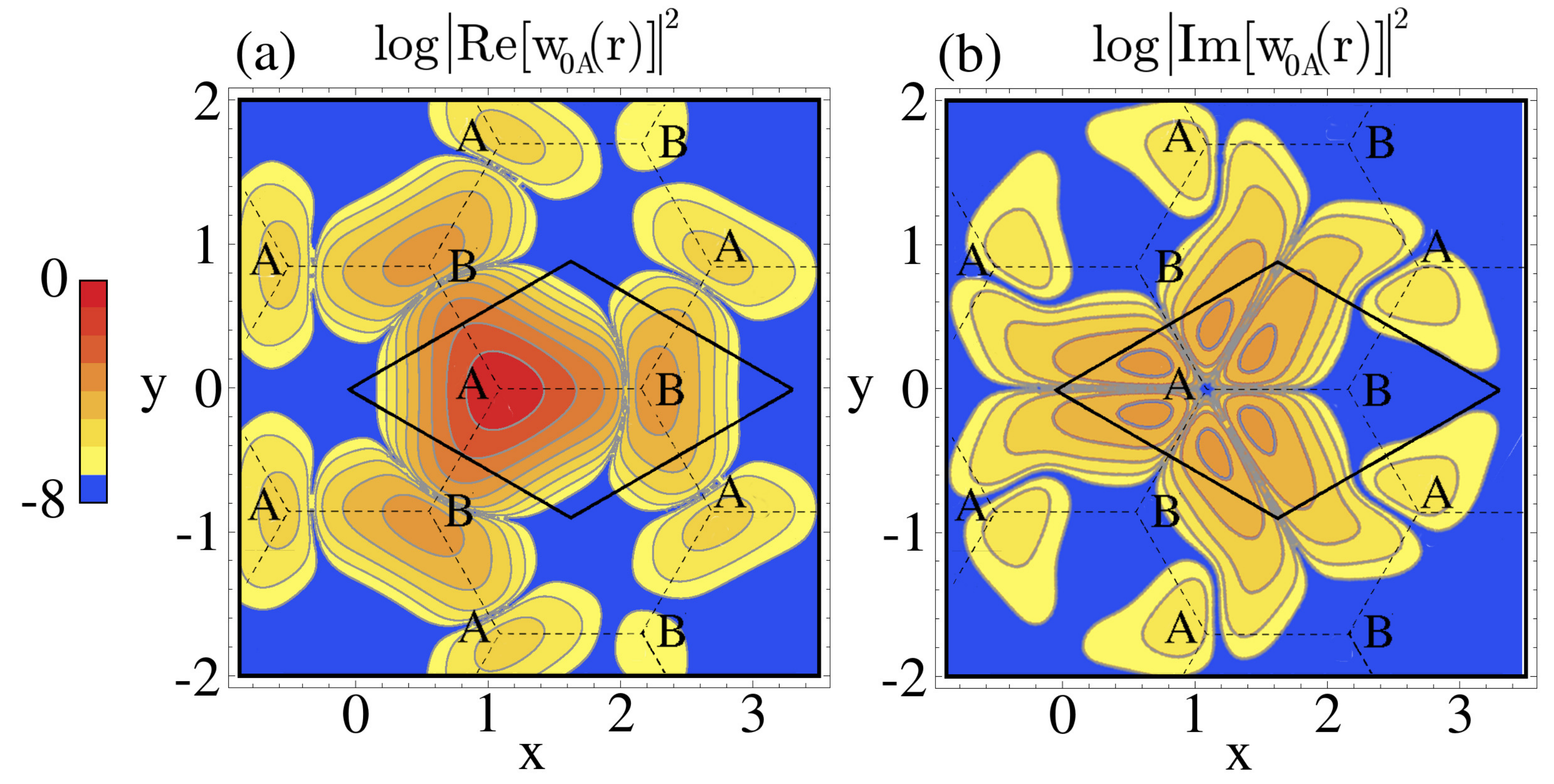}}
\caption{(Color online) Density plot (in logarithmic scale) of the real (left) and imaginary (right) parts of the 
MLWFs for sublattice A of a honeycomb potential in the tight-binding regime (see sect. \ref{sec:twod})\cite{ibanez-azpiroz2015}. 
The solid and dashed lines denote the unit cell and the honeycomb lattice, respectively. Lenghts are in units of the lattice spacing.
}
\label{fig:im-mlwf}
\end{figure}

The first one consists in using the definition of the tight-binding coefficients as expectation values of the single-particle Hamiltonian over Wannier states, as in eqs. (\ref{eq:epsnu})-(\ref{eq:zetanu}). This is an \textit{ab-initio} approach, that requires the determination of the gauge transformation in eq. (\ref{eq:mlwfs}), and gives direct access to the MLWFs as well as to the whole set of tight-binding parameters, at any order of the expansion. In practice, the functional minimization of the spread (see eqs. (\ref{eq:omegad}-\ref{eq:omegaod})) is most conveniently done in k-space. This procedure is implemented in the WANNIER90 software package \cite{mostofi2008}, 
a powerful tool that is largely employed for computing MLWFs of real condensed matter systems\cite{marzari2012}.
The required input quantities are the Bloch spectrum and the matrix elements between Bloch eigenfunctions. For the results presented in this review, these quantities have been calculated with a modified version of the QUANTUM-ESPRESSO package \cite{giannozzi2009} adapted for simulating optical lattices \cite{ibanez-azpiroz2013,ibanez-azpiroz2013a,ibanez-azpiroz2014}\footnote{A different implementation, designed for optical lattice potentials in one and two spatial dimensions, has been discussed in Ref. \cite{walters2013}. For one-dimensional systems it is also possible to write analytically a set of ordinary differential equations for a specific gauge transformation, as we shall see in the next section.
}.
Though the details of the method will not be covered in this review, as an example here we illustrate the properties of the MLWFs obtained for a honeycomb potential in the presence of breaking of time-reversal, that are complex valued \cite{brouder2007,panati2013}, see Figure \ref{fig:im-mlwf}. This specific example will be thoroughly analyzed later on in sect. \ref{sec:twod}.
As shown in the Figure, the MLWFs are localized around the potential minima, and rapidly decay when moving away from their center. 
When the Hamiltonian is invariant under time-reversal, the decay of the tails is exponentially (Figure \ref{fig:im-mlwf}a); in this case both the MLWFs and the tunneling coefficients can be chosen real \cite{brouder2007,panati2013}. The maximal localization of the MLWFs implies that the tunneling amplitudes associated to the hopping between two lattice sites decrease very fast as their distance is increased, offering an optimal choice for the construction of tight-binding models with a minimal set of tunneling coefficients. In addition, the MLWFs are able to adapt and capture diverse features of the system Hamiltonian, such as complex deformations of the honeycomb structure (see sect. \ref{sect:stretched}) and even the presence of a vector potential that breaks the underlying time-reversal symmetry of the lattice (see sect. \ref{sect:haldane}). It is noteworthy that in the latter case the breaking of time-reversal symmetry manifests itself by inducing a finite imaginary part of the MLWFs, as shown in Fig. \ref{fig:im-mlwf}b. Interestingly, this imaginary part is what ultimately determines the topological properties of the system, as it will be thoroughly discussed for the particular case of the Haldane model reviewed in sect. \ref{sect:haldane}.

The second approach instead makes explicit use of the tight-binding expression of the energy spectrum (see eq. (\ref{eq:tbenergies})), and consists in writing down analytical expressions for the tight-binding parameters in terms of specific properties of the exact spectrum \cite{ibanez-azpiroz2014}. In general the functions $z(\bm{k})$ and $f_{\nu}(\bm{k})$ in eqs. (\ref{eq:zetanu}) and (\ref{eq:epsnu}) can be expressed as
\begin{align}
z(\bm{k})&= \sum_{\bm{\alpha}}T^{\alpha}Z_{\alpha}(\bm{k})
\\
f_{\nu}(\bm{k})&= \sum_{\bm{\beta}}J^{\beta}_{\nu}F_{\beta}(\bm{k})
\end{align}
where $Z_{\alpha}$ and $F_{\beta}$ are functions of the quasimomentum $\bm{k}$, and depend only on the geometrical structure of the lattice. Then, from the dispersion relation in eq. (\ref{eq:tbenergies}), it follows
\begin{align}
\epsilon_{\pm}(\bm{k})&= \sum_{\bm{\beta}}\left(J^{\beta}_{A}+J^{\beta}_{B}\right)F_{\beta}(\bm{k})
\\
&\quad\pm\sqrt{\left|\epsilon + \sum_{\bm{\beta}}\left(J^{\beta}_{A}-J^{\beta}_{B}\right)F_{\beta}(\bm{k})\right|^2+
\left|\sum_{\bm{\alpha}}T^{\alpha}Z_{\alpha}(\bm{k})\right|^2}.
\nonumber
\end{align}
Given a certain tight-binding approximation, defined by a finite number of coefficients $T^{\alpha}$ and $J^{\beta}_{\nu}$ (plus $\epsilon$), one can identify a corresponding number of relations evaluated at specific $\bm{k}$ points, to be inverted in order to express the tight-binding parameters in terms of specific properties of the exact spectrum. 
Generally this approach - though it does not give access to the MLWFs - can provide accurate results if the order of the tight-binding expansion is properly chosen, and it has the advantage of requiring a minimal computation effort, as the exact Bloch spectrum can be readily computed by means of a standard Fourier decomposition \cite{lee2009}. 

The two approaches have been explicitly compared for the case of two-dimensional honeycomb potential discussed in sect. \ref{sec:twod}, 
finding a remarkable agreement \cite{ibanez-azpiroz2013,ibanez-azpiroz2014,ibanez-azpiroz2015}.

In order to quantify the degree of accuracy of a given tight-binding model, it is convenient to consider its fidelity in reproducing the exact single-particle Bloch spectrum of the continuous Hamiltonian.
This can be measured by considering the following quantity 
\begin{equation}
\delta \varepsilon_{n} \equiv \frac{1}{\Delta{\varepsilon}_{n}}\sqrt{\frac{d}{2\pi}\int_{\cal{B}} d\bm{k} [\varepsilon_{n}(\bm{k})-\varepsilon_{n}^{tb}(\bm{k})]^{2}},
\label{eq:deltaener}
\end{equation}
which represents the ratio of the quadratic spread between the exact Bloch spectrum $\varepsilon_{n}(\bm{k})$ and the corresponding tight-binding energies\footnote{Here $\varepsilon_{n}^{tb}(\bm{k})$ stands for $\epsilon_{\pm}(\bm{k})$ in eq. (\ref{eq:tbenergies}).} to the bandwidth $\Delta{\varepsilon}_{n}\equiv (\varepsilon_{n}^{max}-\varepsilon_{n}^{min})$. Specific examples will be considered in the following sections.

\section{One-dimensional double-well systems}
\label{sec:oned}

\begin{figure}[H]
\centering
\includegraphics[width=\columnwidth]{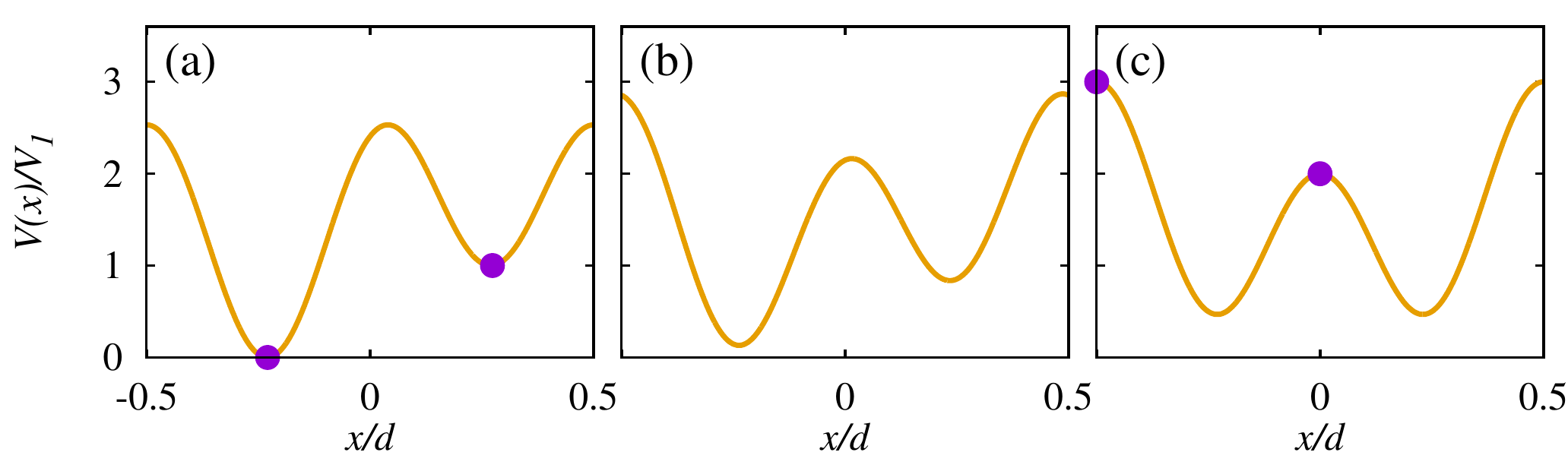}
\caption{(Color online) The three possible configurations for the unit cell of the potential in (\ref{eq:potential0}): (a) two different minima, with the overall potential having two centers of parity, for $\theta_{0}=0$ ($\phi_{0}\simeq \pi/4$); (b) an asymmetric double-well with parity that is broken globally - in this example $\theta_{0}=\pi/4$ ($\phi_{0}\simeq\pi/8$); (c) a symmetric double-well, $\theta_{0}=\pi/2$ ($\phi_{0}=0$). The black dots in (a), (c) represent the parity centers of the whole periodic potential. Here $\epsilon=2$.}
\label{fig:pot}
\end{figure}

As a specific one-dimensional system, here we shall consider an optical lattice with two wells in the unit cell, described by a potential $V(x)$ of the form 
\begin{equation}
V(x)=V_{0}\left[\sin^{2}\left(k_{L}x +\phi_{0}\right)+ \epsilon\sin^{2}\left(2k_{L} x+\theta_{0}+2\phi_{0}\right)\right],
\label{eq:potential0}
\end{equation}
where $V_{0}$ is the overall amplitude of the potential, $\epsilon$ a dimensionless parameter, $\theta_{0}$ and $\phi_{0}$ two arbitrary phases (the latter represents just a rigid shift of the whole potential) and $k_{L}$ the laser wavelength. 
When $\epsilon>0$ the potential has period $d=\pi/k_{L}$, $V(x+d)=V(x)$.
For convenience, here the unit cell is defined as having two (absolute) maxima at the cell borders, and it is centered in $x=0$ by a suitable choice of the phase $\phi_{0}$, $x\in[-d/2,d/2]$.
Different configurations can be realized by varying $\epsilon$ and the phase $\theta_{0}$. They can be divided in three classes according to the value of $\theta_{0}$\footnote{$\theta_{0}$ can be restricted to the range $[0,\pi/2]$ without loss of generality.}, as shown in Figure \ref{fig:pot}. 
(a) $\theta_{0}=n\pi$ ($n\in\mathbb{Z}$): all the maxima are degenerate and the periodic potential has two classes of parity centers, located at the two (inequivalent) minima. 
(b) $\theta_{0}\in(0,\pi/2)+n\pi/2$: the unit cell is an asymmetric double-well with no symmetry centers.
(c) $\theta_{0}= \pi/2+n\pi$: the unit cell is a symmetric double-well, and the potential has two centers of parity placed at the two (inequivalent) maxima. 

Depending on the value of $\epsilon$ and $\theta_{0}$ one can have interesting situations in which the two lowest bands or the the first two excited bands are almost degenerate, making the use of the composite band approach the optimal choice for defining a set of MLWFs.

\subsection{The tight-binding Hamiltonian}

\begin{figure}[H]
\centerline{\includegraphics[width=0.8\columnwidth]{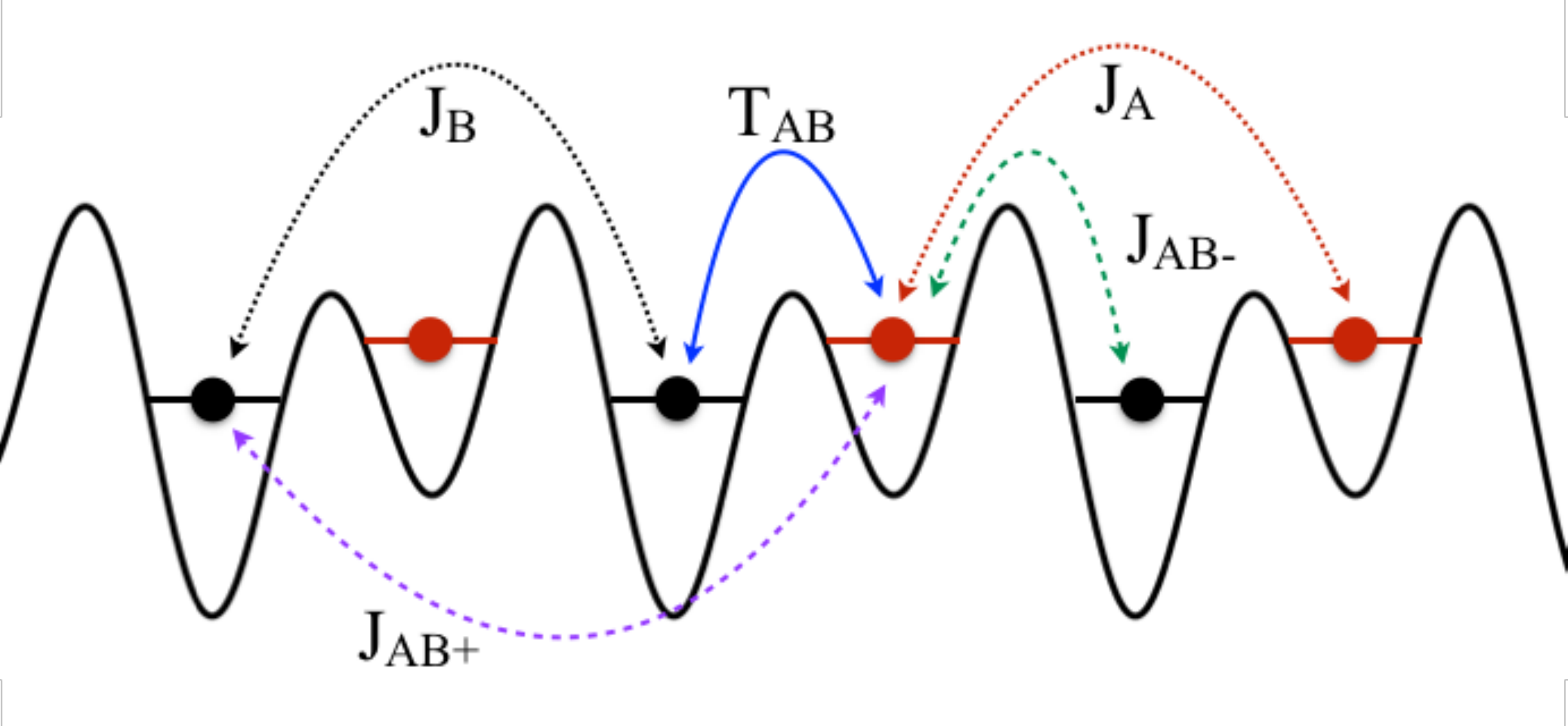}}
\caption{(Color online) A sketch of the double-well structure and of the tunneling coefficients considered here.}
\label{fig:doublewell}
\end{figure}

Here we consider the tight-binding Hamiltonian including all the terms corresponding to nearest-neighboring cells, namely
\begin{align}
\label{eq:fullHam}
\hat{\cal{H}}_{0} &\simeq 
\sum_{\nu=A,B}\sum_{j}E_{\nu}\hat{n}_{j_{\nu}}
-\sum_{\nu=A,B}\sum_{j}J_{\nu}(\hat{a}_{j_{\nu}}^\dagger\hat{a}_{(j+1)_{\nu}} +h.c.)
\\
&-\sum_{j}\left(T_{AB}\hat{a}_{j_{A}}^\dagger\hat{a}_{j_{B}}+
J_{AB_{+}}\hat{a}_{j_{A}}^\dagger\hat{a}_{(j+1)_{B}} +
J_{AB_{-}}\hat{a}_{j_{A}}^\dagger\hat{a}_{(j-1)_{B}} +h.c.\right),
\nonumber
\end{align} 
with the tunneling coefficients as indicated in Figure \ref{fig:doublewell}. This model will be referred to as the (single particle) \textit{extended} tight-binding model. By posing $J_{\nu}=0=J_{AB+}$ one recovers the usual \textit{nearest-neighbor} approximation, commonly used in the literature for both single-well \cite{jaksch1998} and double-well lattices \cite{trebst2006,qian2011,qian2013}. The latter is a reasonable assumption for a single well lattice in the tight-binding regime \cite{he2001,boers2007}, but may not be fully justified in the range of the typical experimental parameters in the double well case \cite{modugno2012}. For this reason, in general it is convenient to consider the \textit{extended} model in eq. (\ref{eq:fullHam}).

Then, according to eqs. (\ref{eq:epsnu})-(\ref{eq:zetanu}), we have 
\begin{align}
\epsilon_{\nu}(k)&\equiv E_{\nu}-2J_{\nu}\cos(kd),
\\
z(k)&\equiv -(T_{AB} + J_{AB_{+}}e^{-ikd} +J_{AB_{-}}e^{ikd}).
\end{align}
The corresponding tight-binding spectrum for \textit{composite bands}, given by eq. (\ref{eq:tbenergies}), can be written as
\begin{equation}
\varepsilon_{\pm}^{tb}(k)=\epsilon_{+}(k)\pm\sqrt{\epsilon_{-}^{2}(k)+|z(k)|^{2}},
\label{eq:ener-tb}
\end{equation}
where $\epsilon_{\pm}(k)\equiv(\epsilon_{A}(k)\pm \epsilon_{B}(k))/2$. 

In the \textit{single band} case we simply have \cite{bloch2008}
\begin{equation}
\label{eq:ener-sb}
\varepsilon_{n}^{sb}(k)= E_{n}^{sb}-2J_{n}^{sb}\cos(kd)
\end{equation}
with (see eq. (\ref{eq:singletunnel})) 
\begin{equation}
E_{n}^{sb}={\frac{d}{2\pi}}\int_{\cal B}\!\!d{k} ~\varepsilon_{n}(k)\,,\quad
J_{n}^{sb}=-{\frac{d}{2\pi}}\int_{\cal B}\!\!d{k} ~\varepsilon_{n}(k) e^{ikd},
\end{equation}
$\varepsilon_{n}(k)$ being the exact Bloch spectrum. Notably, as discussed in sect. \ref{sec:mlwf}, these expressions do not depend on the choice of the Wannier basis.

\subsection{Generalized Wannier functions}
\label{sec:mlwfs}

Here we discuss a specific method - complementary to the standard WANNIER90 approach -- for constructing the MLWFs of a one-dimensional double-well potential. The method consists in deriving a set of ordinary differential equations (with periodic boundary conditions) for the gauge transformation in eq. (\ref{eq:mlwfs}), by using the expressions for $\Omega_D$ and $\Omega_{OD}$ in terms of the Berry connections in eqs. (\ref{eq:omegad})-(\ref{eq:omegaod}) \cite{modugno2012}. As discussed in sect. \ref{sec:mlwf}, in one dimension the spread ${\tilde\Omega}$ can be made strictly vanishing in the \textit{parallel transport} gauge, where the matrix $A_{nm}(k)$ is diagonal, with the diagonal elements being constant and equal to their mean values. In general, the diagonal and off-diagonal spreads $\Omega_{D}$ and $\Omega_{OD}$ can be minimized either simultaneously or independently, but the latter is more convenient from the practical point of view. In particular, the gauge transformation can be decomposed as \begin{equation}
U_{nm}(k)= e^{\displaystyle{i\phi_n(k)}}S_{nm}(k)
\label{eq:decomposition}
\end{equation}
where $S_{nm}\in SU(2)$ is a transformation that makes $\Omega_{OD}$ vanishing (it also affects the diagonal elements $A_{nn}$, but this is not relevant at this stage), and $\exp\{i\phi_n\}$ a diagonal unitary transformation that makes $\Omega_{D}$ vanishing without affecting $\Omega_{OD}$. 

The two transformations are constructed as follows. Let us first consider a diagonal, \textit{single band} transformation $U(n)$ of the form
\begin{equation}
\label{eq:gauge1}
|u_{nk}\rangle\rightarrow |\tilde{u}_{nk}\rangle=e^{ i\phi_{n}(k)} |u_{nk}\rangle,
\end{equation}
with $\phi_{n}(k)$ being a real differentiable function of $k$, such that $\phi_{n}(k+2k_{L})=\phi_{n}(k) + 2\pi \ell$ ($\ell$ integer) in order to have periodic and single valued Bloch eigenstates (here we set $\ell=0$, without loss of generality). Then, since under this transformation $A_{nn}(k)\rightarrow A_{nn}(k)-{\partial_{k} \phi_{n}}(k)$, we have (see eq. (\ref{eq:omegad}))
\begin{equation}
\Omega_{Dn}\rightarrow \tilde{\Omega}_{Dn}=\langle \left(A_{nn}-{\partial_{k} \phi_{n}} -\langle A_{nn}\rangle_{\cal{B}}\right)^{2}\rangle_{\cal{B}},
\end{equation}
that can be made vanishing by imposing
\begin{equation}
{\partial_{k} \phi_{n}}=A_{nn}-\langle A_{nn}\rangle_{\cal{B}}\,.
\label{eq:gauge-a}
\end{equation}
This equation can be readily solved numerically, as discussed in \cite{modugno2012}.
Notice that $\Omega_{OD}$ is not affected by this transformation, as $A_{12}(k)$ just acquires a phase factor.

Let us now turn to a generic \textit{composite band} gauge transformation 
\begin{equation}
|u_{nk}\rangle\rightarrow |\tilde{u}_{nk}\rangle=\sum_{m}U_{nm}(k)|u_{mk}\rangle
\label{gaugematrix}
\end{equation}
with $U_{nm}\left(k+2k_{L}\right)=U_{nm}(k)$. Then, the generalized Berry potentials transform as 
\begin{align}
A_{nm}\rightarrow{\tilde A}_{nm}
&=i\sum_{l}U^{*}_{nl}{\partial_{k} U_{ml} }+\sum_{l,l'}U^{*}_{nl}U_{ml'}A_{ll'}.
\label{eq:atilde}
\end{align}
At this point, it is convenient to use the decomposition\footnote{In general, the group $U(N)$ can be written as a semidirect product $SU(N)\rtimes U(1)$, with $U(1)$ being subgroup of $U(N)$ 
consisting of matrices of the form diag$(1,1,\dots,e^{i\chi})$.}
\begin{equation}
U(k) = \left(\begin{array}{cc}
 z_{1}(k) & -z^{*}_{3}(k) \\
 z_{3}(k) & z^{*}_{1}(k)
 \end{array} \right)
 \left(\begin{array}{cc}
 1 & 0 \\
 0 & r(k)
 \end{array} \right) 
\label{eq:u0matrix}
\end{equation}
with $|z_{1}|^{2} + |z_{3}|^{2}=1$, $r(k)=e^{{i\chi(k)}}$, and the parametrization $S=e^{\displaystyle i\alpha\vec{\sigma}\cdot \hat{n}/2}$ with $\hat{n}=(\cos\varphi\sin\theta,\sin\varphi\sin\theta,\cos\theta)$ 
and $\sigma_i$ being the Pauli matrices (that is valid for any matrix $S\in SU(2)$). Then, one finds
\begin{equation}
U =
 \left(\begin{array}{cc}
\cos\frac{\alpha}{2}+i\sin\frac{\alpha}{2}\cos\theta~ & ie^{\displaystyle{i(\chi-\varphi)}}\sin\theta\sin\frac{\alpha}{2} \\
ie^{\displaystyle{i\varphi}}\sin\theta\sin\frac{\alpha}{2} & e^{\displaystyle{i\chi}}\left(\cos\frac{\alpha}{2}-i\sin\frac{\alpha}{2}\cos\theta\right) 
 \end{array} \right)
 \label{eq:umatrix}
\end{equation}
with $\chi=\chi(k)$, $\varphi=\varphi(k)$, $\alpha=\alpha(k)$ and $\theta=\theta(k)$. 
Since $\Omega$ transforms as
\begin{equation}
\Omega_{Dn}\rightarrow \tilde{\Omega}_{Dn}=
\langle (\tilde{A}_{nn}(k)-\langle \tilde{A}_{nn}\rangle_{\cal{B}})^{2}\rangle_{\cal{B}}
\end{equation}
\begin{equation}
\Omega_{OD}\rightarrow \tilde{\Omega}_{OD}=2\langle |\tilde{A}_{12}|^{2}\rangle_{\cal{B}}
\end{equation}
in order to get $\tilde{\Omega}=0$ one has to impose (see eq. (\ref{eq:atilde}))
\begin{align}
\label{eq:ann}
\tilde{A}_{nn}(k)&\equiv i\sum_{l}U^{*}_{nl}{\partial_{k} U_{nl} }+\sum_{l,l'}U^{*}_{nl}U_{nl'}A_{ll'}=\langle {\tilde A}_{nn}\rangle_{{\cal B}}
\\
\tilde{A}_{12}(k)&\equiv i\sum_{l}U^{*}_{1l}{\partial_{k} U_{2l} }+\sum_{l,l'}U^{*}_{1l}U_{2l'}A_{ll'}=0.
\label{mastereq}
\end{align}
At this point it is worth to notice that the right-hand term in eq. (\ref{eq:ann}) is not known \textit{a priori}, so that this equation is  useless in practice.
However, one can still consider eq. (\ref{mastereq}), that defines a gauge transformation for making $\Omega_{OD}$ vanishing. In fact, the spread $\Omega_{D}$ can be set to zero afterwords, with a diagonal transformation.

By combining eqs. (\ref{eq:umatrix}) and (\ref{mastereq}), one gets a system of four differential equations for $\alpha$, $\theta$, $\varphi$, $\chi$, whose normal form\footnote{
Notice that eq. (\ref{mastereq}) actually represents two real equations, for its real and imaginary components. 
The normal form consists in using a matrix notation, by defining a vector with the four angle derivatives as components. Since initially there are only two equations, only two of the four final equations can be non trivial.} is
\begin{align}
\label{eq:alpha}
\frac{\partial_{k}\alpha}{2}&=-\frac{\cos2\theta}{\sin\theta}\left(A_{12}^{R}\cos\eta +A_{12}^{I} \sin\eta\right) \\
&\quad-\textrm{cotg}\frac{\alpha}{2} \textrm{cotg}\theta \left(A_{12}^{R}\sin\eta
- A_{12}^{I}\cos\eta\right)+\cos\theta (A_{11}-A_{22})
\nonumber
\\
\partial_{k}\theta&=\frac{\cos\theta \sin\alpha}{\sin^2(\alpha/2)} ( A_{12}^{R} \cos\eta + A_{12}^{I}\sin\eta)\label{eq:theta}\\
&\quad+\frac{\cos\alpha}{\sin^2(\alpha/2)} (A_{12}^{R}\sin\eta-A_{12}^{I} \cos\eta)
\nonumber\\
&\quad-\textrm{cotg}\frac{\alpha}{2} \sin\theta (A_{11}-A_{22})
\nonumber
\end{align}
\begin{equation}
\partial_{k}\chi= 0\,,\qquad \partial_{k}\varphi=0
\label{eq:chi-phi}
\end{equation}
where we have defined $\eta\equiv\varphi-\chi$. Form eq. (\ref{eq:chi-phi}) it follows $\partial_{k}\eta=0$,
so that the gauge transformation is determined only by the difference $\eta=\varphi_{0}-\chi_{0}$, that plays the role of a constant parameter. Then, one can safely pose $\chi_{0}\equiv0$, $\eta=\varphi_{0}$, without loss of generality. This simplifies the expression (\ref{eq:u0matrix}), $r(k)\equiv1$, so that only the $SU(2)$ component $S_{nm}$ survives. 

Finally, we remind that \textit{single band} MLWFs can be obtained by using just the diagonal gauge transformation, and correspond to the exponentially decaying Wannier functions discussed by Kohn \cite{kohn1959,marzari1997}. 
Both gauge transformations can be solved by using the representation of Bloch functions in $k$-space; the reader is referred to Ref. \cite{modugno2012} for the details of the numerical implementation.

\subsection{Tunneling coefficients}

By using eqs. (\ref{eq:mlwfs}), (\ref{eq:epsnu})-(\ref{eq:zetanu}), and (\ref{eq:decomposition}), the onsite energies and the tunneling coefficients can be conveniently expressed as 
\begin{align}
E_{\nu}&=
{\frac{d}{2\pi}}\int_{\cal B} d{k}\sum_{m=1}^{2}|S_{\nu m}(k)|^{2}\varepsilon_{m}(k)
\\
J_{\nu}&=
-{\frac{d}{2\pi}}\int_{\cal B} d{k}e^{-ikd}\sum_{m=1}^{2}|S_{\nu m}(k)|^{2}\varepsilon_{m}(k)
\\
T_{AB}&=
-{\frac{d}{2\pi}}\int_{\cal B} d{k} 
~ e^{i\Delta\phi(k)}\sum_{m=1}^{2}S_{1m}^{*}(k)S_{2m}(k)\varepsilon_{m}(k)
\\
J_{AB_{\pm}}&=
-{\frac{d}{2\pi}}\int_{\cal B} d{k} 
~ e^{i(\Delta\phi(k)\mp kd)}\sum_{m=1}^{2}S_{1m}^{*}(k)S_{2m}(k)\varepsilon_{m}(k)
\end{align}
with $\Delta\phi(k)=\phi_{2}(k)-\phi_{1}(k)$.
Notice that the terms $E_{\nu}$ and $J_{\nu}$, that involve sites of the same types, depend just on the $S_{nm}$ transformation. Instead, those connecting sites of type $A$ and $B$, namely $T_{AB}$ and $J_{AB_{\pm}}$, also depend on the diagonal transformation. 

The above formulas can be used also in the \textit{single band} case,
by replacing $S_{nm}$ with $\delta_{nm}$. Remarkably, in this case the tunneling amplitudes between $A$ and $B$ sites are vanishing at any order ($T_{AB}$, $J_{AB_{\pm}}$, and so on) owing to the orthogonality of states belonging to different Bloch bands. Therefore, the description in terms of single band MLWFs
contains just hopping terms between homologous sites (either of type $A$ or $B$) belonging to different cells, and cannot be used to describe hopping between sites of type $A$ and $B$. 

A number of specific applications of the present formalism are presented in the following section.

\subsection{Applications}

\subsubsection{A conventional case} 

\begin{figure}[H]
\centering
\includegraphics[width=0.9\columnwidth]{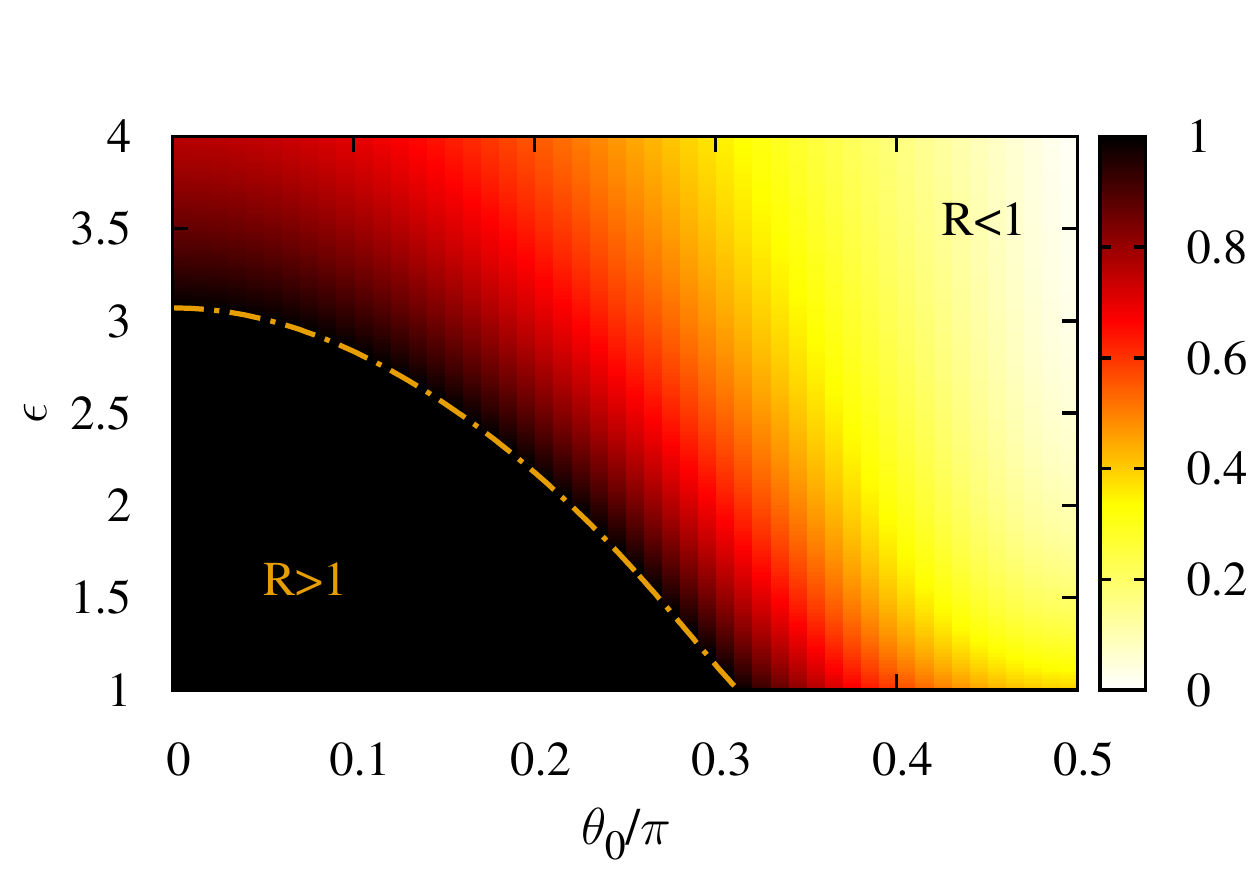}
\caption{(Color online) Density plot of the ratio between the first and second band gaps, $R\equiv\delta_{12}/\delta_{23}$ as a function of $\theta_{0}$ and $\epsilon$. The dashed dotted line corresponds to $R=1$. The color scale is saturated at $R=1$.}
\label{fig:gap}
\end{figure}

As a first example, we consider the case in which the two lowest bands are almost degenerate, as discussed in Ref. \cite{modugno2012}; here we fix $V_{0}=10E_{R}$. In order to characterize the band structure one can consider the quantity $R\equiv\delta_{12}/\delta_{23}$, with $\delta_{12}$ ($\delta_{23}$) being the band gap between the first and second (second and third) band. Its behavior as a function of $\theta_{0}$ and $\epsilon$ is shown as a density plot in Figure \ref{fig:gap}. In general, the composite band approach provides an optimal basis of MLWFs up to $R\approx1$ \cite{modugno2012}.

\begin{figure}[H]
\centerline{\includegraphics[width=0.99\columnwidth]{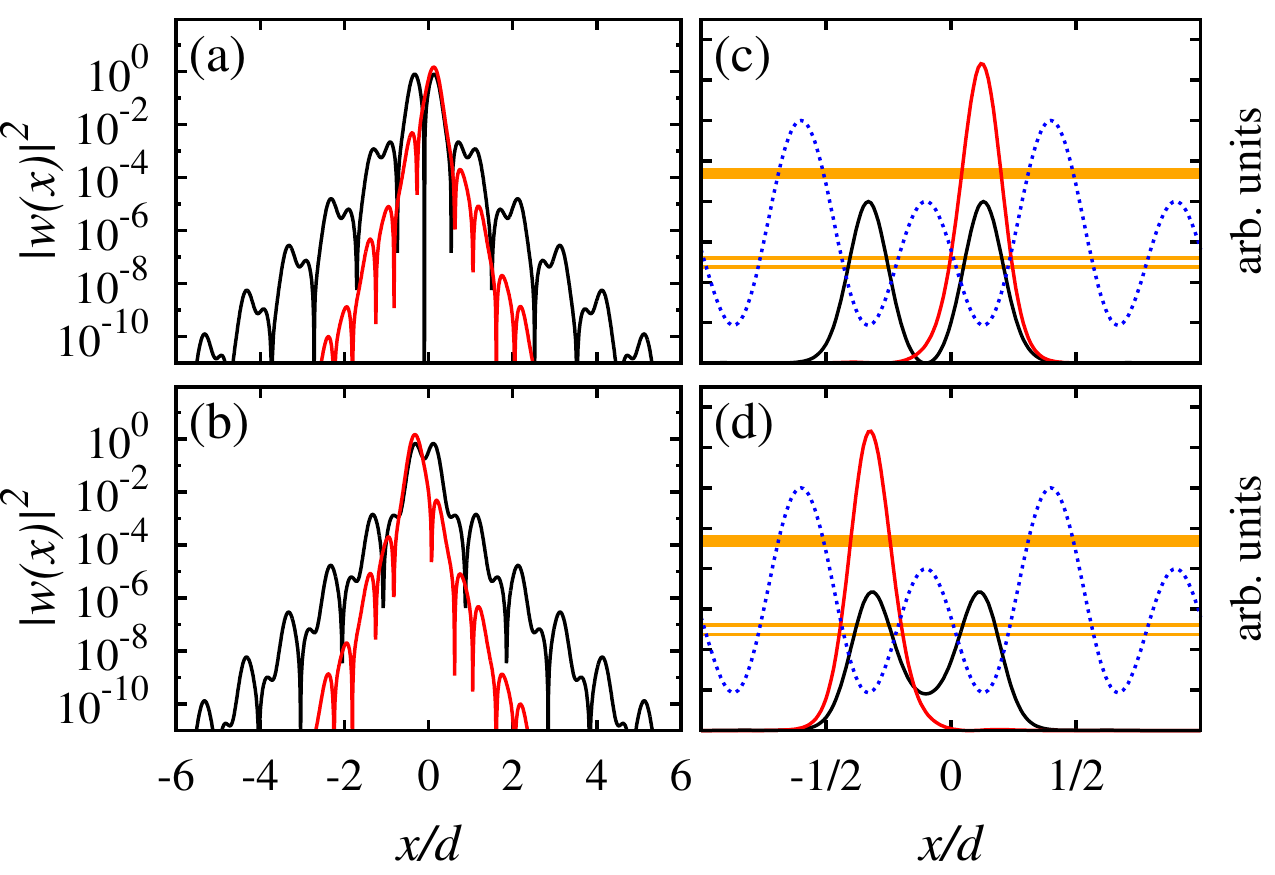}}
\caption{(Color online) Plot of the density of the two lowest \textit{single band} (black lines) and \textit{generalized} (red lines) MLWFs, in log (a,b) and linear scale (c,d). The dotted line in (c,d) represents the potential, while the horizontal orange stripes are the first three Bloch bands (on the same scale of the potential). Here $\epsilon =2$, $\theta_{0}=\pi/2$.}
\label{fig:wannier}
\end{figure}

\begin{figure}[H]
\centerline{\includegraphics[width=0.99\columnwidth]{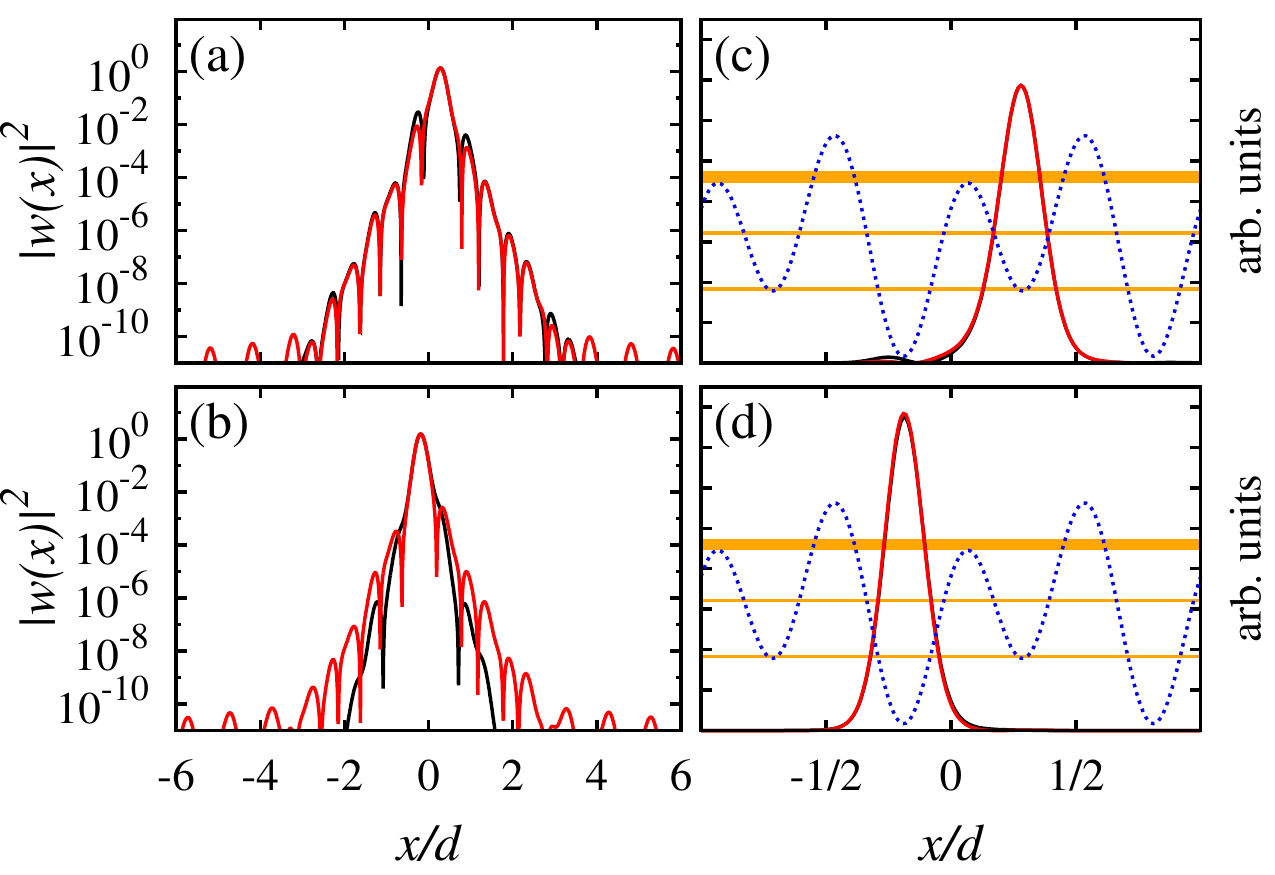}}
\caption{(Color online) Same as Figure \ref{fig:wannier} for $\theta_{0}=0.2\pi$.}
\label{fig:wannier0-10}
\end{figure}

\textit{MLWFs.} A comparison between the \textit{single} and \textit{composite band} MLWFs is shown in Figures \ref{fig:wannier},\ref{fig:wannier0-10}, for $\epsilon =2$.
Figure \ref{fig:wannier} refers to a symmetric double well, $\theta_{0}=\pi/2$. In this case, the \textit{single band} MLWFs have the same symmetry of the potential \cite{kohn1959}, and they occupy both wells of the unit cell. Instead, each of the \textit{composite band} MLWFs nicely localizes in one of the sub-wells. For this symmetric case, a reasonable estimate for the bulk properties of the composite band MLWFs, and for the nearest-neighbor tunneling coefficients, can also be obtained by considering symmetric and antisymmetric combinations of the single band MLWFs. 
This approach is particularly effective when each unit cell can be regarded as a single double-well, namely when there are large barriers at the cell borders. However, since the tails of the proper MLWFs decay much faster, this approximation fails in reproducing higher order tunneling coefficients. In fact, it can proved analytically that one would get $J_{AB-}=J_{AB+}$, that is manifestly incorrect (see Figure \ref{fig:doublewell}).
In Figure \ref{fig:wannier0-10}, it is shown the case of an asymmetric double well, for $\theta_{0}=0.2\pi$. Here the gap between the first two bands is of the order of the one between the second and third band. In this case even the \textit{single band} MLWFs are almost localized within a single well. However, we remind that they cannot be used to describe hopping between sites of type $A$ and $B$, as discussed in the previous section.
We also remark that the exponential decay of the Wannier functions for a given band is controlled by a parameter of the order of the smallest band gap separating that band from the neighboring ones, see Refs. \cite{kohn1959,he2001,cloizeaux1963,cloizeaux1964,blount1962,anderson1968,kivelson1982}. For the present situation, the relevant (minimal) gap in the single band picture is that between the two bands, and this explains why this approach fails when the two bands are close to each other (the gap vanishes in the degenerate limit). Instead, the localization properties of the composite band MLWFs are controlled by the gaps with the outer bands, making their use the right choice when the internal gap vanishes.

\begin{figure}[H]
\centering
\includegraphics[width=\columnwidth]{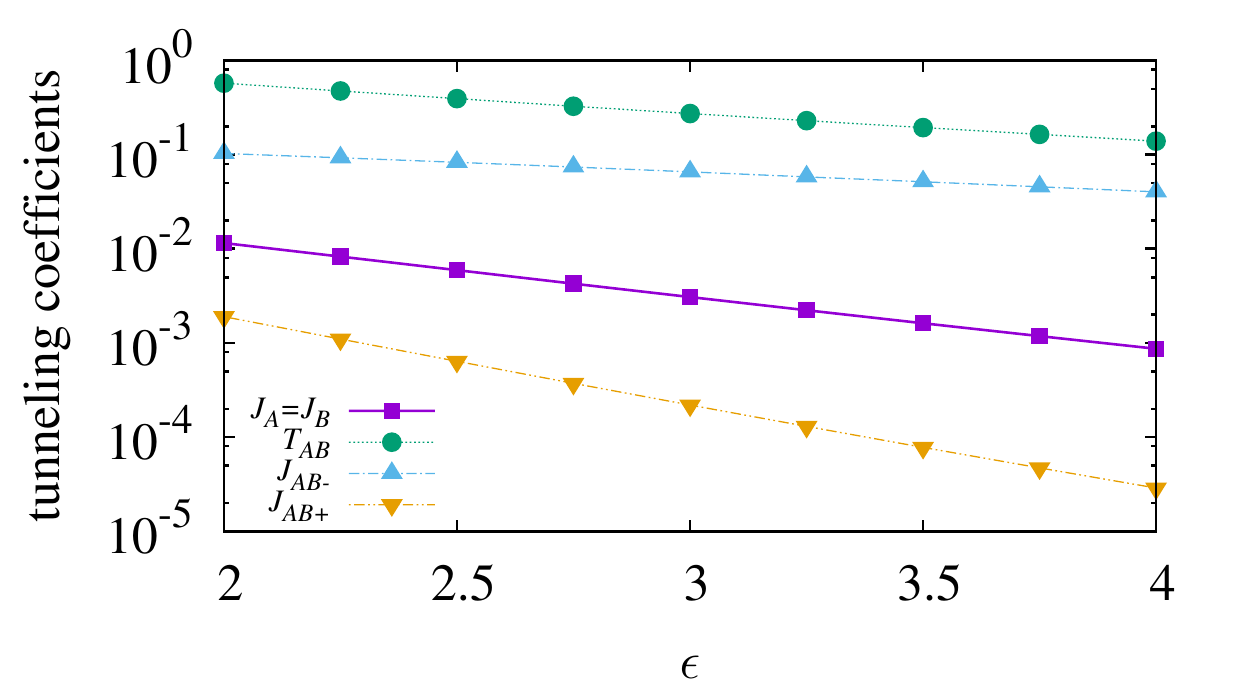}
\caption{(Color online) Plot of the tunneling coefficients (in modulus), as a function of $\epsilon$, for $\theta_{0}=\pi/2$.}
\label{fig:tunnel}
\end{figure}

\textit{Tunneling coefficients.} Let us now turn to the tunneling coefficients. 
They have a weak dependence on $\theta_{0}$, the only notable effect being that for $\theta_{0}=0$ (where all the maxima are degenerate) $T_{AB}=J_{AB-}$, whereas at $\theta_{0}=\pi/2$ we have $J_{A}=J_{B}$ \cite{modugno2012}.
Their behavior as a function of $\epsilon$ is shown in Figure \ref{fig:tunnel}, for $\theta_{0}=\pi/2$. This Figure reveals that the nearest-neighbor approximation improves when $\epsilon$ is increased, as one may expect owing to the increased localization of the MLWFs.

\begin{figure}[H]
\centerline{
\includegraphics[width=0.99\columnwidth]{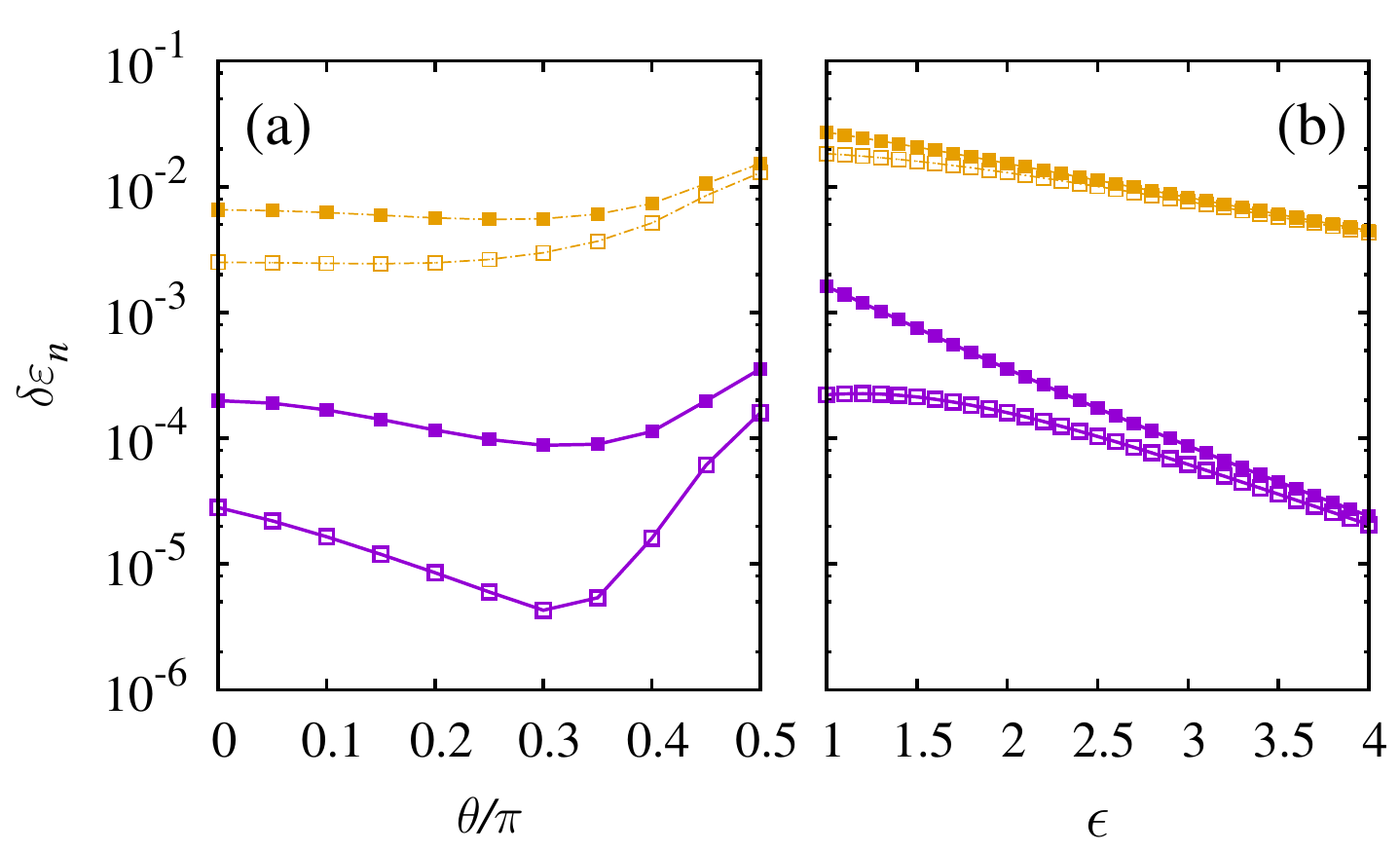}
}
\caption{(Color online) Plot of the quantity $\delta \varepsilon_{n}$ (see text), as a function of $\theta_{0}$ for $\epsilon=2$ (a), and as a function of $\epsilon$ for $\theta_{0}=\pi/2$ (b). Solid line: \textit{extended} tight-binding model; Dotted-dashed line: \textit{nearest-neighbor} approximation. 
Empty squares: first band; Solid squares: second band. }
\label{fig:de}
\end{figure}

\textit{Accuracy.}
As anticipated, the accuracy of a given tight-binding approximation can be measured by the average energy mismatch $\delta \varepsilon_{n}$ defined in eq. (\ref{eq:deltaener}).
This quantity is shown in Figures \ref{fig:de}a,b as a function of $\theta_{0}$ ($\epsilon=2$) and $\epsilon$ ($\theta_{0}=\pi/2$), respectively. These Figures reveal that in the regime $R\lesssim1$ (cfr. Figure \ref{fig:gap}) the \textit{extended} tight-binding model of eq. (\ref{eq:fullHam}) reproduces the exact energies with great accuracy. Instead, the commonly used \textit{nearest neighbor} model is less accurate, and its use may not always be justified.

\subsubsection{``s-p'' resonance} 

\begin{figure}[H]
\centerline{\includegraphics[width=0.8\columnwidth]{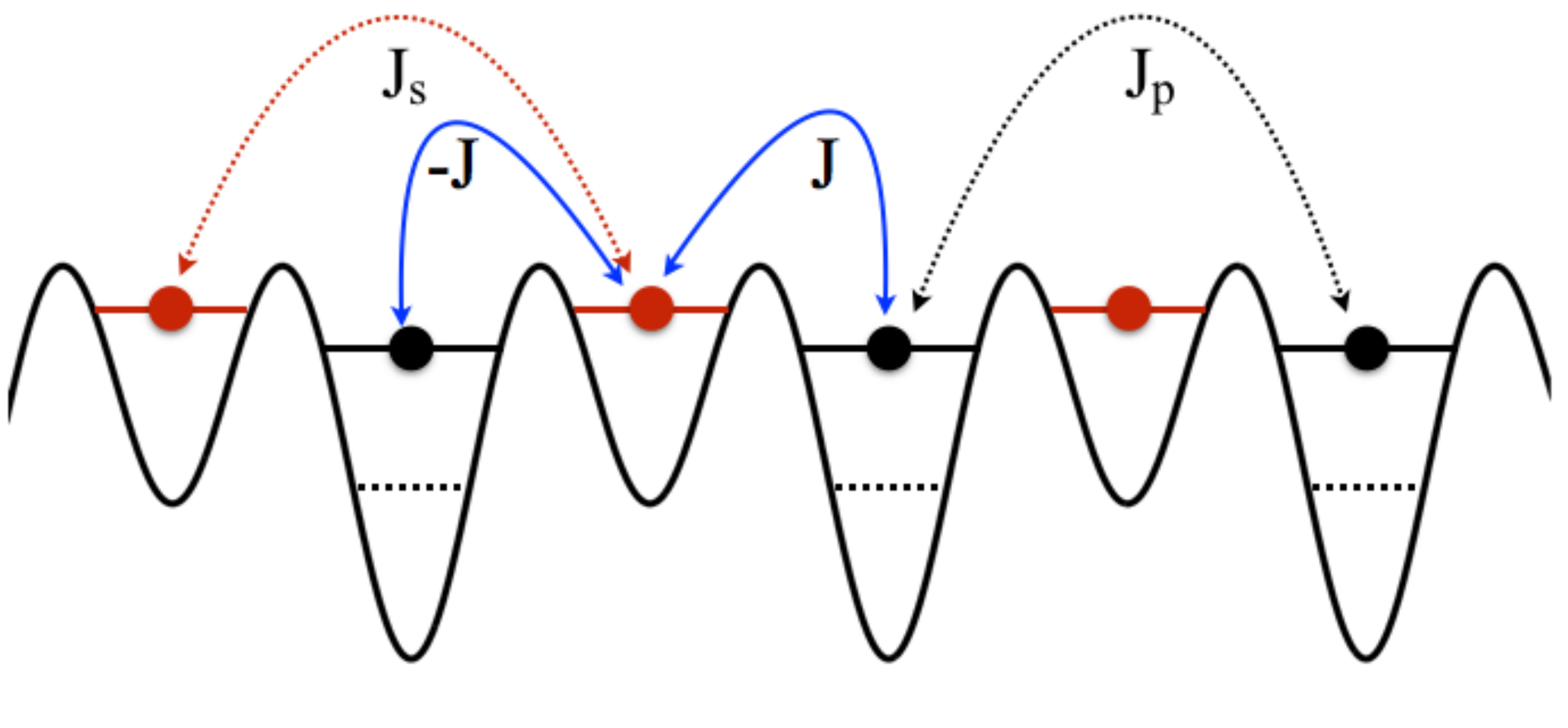}}
\caption{(Color online) A sketch of the various tunneling terms of the extended Bose-Hubbard model in case of $s-p$ resonance. Due to the parity properties of the $p$ orbital, the nearest-neighbor tunneling amplitudes from a given site to the left and to the right have the same magnitude $J$, but opposite sign. We remind that this term is strictly vanishing within the single band approach. 
}
\label{fig:sketch}
\end{figure}

Another interesting situation emerges when the first two excited bands become resonant, a situation that occurs when the $p$-like orbital in the deepest well and the $s$-like orbital in the other one are almost degenerate \cite{wirth2011,ganczarek2014}.
This can be realized e.g. with $\theta_{0}=0$ and $V_{0}=32E_{R}$, for which the $s-p$ resonance occurs at $\epsilon=2$ \cite{ganczarek2014}. 
This configuration is also relevant for the realization of an effective Dirac dynamics \cite{witthaut2011,salger2011,lopez-gonzalez2014}, that will be considered in the following section.
We recall that for $\theta_{0}=0$ all the maxima are degenerate, so that $T_{AB}=J_{AB-}$. Moreover, in this regime the term $J_{AB+}$ can be safely neglected (for $\epsilon\gtrsim1.5$, see later on). For convenience, here it is also natural to change the notation from $A,B$ to $s,p$, as indicated in Figure \ref{fig:sketch}.

\begin{figure}[H]
\centerline{\includegraphics[width=0.9\columnwidth]{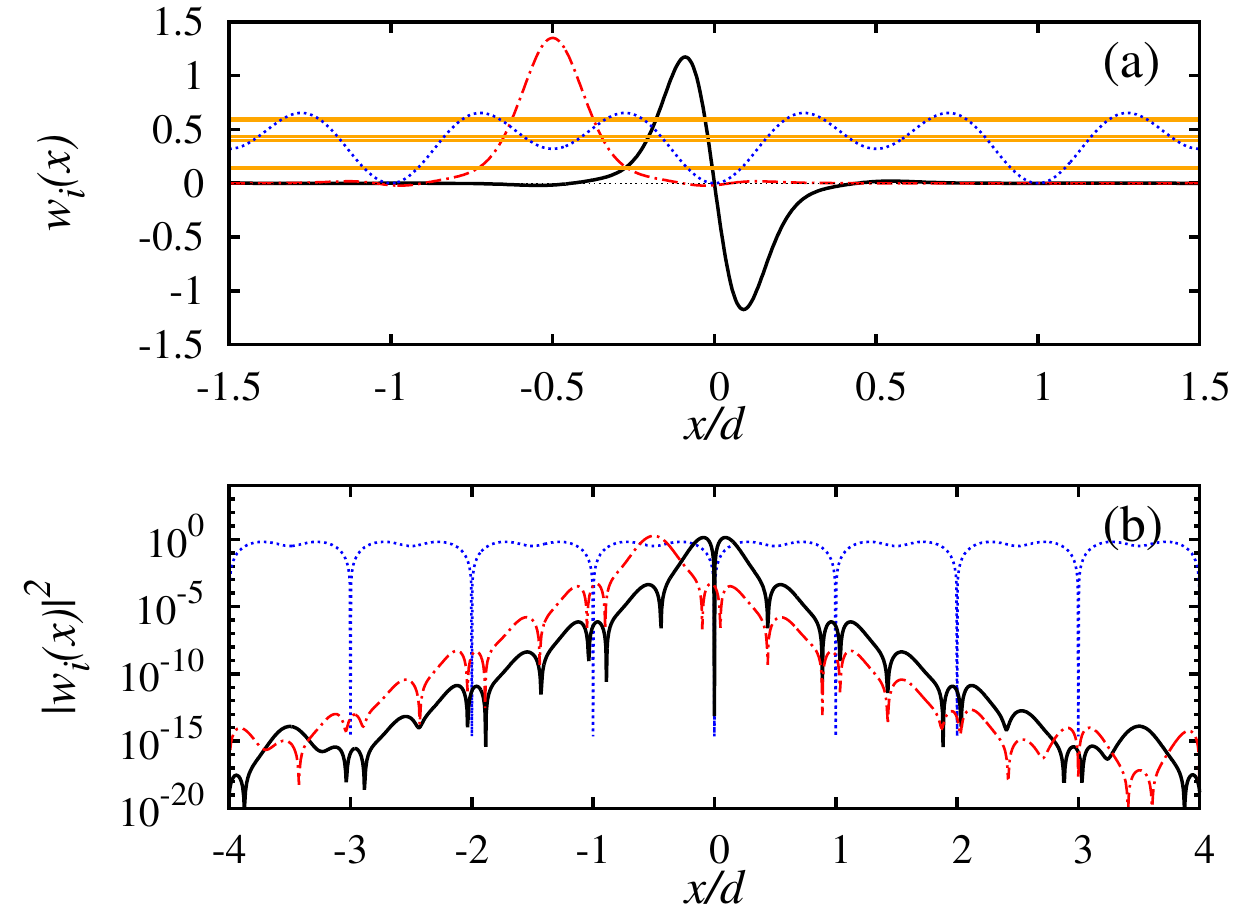}}
\caption{(Color online) (a) Plot of the $s-$ (red dotted dashed line) and $p-$like
 (black solid line) MLWFs for $\epsilon=1.5$.
 The potential is represented by the dotted (blue) line, whereas the horizontal orange stripes represent the lowest four Bloch bands (on the same scale as the potential). (b) The density plot of the same composite-band MLWFs is shown here in
 logarithmic scale. Note the exponential decay of the tails.}
\label{fig:wan-off1}
\end{figure}

\begin{figure}[H]
\centerline{\includegraphics[width=0.9\columnwidth]{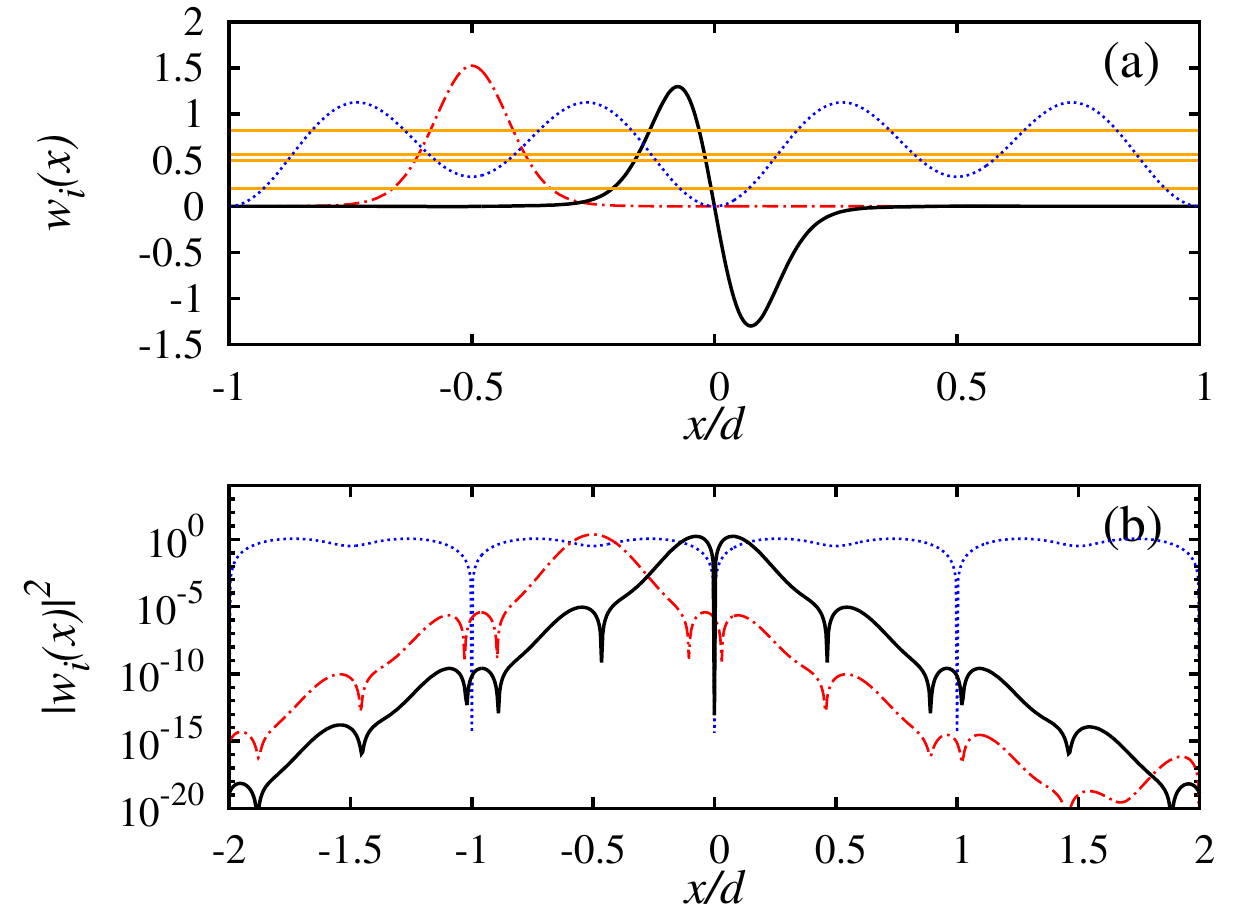}}
\caption{(Color online) Same as above, but for $\epsilon=3$.}
\label{fig:wan-off3}
\end{figure}

\begin{figure}[H]
\centering
\includegraphics[width=0.8\columnwidth]{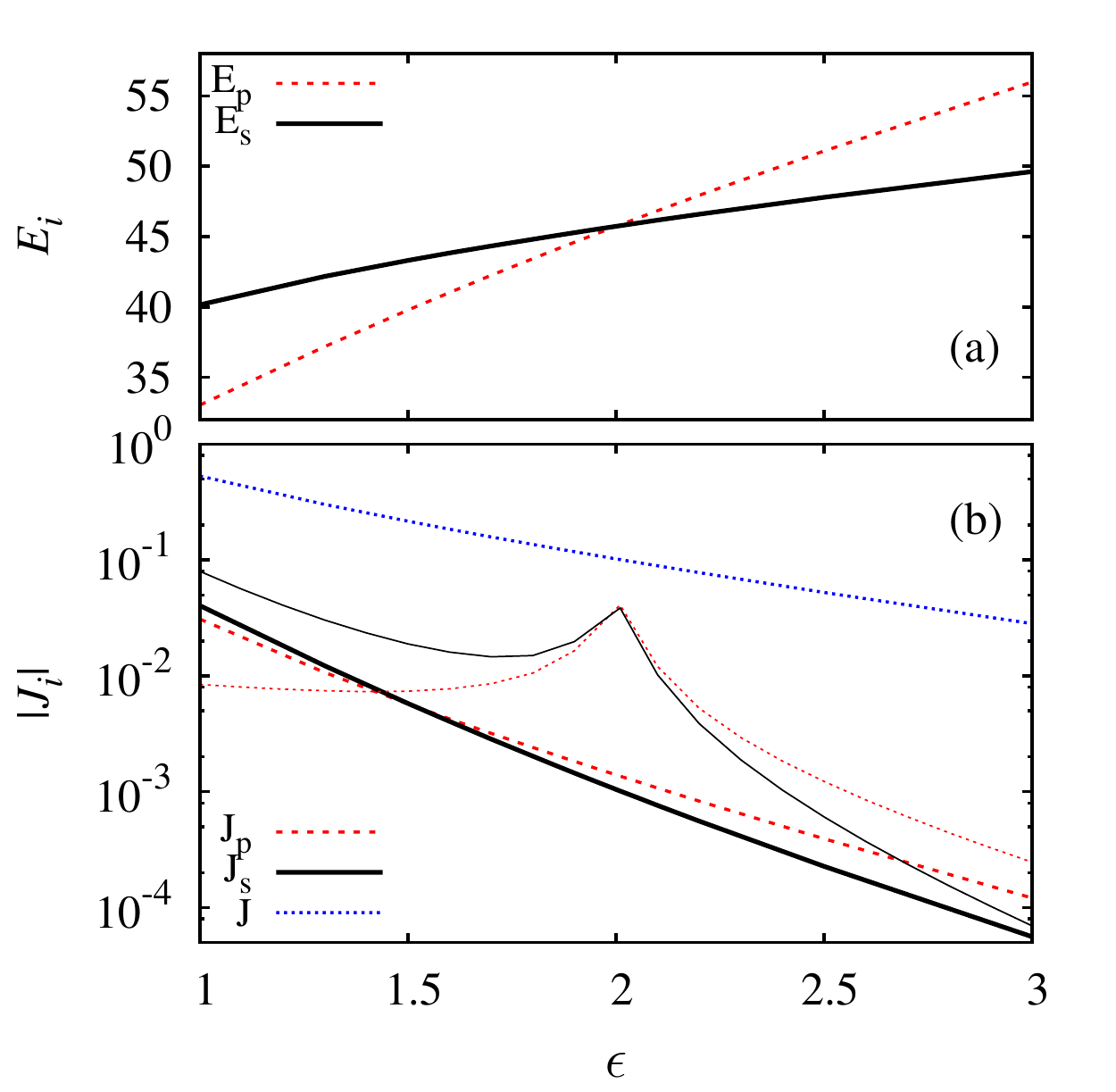}
\caption{(Color online) 
(a) Onsite energies and (b) tunneling amplitudes as a function of $\epsilon$.
The thin lines in panel (b) correspond to the single band values for $J_{s}$ and $J_{p}$ (see text). 
}
\label{fig:tun-32}
\end{figure}

\textit{MLWFs.}
An example of the shape of the MLWFs below and above the resonance is shown in Figures \ref{fig:wan-off1}, \ref{fig:wan-off3} for $\epsilon=1.5$ and $\epsilon=3$ respectively. As anticipated, the MLWFs have the form of an $s$-like state in the shallow well, and a $p$-like state in the deeper one. At $\epsilon=2$ the energies of the two states become degenerate, see Figure \ref{fig:tun-32}a. 
It is also interesting to note that, at resonance, almost optimally localized states can be built from a simple analytic approach that captures the essential features of the system and leads to simple analytic expressions for the tunneling coefficients, see Ref. \cite{ganczarek2014}.

\textit{Tunneling coefficients.}
The behavior of the tunneling amplitudes as a function of $\epsilon$ is shown Figure \ref{fig:tun-32}b.
 They present a monotonic decrease for increasing $\epsilon$, as a consequence of the deepening of the potential wells. Notice that the degeneracy between $J_{s}$ and $J_{p}$ takes place below the resonance, at $\epsilon=1.5$\cite{ganczarek2014}. This figure also shows the behavior of the tunneling coefficients $J_{s}$ and $J_{p}$ obtained from the single band approach (thin lines), for comparison. We recall that the nearest neighbors tunneling amplitude $J$ -- that is the dominant term in the composite band approach (blue dotted line in the Figure) -- is strictly vanishing in this case, i.e. $J\equiv0$.
In fact, the single band approach is expected to be reliable only far from the resonance, where it may be convenient to adopt a picture with the two sublattices $s$ and $p$ being completely decoupled (except for the effect of external forces or interaction terms). 

\begin{figure}[H]
\centering
\includegraphics[width=0.9\columnwidth]{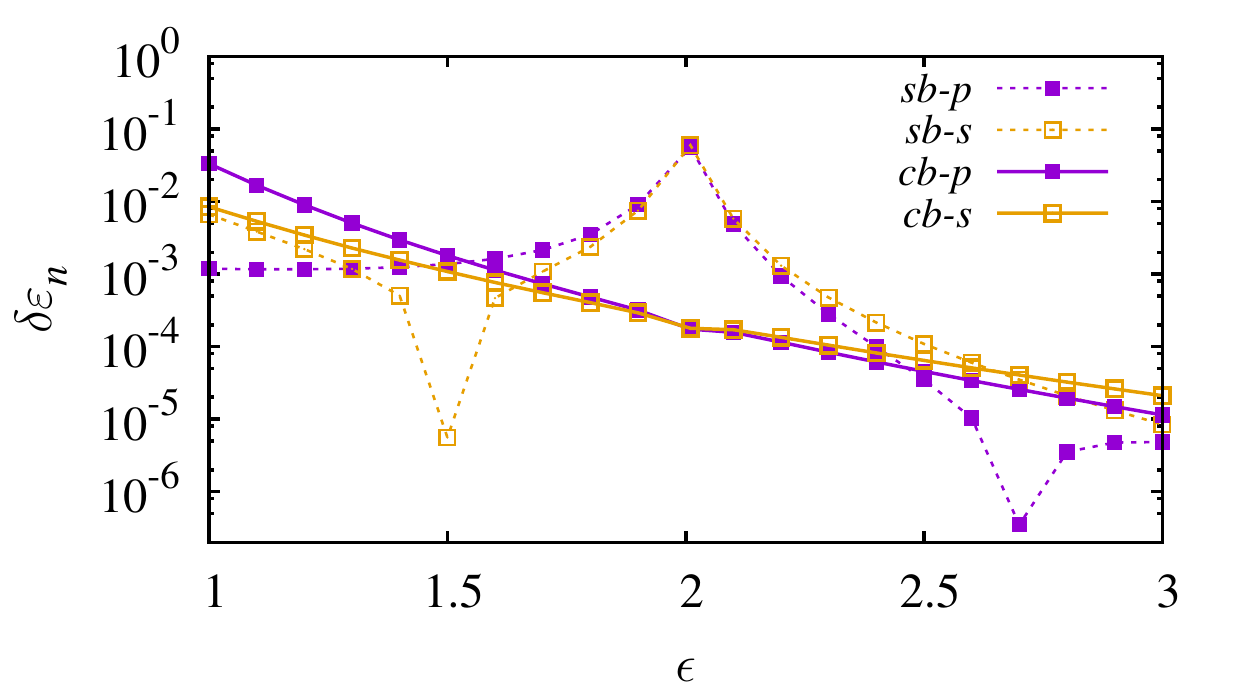}
\caption{(Color online) Plot of $\delta \varepsilon_{n}$ ($n=s,p$, see text) as a function of $\epsilon$, as obtained from the single- and composite-band approaches (indicated in the label as $sb$ and $cb$, respectively).}
\label{fig:enerdiff}
\end{figure}

\textit{Accuracy.}
The accuracy in reproducing the single particle spectrum of the single and composite band approaches are compared in Figure \ref{fig:enerdiff}, where the energy mismatch $\delta \varepsilon_{n}$ (see eq. \ref{eq:deltaener}) is shown as a function of $\epsilon$. The accuracy of the composite-band approach increases monotonically with $\epsilon$, and provides a very good approximation starting from $\epsilon\approx1.5$. For lower values, additional tunneling terms, or even a different band mixing (the fourth band approaches to the third one), may become necessary. As for the single-band approach, it fails in the resonance region, as expected. Remarkably, close to $\epsilon=3$ both the single and composite band approaches provide an accurate description of the system, even though they correspond to very different pictorial representations.

\subsection{Semiclassical dynamics}

The Wannier functions play a relevant role also in the derivation of effective dynamical equations in the semiclassical regime, obtained by a \textit{corse-graining} procedure \cite{morandi2005}. To illustrate this, let us consider the case of a single particle in the presence of a periodic potential $V_L({x})$ (of period $d$) and an additional slowly varying potential $V({x})$. The Schr\"odinger equation reads
\begin{equation}
i\hbar\partial_t\Psi({x},t)=\left[
 H_L({x})+V({x})\right]
\Psi({x},t),
 \label{eq:schrod}
\end{equation}
where $H_L(x)=-(\hbar^2/2M)\nabla^2+V_L(x)$ is the unperturbed lattice Hamiltonian, whose eigenvectors 
are Bloch functions
$\psi_n({k},{x})={\rm
e}^{i{k}{x}}u_n({k},{x}) \equiv
\langle{x}|n,{k}\rangle$.
The above equation can be mapped onto quasimomentum space as
 (see e.g. \cite{callaway1964,morandi2005})
\begin{align}
i\hbar\partial_t\varphi_n({k},t)&=E_n({k})\varphi_n({k},t)
\nonumber\\
&\quad+
\sum_{n'}\int_{k'}\langle
n,{k}|V|n',{k}'\rangle\varphi_{n'}({k}',t)
\end{align}
where $\varphi_n({k},t)$ represent the expansion coefficients of a generic wave-packet $\Psi({x},t)$
on the Bloch basis, namely $\Psi({x},t)=
\sum_n\int_{{k}}\varphi_n({k},t)\psi_n({k},{x})$, and ${k}$ runs over the first Brillouin zone (the dependence on $t$ will be omitted in the following). 
The above equation can be written in vectorial form as 
\begin{equation}
i\hbar\partial_t\underline{\varphi}({k})=H_{L}({k})\underline{\varphi}({k})+
\int_{k'}\tilde{V}({k},{k}')\underline{\varphi}({k}'),
\label{eq:vec}
\end{equation}
with $H_{L}({k})=E_{n}({k})\delta_{nn'}$, $\tilde{V}({k},{k}')=
\langle n,{k}|V|n',{k}'\rangle$.
Let us now consider a subset of two bands. In the following section we shall discuss an effective Dirac dynamics, for which it is convenient to introduce a $SO(2)$ rotation $R(\theta(k))$ \cite{witthaut2011}
\begin{equation}
R(\theta(k))= \begin{pmatrix} \cos\theta(k) & -\sin\theta(k) \\ \sin\theta(k) & \cos\theta(k) \end{pmatrix},
\label{eq:mixing}
\end{equation}
where $\theta(k)$ will be specified later on.
Then, eq. (\ref{eq:vec}) can be written as
\begin{equation}
i\hbar\partial_t \underline{\varphi}'({k})=
H_{L}'({k})\underline{\varphi}'({k})
+\int_{k'}R({k})\tilde{V}({k},{k}') R^{T}({k}')\underline{\varphi}'({k}')
\label{eq:rotated}
\end{equation}
with ${\varphi}'=R\underline{\varphi}$, and 
\begin{equation}
H_{L}'({k})=R({k})H_{L}({k})R^{T}({k})
= \begin{pmatrix} c\hbar k & -m c^{2} \\ -m c^{2} & - c\hbar k \end{pmatrix}.
\label{eq:rotatedH}
\end{equation}
Eq. (\ref{eq:rotated}) can be transformed back in coordinate space 
by projection on a basis of Wannier functions, as discussed in the following. We recall that a
generic wave packet $\Psi({x})$ can be expanded as $\Psi({x})=\sum_{n,i}
\chi_n({R}_i)w_{n}({x}\!-\!{R}_i)$, where the amplitudes $\chi_n({R}_i)$ can be
obtained from the Bloch coefficients by a
simple Fourier transform\footnote{The same relation holds in the rotated basis \cite{modugno2009}.}
\begin{equation}\label{eq:def_env_f}
\chi_n({R}_i)=\sqrt{\frac{d}{2\pi}}
\int_k\varphi_n({k}){\rm e}^{i{k}{R}_i}.
\end{equation}
When the Wannier functions in the \textit{rotated} basis are sufficiently localized in each cell, 
the rotated amplitudes $\chi'_n({R}_i)$ play the role of envelope functions associated to the site ${R}_i$, corresponding to a \textit{corse graining} on the scale of a single cell \cite{adams1952,morandi2005}. 
In general, the coefficients 
$\underline{\chi}'({R}_i)$ can be supposed to be differentiable functions of ${R}_i$, the latter being considered as a continuous variable. This holds when $\underline{\chi}'({R}_i)$ is slowly varying on the scale of the lattice period, namely in case of a ``smooth'' wave packet. Then, by using the properties of the Fourier transform \cite{adams1952,morandi2005}, 
the Hamiltonian $H_{L}'$ in coordinate space can be obtained by the replacement 
${k}\to -i{\nabla}_{{R}_i}$, so that eq. (\ref{eq:rotated}) can be mapped in coordinate space as
\begin{align}
\label{eq:chiprime}
&i\hbar\partial_t \underline{\chi}'({R}_i)=H_{L}'(-i{\nabla}_{{R}_i})\underline{\chi}'({R}_i)
\\
&+\sum_{j}\int_{k}\int_{k'}e^{i{k}\cdot{R}_i}R({k})\tilde{V}({k},{k}')R^{T}({k}')e^{-i{k}'\cdot{R}_j}
\nonumber\underline{\chi}'({R}_j).
\end{align}
In addition, it is easy to show that
\begin{align}
&\left.\int_{k}\int_{k'}R({k})\tilde{V}({k},{k}')R^{T}({k}')e^{i{k}\cdot{R}_i}e^{-i{k}'\cdot{R}_j}\right|_{nn'}
\nonumber\\
&=\int_{x}{w'_{n}}^{*}({x}-{R}_{i})V({x})w'_{n'}({x}-{R}_{j})\equiv \langle V\rangle^{ij}_{nn'},
\label{eq:10}
\end{align}
yielding
\begin{equation}
i\hbar\partial_t \underline{\chi}'({R}_i)=H_{L}'(-i{\nabla})\underline{\chi}'({R}_i)+
\sum_{j}\langle V\rangle^{ij}\underline{\chi}'({R}_j).
\label{eq:fulleq}
\end{equation}
Moreover, if the potential $V(x)$ is slowly varying on the lattice scale, one has 
\begin{equation}
\langle V\rangle^{ij}_{nn'}
\approx V({R}_{i})\delta_{nn'}\delta_{ij},
\label{eq:V-approx}
\end{equation}
so that we eventually get ($R_{i}\to x$)
\begin{equation}
i\hbar\partial_t \underline{\chi}'(x)=\left[ H_{L}'(-i{\nabla})+V(x)\right]
\underline{\chi}'(x).
\label{eq:chi}
\end{equation}

In order to check how the approximation (\ref{eq:V-approx}) behaves under rotation, one can perform a series expansion around $x=R_{j}$, yielding the following result for the first order correction \cite{ganczarek2014}
\begin{equation}
\frac{\delta V_{nn'}^{(\ell)}(R_{j})}{E_{R}}\approx\frac{d}{E_{R}}\left.\frac{\partial V}{\partial x}\right|_{R_{j}}\!\!\!\cdot\frac{1}{d}\left(\langle x\rangle^{(\ell)}_{nn'}-R_{j}\delta_{nn'}\delta_{\ell0}\right),
\end{equation}
with $l=j'-j$. The term $(d/E_{R})(\partial V/\partial x)|_{R_{j}}$ represents the variation of the potential on the scale of the lattice spacing $d$, divided by the characteristic energy scale $E_{R}$ of the lattice. Since this term is small under the assumption of a slowly varying potential, one has to check that the remaining term, $\Delta^{(\ell)}_{nn'}\equiv(\langle x\rangle^{(\ell)}_{nn'}-R_{j}\delta_{nn'}\delta_{\ell0})/d$, 
is sufficiently smaller than unity. 
In principle, this condition is expected to be satisfied when the Wannier functions are sufficiently localized within each lattice cell.

At this point we remark that in the presence of a constraint fixing the mixing angle $\theta(k)$, the general approach for defining the MLWFs cannot be applied. In fact, in this case the only freedom left is the choice of the phases of the single band Bloch functions before the rotation, that should be determined in order to minimize the diagonal\footnote{The the off diagonal spread is fixed by the rotation $R(\theta)$.} spread $\tilde{\Omega}_{D}$ in eq. (\ref{eq:ann}), for a gauge transformation of the form
$U(k)=R(\theta(k))\times\textrm{diag}(e^{i\phi_{1}(k)},e^{i\phi_{2}(k)})$.
Unfortunately, in general this results in complicated integro-differential expression, whose solution is not viable. So, a minimal approach that one may adopt consists in starting with the single band MLWFs for the original Bloch bands, and verify that the rotation (\ref{eq:mixing}) does not affect substantially their localization properties.

\textit{Effective Dirac equation.}
As an application, here we consider the case of an $s-p$ resonance between the second and third Bloch band, that gives rise to an effective Dirac dynamics owing to the ``relativistic'' form of the dispersion relation around $k=0$, $E_{\pm}(k)=\pm\sqrt{m^{2} c^{4} + c^{2}(\hbar k)^{2}}$ \cite{witthaut2011,salger2011,lopez-gonzalez2014}. By choosing the rotation angle as\footnote{This corresponds to an inverse free-particle Foldy-Wouthuysen transformation in the momentum representation \cite{bjorken1964,greiner2000,lopez-gonzalez2014}.} 
\begin{equation}
\tan\theta(k)=-\frac{m c^{2}}{c\hbar k + \sqrt{m^{2} c^{4} + c^{2}(\hbar k)^{2}}}
\label{eq:tantheta}
\end{equation}
and applying the $U(2)$ transformation 
\begin{equation}
U=\frac{1}{\sqrt{2}}\begin{pmatrix} 1 & -1 \\ 1 & 1 \end{pmatrix}
\label{eq:U}
\end{equation}
to the vector $\underline{\chi}'$, $\underline{\psi}\equiv U\underline{\chi}'$, eq. (\ref{eq:chi}) becomes
\begin{equation}
i\hbar\partial_t \underline{\psi}(x)= \begin{pmatrix} V(x) + m c^{2} & c \hat{p} \\ c \hat{p} & V(x) - m c^{2} \end{pmatrix}\underline{\psi}(x),
\label{eq:simildirac}
\end{equation}
corresponding to the canonical form of the Dirac equation in $1+1$ dimensions, in the presence of a scalar potential $V(x)$ \cite{witthaut2011}.

\begin{figure}[H]
\centerline{\includegraphics[width=0.99\columnwidth]{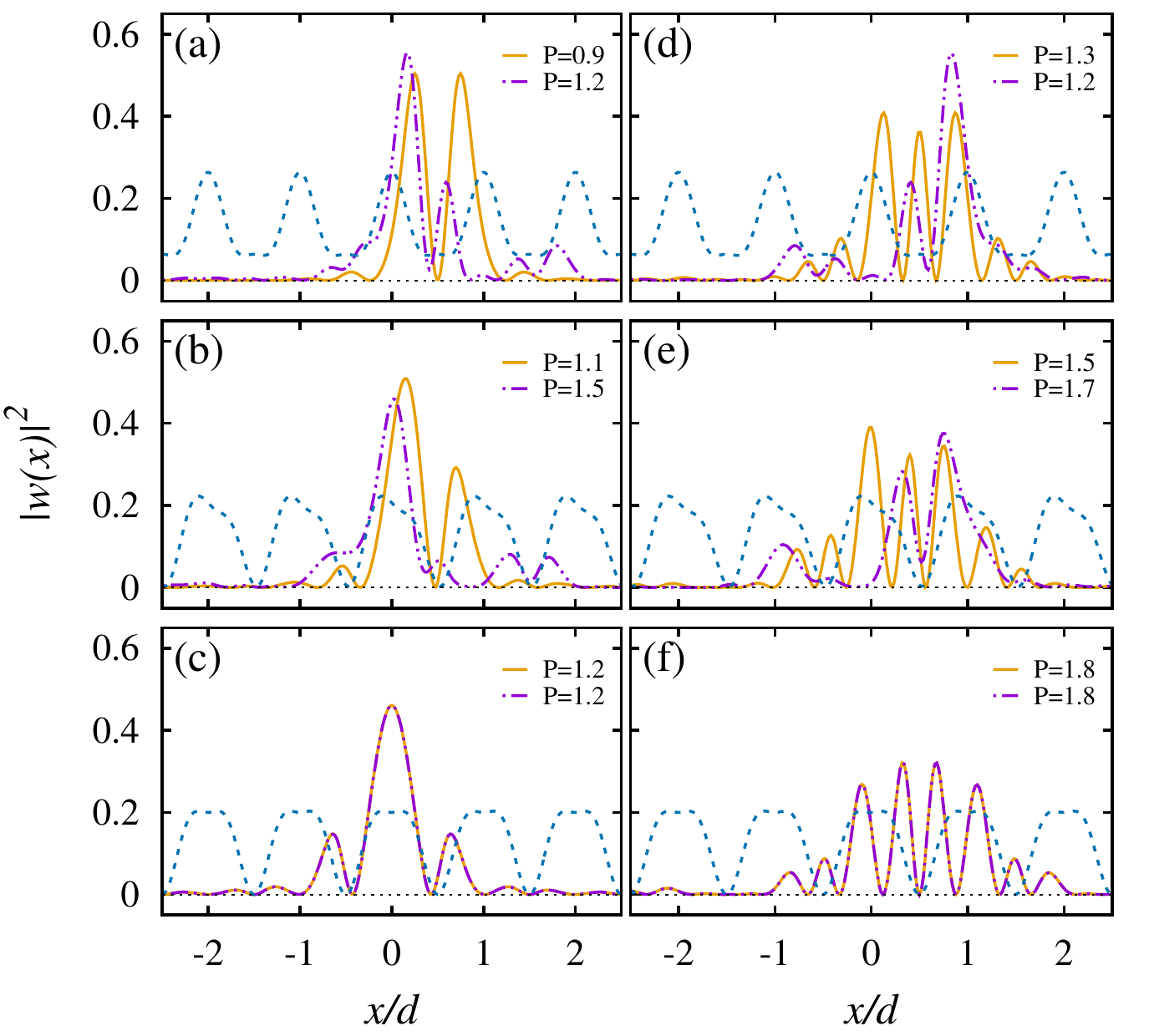}}
\caption{(Color online) Plot of the density of the Wannier functions, for the first and second excited bands (left and right, respectively) and $\phi=0,0.8\pi,\pi$ (from top to bottom). The MLWFs for the original Bloch bands are shown  as solid line, those rotated as dotted-dashed line; the dotted line represents the lattice potential.
At resonance ($\phi=\pi$), the mass $m$ vanishes, and there is no rotation, $R=I$.
 The numbers in the legend correspond to the values of the participation ratio $P$ (see text).}
\label{fig:mlwf-s}
\end{figure}

\textit{The case of Ref. \cite{salger2011}.} As a specific example we consider the case of the experiment in Ref. \cite{salger2011}, namely $V_{0}=-5E_{R}$, $\epsilon=1.6$. The single-band MLWFs, and the corresponding \textit{rotated} MLWFs\footnote{The parameters $mc^{2}$ and $c$ are obtained from a fit of the energy dispersion around $k=0$ \cite{witthaut2011,lopez-gonzalez2014}.} are shown in Figs. \ref{fig:mlwf-s}, for different values of the angle $\phi$.
This Figure shows that the rotation does not affect dramatically their localization properties, the rotated Wannier functions having a behavior similar to the MLWFs for the original Bloch band. As a matter of fact, though the two sets of Wannier functions have a different ÒmicroscopicÓ structure, the corresponding values of the participation ratio $P=\left(d\int dx |w_{n}|^{4}\right)^{-1}$ -- that measure of the extent of the Wannier functions $w_{n}(x)$, in units of the lattice period $d$ -- do not differ too much. Finally, for the ``slowly varying'' potential used in Refs. \cite{salger2011,witthaut2011}, $V(x)=V_{0}\exp[-2(x/x_{0})^{2}] -Fx$, it is possible to verify that 
$|\Delta^{(\ell)}_{nn'}|<0.5$ for $\ell=0,\pm1,\pm2$ for all the cases in Figs. \ref{fig:mlwf-s}. 

\section{Two dimensional honeycomb lattices}
\label{sec:twod}

\begin{figure}[H]
\centerline{\includegraphics[width=0.48\columnwidth]{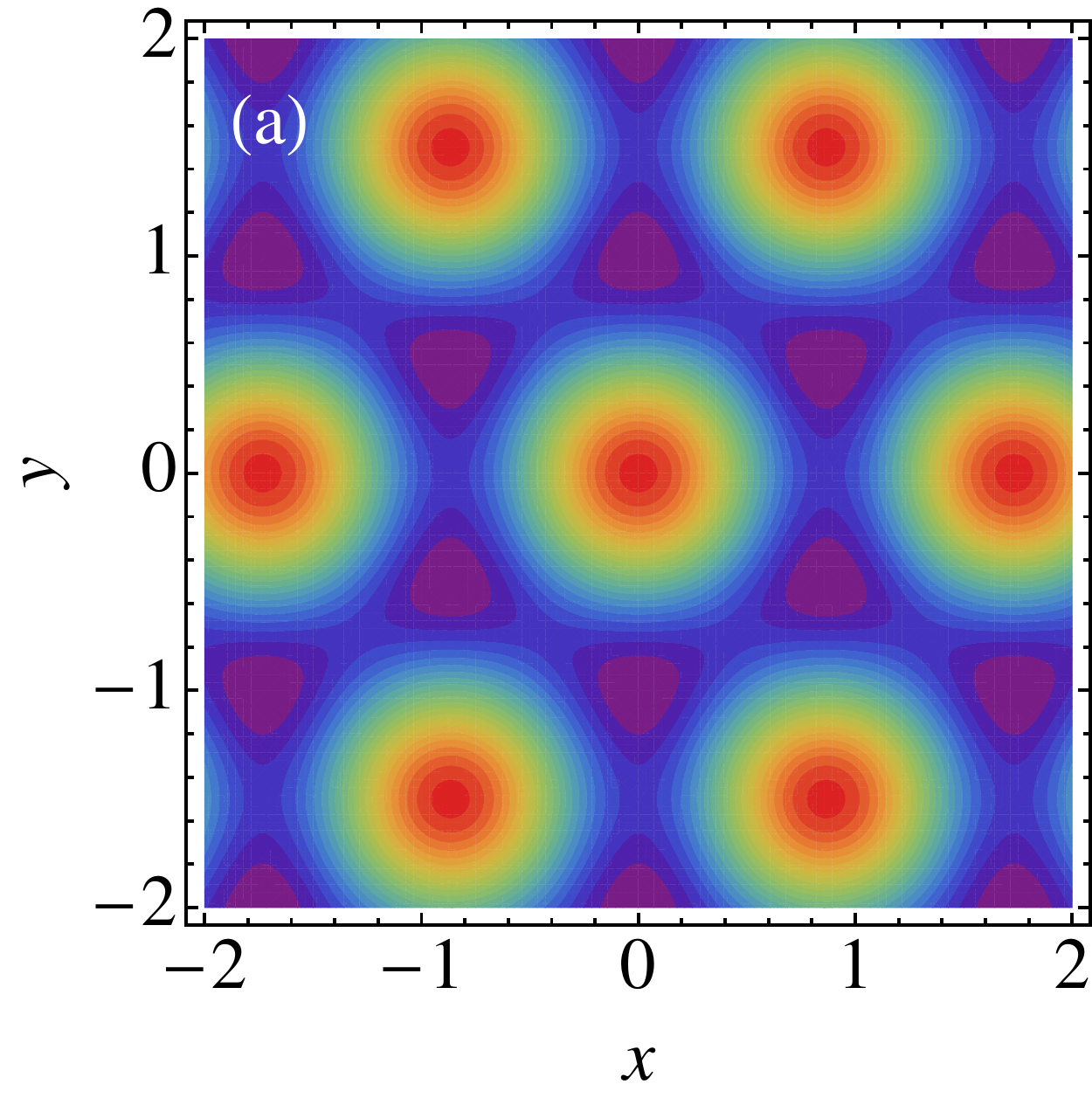}
\includegraphics[width=0.52\columnwidth]{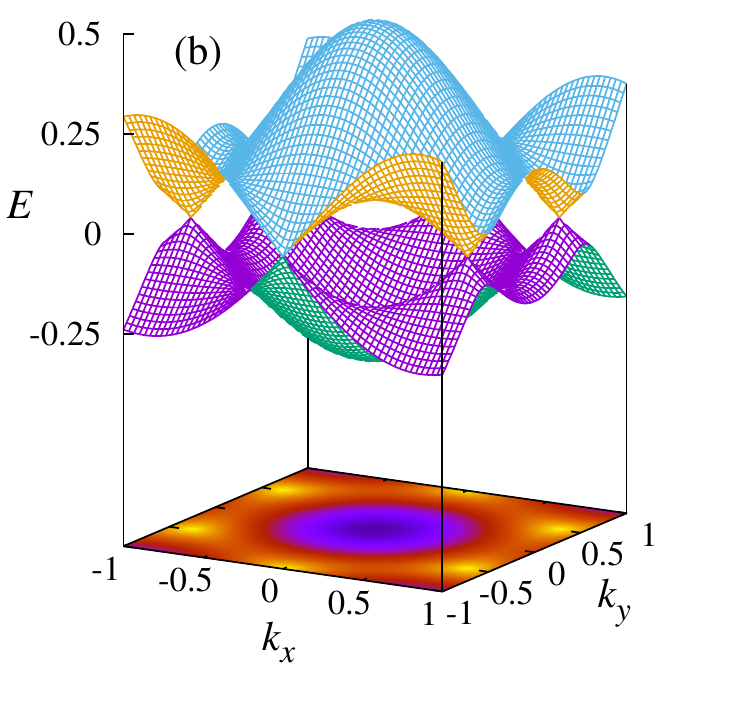}}
\caption{(Color online) (a) Density plot of the honeycomb potential in eq. (\ref{eq:pot}) for $\chi_{A}=0$. Hot and cold colors correspond to maxima and minima of the potential, respectively. 
(b) Bloch spectrum $E_{\pm}(\bm{k})$ of the lowest two bands for $s=5$. The Dirac points (where the two bands are degenerate) represents the vertices of the first Brillouin zone (a regular hexagon). Lengths in units of $k_L^{-1}$, momenta in units of $k_L$, and energies in units of $E_R$.}
\label{fig:honeycombpot}
\end{figure}

\begin{figure}[H]
\centerline{\includegraphics[width=0.9\columnwidth]{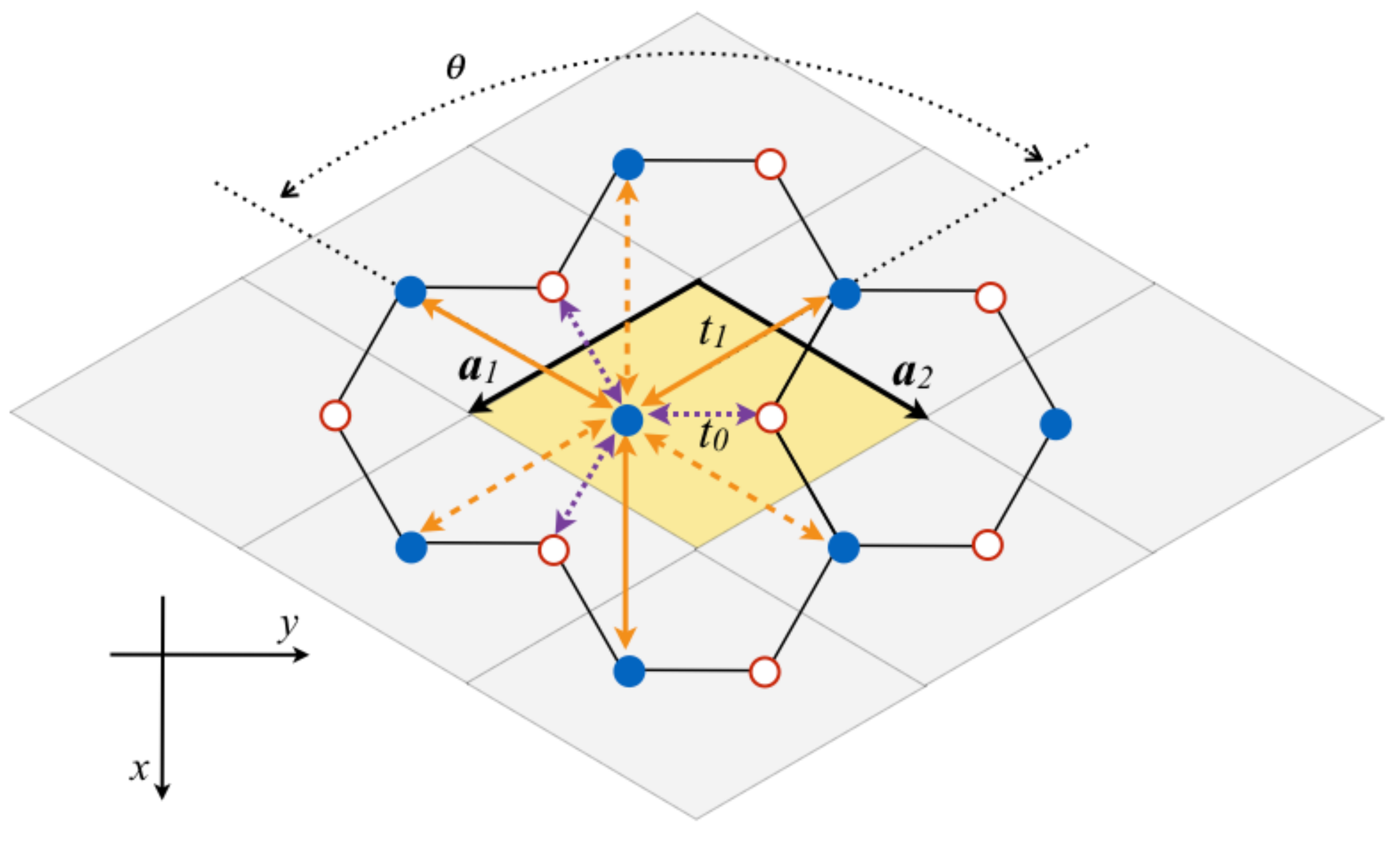}}
\caption{(Color online) Bravais lattice associated to the honeycomb potential in eq. (\ref{eq:pot}). Filled and empty circles refer to minima of type $A$ and $B$, respectively. The elementary cell is highlighted in yellow. The tunneling coefficients up to next-to-nearest neighbors ($t_{0}$, $t_{1}$) are indicated for the site of type $A$ in the central cell. The length of each side of the hexagon is $a=4\pi/(3\sqrt{3}k_{L})$.
The system is invariant under discrete translations generated by the Bravais vectors $\bm{a}_{1/2}$ and under rotations by $\theta=2\pi/3$ radians around any vertex of the lattice.
The former implies that next-to-nearest tunneling amplitudes $t_{1}$ along the same direction are conjugate pairs (solid and dashed lines); from the latter follows the equivalence of the hopping amplitudes separated by 
$2\pi/3$ radians. When the sites $A$ and $B$ are degenerate, the system is also invariant 
under rotations by $\pi$ radians around the center of any elementary cell; this implies that $t_{0}$ is real.}
\label{fig:honeycomb}
\end{figure}

Ultracold atoms in honeycomb optical lattices have been the subject of intense research activity in recent years, due to their analogies with graphene, and the possibility of simulating the physics of Dirac points\cite{zhu2007,wu2007,wunsch2008,lee2009,montambaux2009,montambaux2009b,stanescu2009,soltan-panahi2011,sun2012,tarruell2012,lim2012,hasegawa2012,uehlinger2013a,ibanez-azpiroz2013,ibanez-azpiroz2014}.
Honeycomb lattices can be generated by a potential of the form \cite{lee2009,ibanez-azpiroz2013,ibanez-azpiroz2014} 
\begin{align}
\label{eq:pot}
V_{L}(\bm{r})=&2sE_R\Big\{\cos\left[(\bm{b}_{1}-\bm{b}_{2})\cdot\bm{r}\right] 
\\
& 
+ \cos\left(\bm{b}_{1}\cdot\bm{r}-\frac{\pi}{3}\chi_{A}\right)
 +\cos\left(\bm{b}_{2}\cdot\bm{r}\right)\Big\}
 \nonumber
\end{align}
with $\bm{r}=(x,y)$, $\bm{b}_{1/2}=({\sqrt{3}}k_{L}/{2}) (\sqrt{3}\bm{e}_{x}\mp{\bm{e}}_{y})$, and $\chi_{A}$ being a phase associated to the breaking of parity. The dimensionless parameter $s$ represents the amplitude of the potential in units of the recoil energy $E_{R}$. This potential is characterized by two minima per unit cell, arranged at the vertices of a regular honeycomb, see Figure \ref{fig:honeycombpot}a. 
The Bravais lattice is generated by the two basis vectors 
$\bm{a}_{1/2} =({2\pi}/{3k_{L}}) (\bm{e}_{x},\mp\sqrt{3}\bm{e}_{y})$, obeying $\bm{a}_i\cdot\bm{b}_j=2\pi\delta_{ij}$, with a diamond-shaped elementary cell with basis $A$ and $B$, as shown in Figure \ref{fig:honeycomb}.
The vectors $\bm{b}_{1/2}$ generate the corresponding reciprocal space, whose first Brillouin zone is a regular hexagon as well.
When $\chi_{A}=0$ the two minima are degenerate, and the spectrum is characterized by Dirac points at the six vertices $\bm{k_D}$ of the Brillouin zone, where the two lowest bands $E_{\pm}(\bm{k})$ are degenerate, and their local dispersion is linear (corresponding to relativistic particles with vanishing mass) \cite{lee2009}, see Figure \ref{fig:honeycombpot}b. These points are defined by $z(\bm{k}_{D})=0$ (see eq. (\ref{eq:degeneratecase}))\footnote{This is valid at any order of the tight-binding expansion.} and come in pairs in the presence of time-reversal invariance, that implies $z^{*}(\bm{k}_{D})=z(-\bm{k}_{D})$ \cite{montambaux2009,montambaux2009b,hasegawa2012}.

\textit{The tight-binding model.}
When one considers MLWFs as basis functions, one can restrict the tight-binding expansion up to next-to-nearest neighbors, corresponding to the tunneling coefficients $t_{0}$ and $t_{1}$ shown in Figure \ref{fig:honeycomb}. In fact, the analysis of Ref. \cite{ibanez-azpiroz2013} proves that the successive terms are negligible  for $s>3$. 
The tight-binding Hamiltonian reads
\begin{equation}
{\hat{\cal{H}}}_0 = \sum_{\nu=A,B}\sum_{\bm{j}}E_{\nu}\hat{n}_{\bm{j}\nu}
+ 
t_{0}\sum_{\bm{j}}\left({\hat{a}}^{\dagger}_{\bm{j}A}{\hat{a}}_{\bm{j}B}+h.c.\right)
+
t_{1}\sum_{\nu=A,B}\sum_{\langle\bm{j,j'}\rangle}{\hat{a}}^{\dagger}_{\bm{j}\nu}{\hat{a}}_{\bm{j}'\nu}
\end{equation}
where both $t_{0}$ and $t_{1}$ can be chosen to be real \cite{ibanez-azpiroz2014,ibanez-azpiroz2015}.
 From the general definitions in eqs. (\ref{eq:epsnu}), (\ref{eq:zetanu}),  we have
\begin{align}
f(\bm{k})&=t_{1}\left[2 \cos\left(\bm{k}\cdot(\bm{a}_{1}+\bm{a}_{2})\right) + 2\sum_{i=1,2}\cos\left(\bm{k}\cdot\bm{a}_{i}\right)\right]\equiv t_{1}F(\bm{k})
\\
z(\bm{k})&=t_{0}\left(1+e^{i\bm{k}\cdot\bm{a}_{1}}+e^{-i\bm{k}\cdot\bm{a}_{2}}\right)\equiv t_{0}Z(\bm{k}).
\label{eq:zzero}
\end{align}

For the specific case of degenerate minima (the general case will be considered in the next section) the expression for the tight-binding spectrum follows from eq. (\ref{eq:tbenergies})
\begin{equation}
\bar{\epsilon}_{\pm}(\bm{k})=t_{1}F(\bm{k})\pm |t_{0}Z(\bm{k})|.
\label{eq:enerdeg}
\end{equation}
Then, it is possible to express $t_{0}$ and $t_{1}$ as \begin{align}
\label{eq:t0eq}
t_{0}&=(\bar{\epsilon}_{+}(\bm{0})-\bar{\epsilon}_{-}(\bm{0}))/6
\\
t_{1}&=(\bar{\epsilon}_{+}(\bm{0})+\bar{\epsilon}_{-}(\bm{0}))/18,
\label{eq:t1eq}
\end{align} 
where the value of $\bar{\epsilon}_{+}(\bm{0}=\bm{0})$ can be easily extracted from the numerical computation of the Bloch spectrum at $\bm{k}=\bm{0}$. This is an extremely effective method, that provides a remarkable agreement with the prediction of the ab-initio approach based on the MLWFs, as discussed at the end of sect. \ref{sec:mlwf}. 

\begin{figure}[H]
\centerline{\includegraphics[width=0.9\columnwidth]{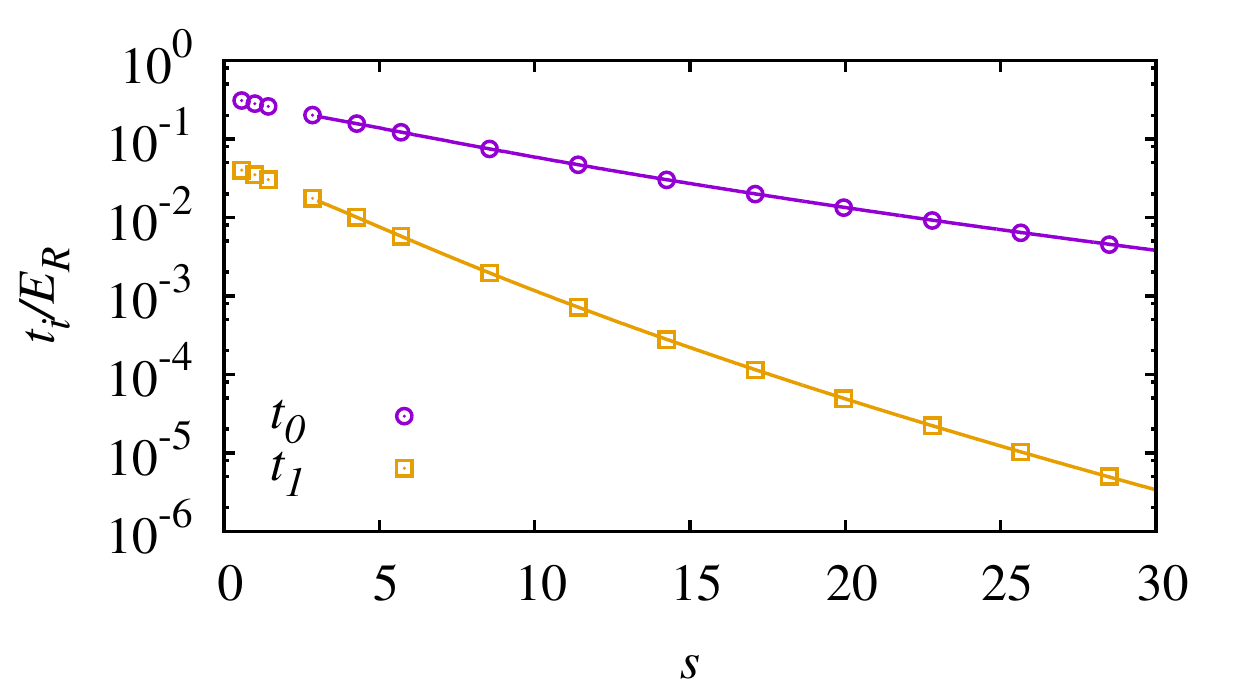}}
\caption{(Color online) Behavior of the tunneling coefficients as a function of the lattice amplitude $s$. The lines are the result of a fit of the numerical data, and coincide with the values extracted from the Bloch spectrum (points), see eqs. (\ref{eq:t0eq}),(\ref{eq:t1eq}).}
\label{fig:tunnel1}
\end{figure}

In Figure \ref{fig:tunnel} we plot the tunneling coefficients as a function of the lattice amplitude\footnote{Notice that for this specific case, $\chi_{A}=0$, $s$ is the only free parameter of the system.
} $s$. Their behavior can be modeled with an analytic expression of the form $t_{i}/E_R= A s^{\alpha} e^{-\beta\sqrt{s}}$ ($i=0,1,2$), where $A$, $\alpha$, and $\beta$ are parameters
to be extracted from a numerical fit. One has \cite{ibanez-azpiroz2013}
\begin{align}
t_{0}/E_R&= 1.16 s^{0.95} e^{-1.634\sqrt{s}},
\\
t_{1}/E_R &= 0.78 s^{1.85} e^{-3.404\sqrt{s}},
\end{align}
corresponding to the lines in Figure \ref{fig:tunnel1}. 
As shown in Ref. \cite{ibanez-azpiroz2013} these estimates permit to reproduce the exact spectrum with great accuracy\footnote{Note that the inclusion of $t_{1}$ is crucial for reproducing the band asymmetry in the low $s$ regime.}, with the energy mismatch $\delta\varepsilon_{n}$ in eq. (\ref{eq:enerdeg}) being well below $1\%$ for $s>3$. 

\subsection{The Haldane model}
\label{sect:haldane}

By adding a microscopic vector potential $\bm{A}$
to the configuration considered in the previous case,
it is possible to realize the celebrated Haldane model\cite{haldane1988,shao2008}.
For such purpose, the vector potential must have the same periodicity of the main lattice and zero magnetic flux through each unit cell \cite{haldane1988}.
This model, originally proposed for electrons in a two-dimensional crystal lattice, describes a Chern insulator~\cite{qahe}, characterized by the presence of the quantum Hall effect (QHE)~\cite{qhe} in the absence of a macroscopic magnetic field. In this case, the tunneling amplitude $t_{1}$ acquires a complex phase due to the breaking of time-reversal symmetry. Interestingly, an effective experimental realization of the Haldane model has been recently reported in \cite{jotzu2014}.

In order to derive the Haldane model from first principles \cite{ibanez-azpiroz2014,ibanez-azpiroz2015}, let us consider the minimal-coupling Hamiltonian 
\begin{equation}
\hat{H_{0}}=\frac{1}{2m}\left[\hat{\bm{p}}-\bm{A}(\bm{r})\right]^{2} + V_{L}(\bm{r})
\label{eq:hamiltonian}
\end{equation}
with $V_{L}(\bm{r})$ being the honeycomb potential in eq. (\ref{eq:pot}). The vector potential $\bm{A}(\bm{r})$ is assumed to have the same periodicity of $V_{L}(\bm{r})$, with the flux across the unit cell of the corresponding magnetic field $\bm{B}=\nabla\times \bm{A}$  being null.
As a specific realization, we adopt the Coulomb gauge, $\nabla\!\cdot\!\bm{A}(\bm{r})=0$, and consider the case of Refs. \cite{shao2008,stanescu2009} 
\begin{align}
&\bm{A}(\bm{r})=\alpha k_{L}\left[\left(\sin((\bm{b}_{2}-\bm{b}_{1})\!\cdot\!\bm{r}) 
+ \frac12\sin(\bm{b}_{2}\!\cdot\!\bm{r})
-\frac12\sin(\bm{b}_{1}\!\cdot\!\bm{r})\right)\bm{e}_{x}
\right.
\nonumber
\\
&\qquad\qquad\qquad
\left.
-\frac{\sqrt{3}}{2}\left(\sin\left(\bm{b}_{1}\!\cdot\!\bm{r}\right)
+\sin\left(\bm{b}_{2}\!\cdot\!\bm{r}\right)\right)\bm{e}_{y}\right],
\end{align}
shown in Figure \ref{fig:honeycomb}. The parameter $\alpha$ represents the amplitude of the vector potential in units of $ k_{L}$. 

\begin{figure}[H]
\centerline{\includegraphics[width=0.5\columnwidth]{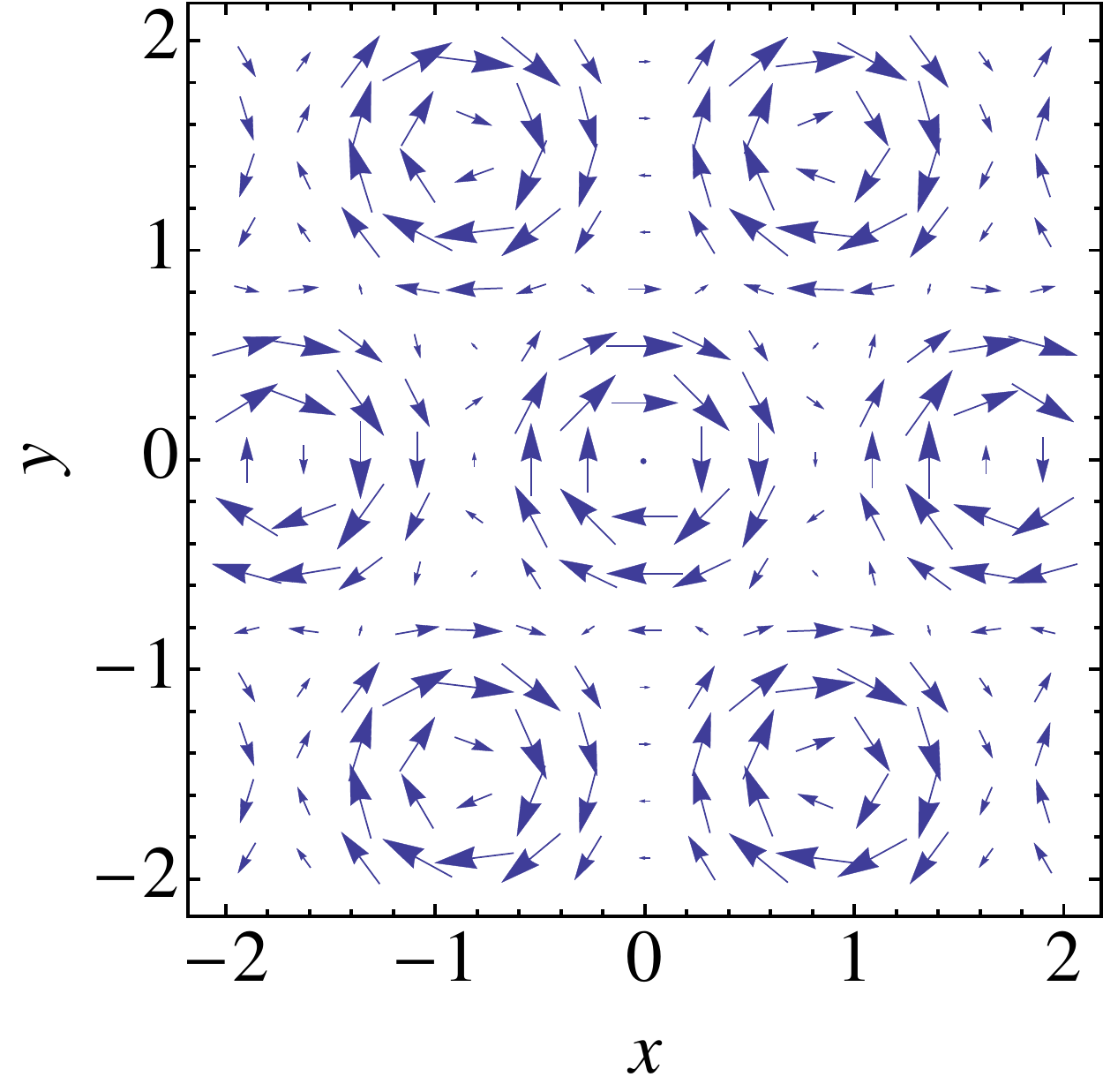}}
\caption{(Color online) Structure of the vector potentials $\bm{A}$.  Lengths in units of $k_L^{-1}$}
\label{fig:honeycombv}
\end{figure}

\textit{The tight-binding model.}
The Haldane model is obtained by considering a tight-binding expansion up to next-to-nearest neighbors, as in the previous section. Even in the presence of the vector potential, the nearest-neighboring tunneling coefficients are all equal and can be chosen to be real\footnote{Although the Wannier functions $w_{\bm{j}\nu}(\bm{r})$ are complex, $t_{0}$ can be chosen real by means of a suitable global gauge fixing.}. Instead, the
next-to-nearest tunneling coefficients acquire a complex phase, one for each site type ($\nu=A,B$), namely $t^{\pm}_{1\nu}=|t_{1\nu}|e^{\pm i\varphi_{\nu}}$ ($j = 1,2,3$). These properties follow from the symmetries of the microscopic Hamiltonian, see Figure \ref{fig:honeycomb}, and no additional hypothesis (like the use of the Peierls substitution \cite{haldane1988,shao2008}) are required \cite{ibanez-azpiroz2014,ibanez-azpiroz2015}.
Then, the expression for $Z(\bm{k})$ is that of eq. (\ref{eq:zzero}), whereas $F_{\nu}(\bm{k})$ becomes 
\begin{align}
F_{\nu}(\bm{k})
&=2\cos\left[\bm{k}\!\cdot\!(\bm{a}_{1}+\bm{a}_{2})
+\varphi_{\nu}\right]+
2\sum_{i=1,2}\cos\left(\bm{k}\!\cdot\!\bm{a}_{i}-\varphi_{\nu}\right).
\end{align}
In addition, in order to recover the original model proposed by Haldane \cite{haldane1988},
some approximations are needed \cite{ibanez-azpiroz2015}. In particular, one should pose $|t_{1A}|=|t_{1B}|\equiv|t_{1}|$, and $\varphi_A=-\varphi_B\equiv\varphi$. Both assumptions are reasonable in the tight-binding regime, as we shall see later on.

\textit{Dirac Points.} As discussed in the previous section, for $\alpha=0$ and $\chi_A=0$ the two lowest energy bands are degenerate at the Dirac points, located at $\bm{k^{\pm}}_D=\pm(1,0)k_{L}$\footnote{In the full model, where in case of parity breaking it is $|t_{1A}|\neq|t_{1B}|$ , the position of the Dirac points may be slightly shifted from $\pm(1,0)k_{L}$ \cite{ibanez-azpiroz2015}. }. 
In general, when the time-reversal or the inversion symmetry are broken ($\alpha\neq 0,\chi_A\neq 0$) two inequivalent gaps open at $\bm{k^{+}}_D$ and $\bm{k^{-}}_D$, see eq. (\ref{eq:tbenergies})
\begin{equation}
 \delta_{\pm}= 2|h_{3}(\bm{k}^{\pm}_{D})|=2\left|\epsilon\pm 3\sqrt{3}|t_1|\sin\varphi\right|.
 \label{eq:gaphaldane}
 \end{equation}
For certain values of $\alpha$ and $\chi_A$ one of the two gaps may close again, namely when ${\bm h}(\bm{k}_{D}^{\pm};\alpha,\chi_{A})=0$ (see eq. (\ref{eq:tbenergies})). This relation identifies the boundary between the normal and topological insulator phases \cite{haldane1988}, as discussed later on in sect. \ref{sec:topo}.

\subsubsection{Tight-binding parameters}

As anticipated in sect. \ref{sec:mlwf}, it is possible to derive a closed set of analytical expressions in terms of specific properties of the spectrum, namely the gaps at the Dirac points $\delta_{\pm}$ in eq. (\ref{eq:gaphaldane}), and the following bandwidths \cite{ibanez-azpiroz2015}
\begin{align}
\Delta^{\pm}_{+}&=+[\epsilon_{+}(\bm{0})-\epsilon_{+}(\bm{k}^{\pm}_D)],\\
\Delta^{\pm}_{-}&=-[\epsilon_{-}(\bm{0})-\epsilon_{-}(\bm{k}^{\pm}_D)].
\label{eq:bandwidths}
\end{align}
Then, by considering e.g. $\epsilon,\varphi\geq 0$ \footnote{The solutions corresponding to a different regime can be obtained straightforwardly from symmetry considerations, by exchanging the role of the two basis points $A,B$ and/or of the two Dirac points.}, one has
\begin{align}
\label{eq:eps-spc}
\epsilon&=\frac{\delta_{+}\pm\delta_{-}}{4},\\
\label{eq:t0-spc}
t_{0}&=\frac{1}{6}\sqrt{\left(\Delta_+^++\Delta_-^++\delta_{+}\right)^2-\frac{\left(\delta_{+}\pm\delta_{-}\right)^2}{4}},
\\
\label{eq:t1-spc}
|t_{1}|&=\frac{1}{18}\sqrt{\left(\Delta_+^+-\Delta_-^+\right)^2+\frac{3}{4}\left(\delta_{+}\mp\delta_{-}\right)^2},\\
\label{eq:phi-spc}
\varphi&=\rm{tg}^{-1}\left[\frac{\sqrt{3}}{2}\frac{\delta_{+}\mp\delta_{-}}{\Delta_+^+-\Delta_-^+}\right].
\end{align}
where the signs in $\pm$ refer to the normal and topological insulator phases, respectively.
The behavior of the various tight-binding parameters as a function of the intensity of the vector potential  $\alpha$, and of the lattice amplitude $s$, will be discussed in the following. Three regimes can be identified, according to the symmetries of the system.

\begin{figure}[H]
\centerline{\includegraphics[width=0.9\columnwidth]{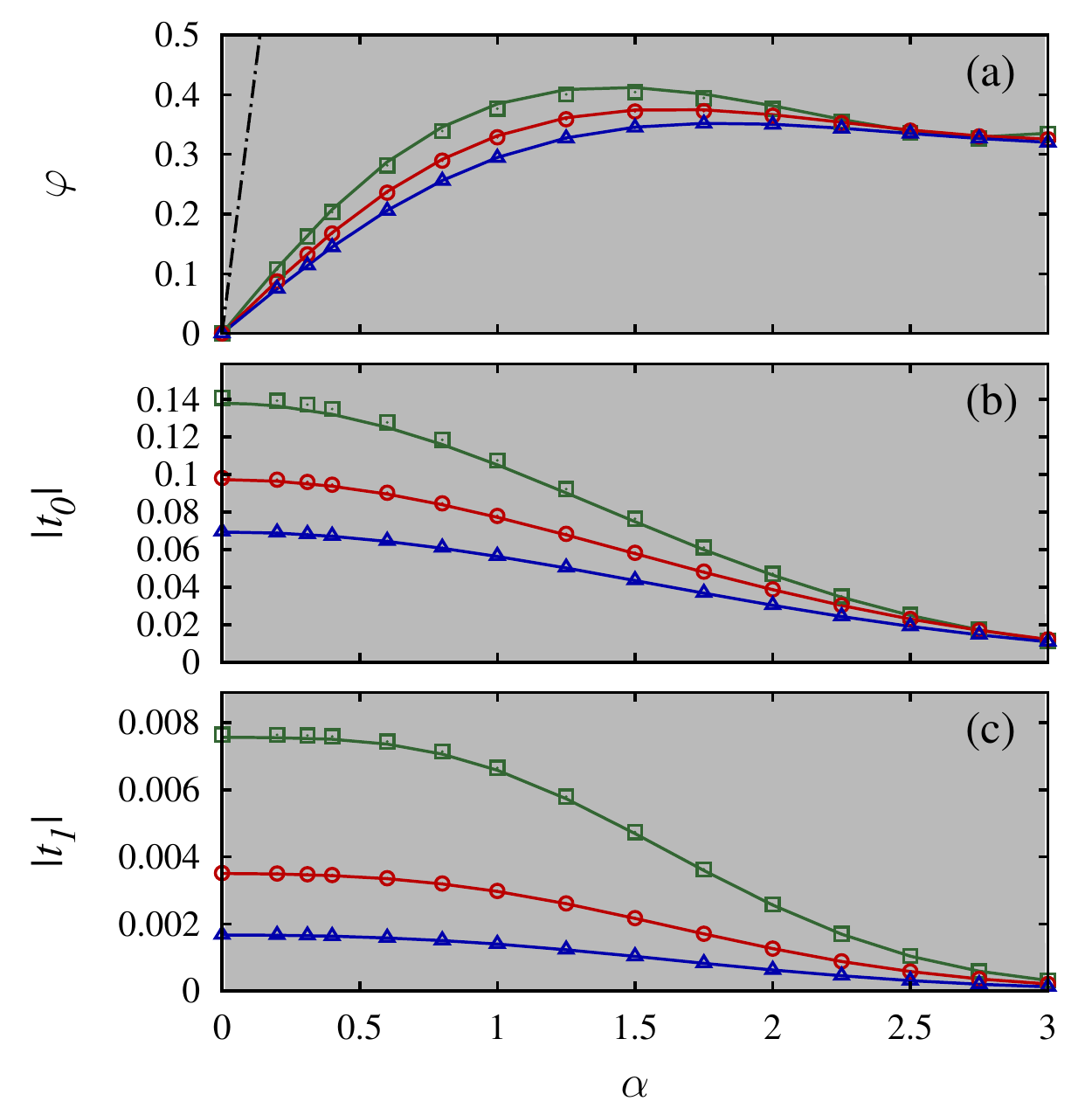}}
\caption{(Color online) 
Plot of the tight-binding parameters as a function of $\alpha$
for the parity-symmetric case ($\chi_{A}=0$), with $s = 5$ (green, squares), $7$ (red, circles) and $9$ (blue, triangles). The solid lines correspond to the \textit{ab initio} calculation, whereas symbols are obtained from the analytic formulas (\ref{eq:t0-spc})-(\ref{eq:phi-spc}). The gray background indicates that the system is in the topological insulating phase. (a) Phase $\varphi$. The dotted-dashed line represents the phase $\varphi_{P}$ obtained from the Peierls substitution.
(b,c) tunneling coefficients $t_{0}$ and $|t_{1}|$, respectively. }
\label{fig:phi-alpha}
\label{fig:t0-t1}
\end{figure}

\textit{Parity conserving, time-reversal breaking case} ($\alpha\neq 0$, $\chi_A=0$)\footnote{
In this regime $|t_{1A}|=|t_{1B}|$ and $\varphi_A=-\varphi_B=\varphi$,
so that the Haldane model with $\epsilon=0$ is strictly recovered. }. 
In this regime we have $\epsilon=0$, so that the energy gaps in eq. (\ref{eq:gaphaldane}) 
become degenerate, $\delta_+=\delta_-\equiv\delta_D$, and 
the four bandwidths in (\ref{eq:bandwidths}) merge into two, namely $\Delta^{+}_{+}= \Delta^{-}_{+}\equiv\Delta_{+}$
and $\Delta^{+}_{-}= \Delta^{-}_{-}\equiv\Delta_{-}$. 
The behavior of the tight-binding parameters as functions of the amplitude $\alpha$ of the vector potential
is shown in Figure \ref{fig:phi-alpha} for different values of the lattice amplitude $s$. 
From these figures we can identify two regimes: \textit{(i)} $\alpha\lesssim0.5$, where $t_{0}$ and $|t_{1}|$ are almost constant and the phase $\varphi$ is linear in $\alpha$; \textit{(ii)} $\alpha\gtrsim0.5$, with the phase $\varphi$ deviating from the linear behavior, and the tunneling amplitudes $t_{0}$ and $|t_{1}|$ being largely suppressed. 
\begin{figure}[H]   
\psfrag{epsilon}{$\epsilon$ (mm/mm)}
\centerline{\includegraphics[width=0.9\columnwidth]{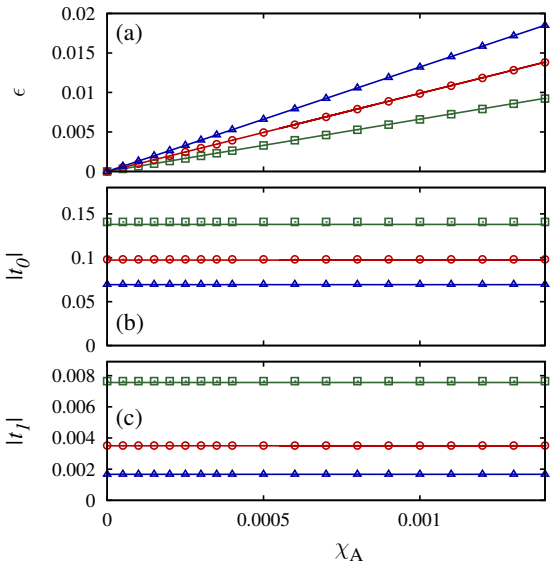}}
\caption{(Color online) 
Plot of the tight-binding parameters as a function of $\chi_{A}$, for $\alpha=0$ and $s = 5$ (green, squares), $7$ (red, circles) and $9$ (blue, triangles).
The solid lines correspond to the \textit{ab initio} calculation, whereas symbols are obtained from the analytic formulas (\ref{eq:eps-spc})-(\ref{eq:t1-spc}).
(a) On site energy difference $\epsilon=\delta_{D}/2$.
(b,c) tunneling coefficients $t_{0}$ and $|t_{1}|$, respectively. 
}
\label{fig:gappr}
\label{fig:t1AB}
\end{figure}
\textit{Time-reversal conserving, parity breaking case} ($\chi_A\neq 0$, $\alpha=0$).
This situation corresponds to a honeycomb lattice with non-degenerate minima, with the energy gaps at the Dirac points still being degenerate, so that there are only two bandwidths, $\Delta_{\pm}$, as in the previous case. 
Notably, now we have
$\varphi_A=\varphi_B=\varphi=0$, implying that 
the system behaves as a normal insulator.
The behavior of $\epsilon$, $t_{0}$ and $|t_{1}|$ as a function of $\chi_{A}$ is shown in Figure \ref{fig:gappr}. Notice that both tunneling coefficients, $t_{0}$ and $|t_{1}|$ are barely affected by parity breaking in this range of parameters. We also remarks that, though in principle the degeneracy between $|t_{1A}|$ and $|t_{1B}|$ is formally broken, their difference is actually negligible (see later on).

\begin{figure}[H]
\centerline{\includegraphics[width=0.9\columnwidth]{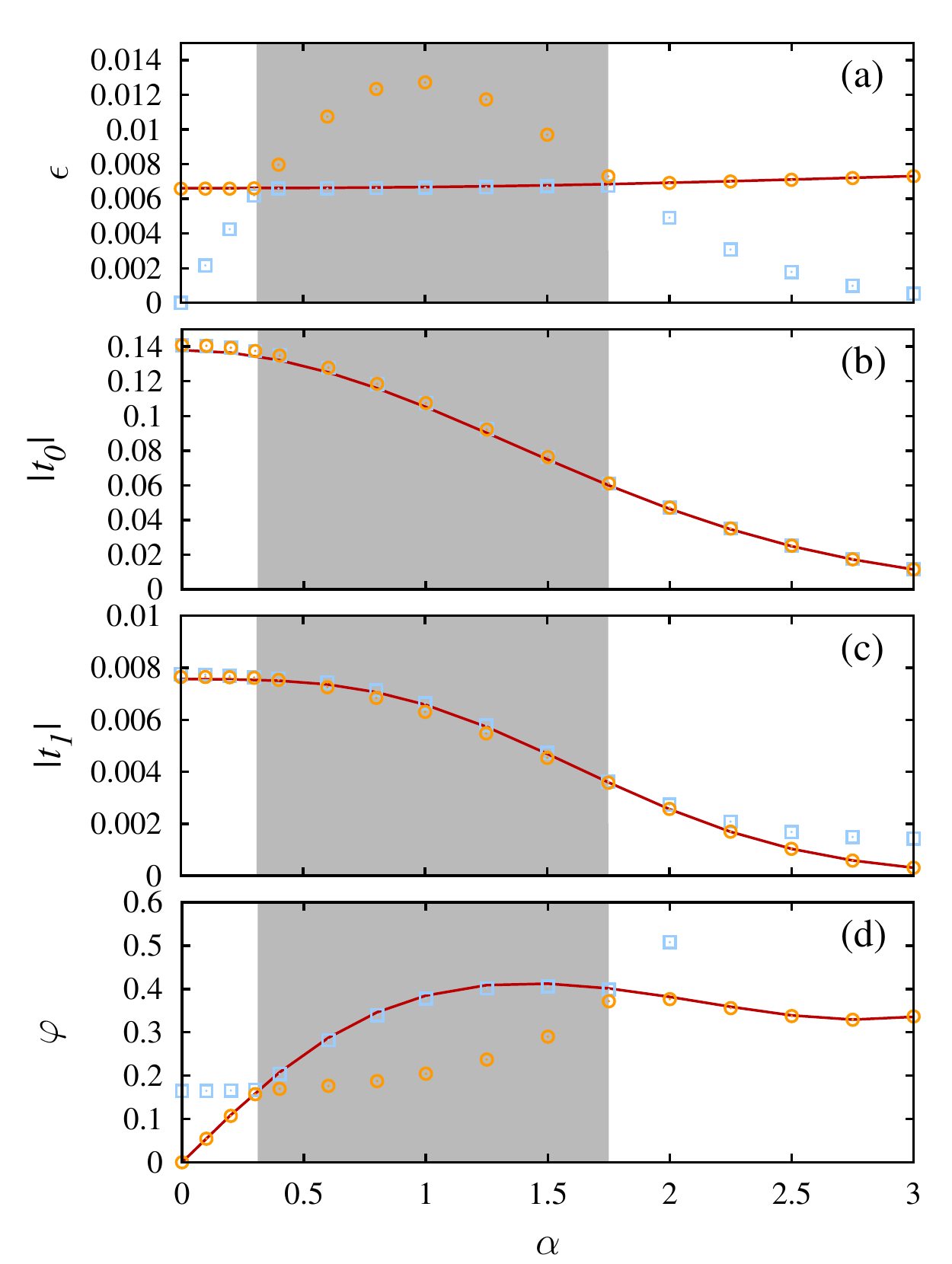}}
\caption{(Color online) Plot of the tight-binding coefficients as a function of $\alpha$, for $s=5$ and $\chi_{A}=0.001$. The solid line corresponds to the \textit{ab-initio} calculations from the MLWFs, whereas the points (squares and circles) are obtained from the the analytical expressions in eqs. (\ref{eq:eps-spc})-(\ref{eq:phi-spc}).
The grey area denotes the topological insulator regime.}
\label{fig:t0-t1-phi}
\end{figure}

\textit{General case} ($\alpha\neq0$, $\chi_{A}\neq0$).
 This is the most interesting case, in which both time-reversal and inversion symmetry are broken.
 The corresponding tight-binding parameters are shown in Figure \ref{fig:t0-t1-phi} as a function of $\alpha$, for $s=5$ and $\chi_{A}=0.001$. 
 It is interesting to note that the two set of solutions corresponding to the two sign choices in eqs. (\ref{eq:eps-spc}-\ref{eq:phi-spc}) connect smoothly across the boundary between normal and topological insulator regimes.

\begin{figure}[H]
\centerline{\includegraphics[width=0.9\columnwidth]{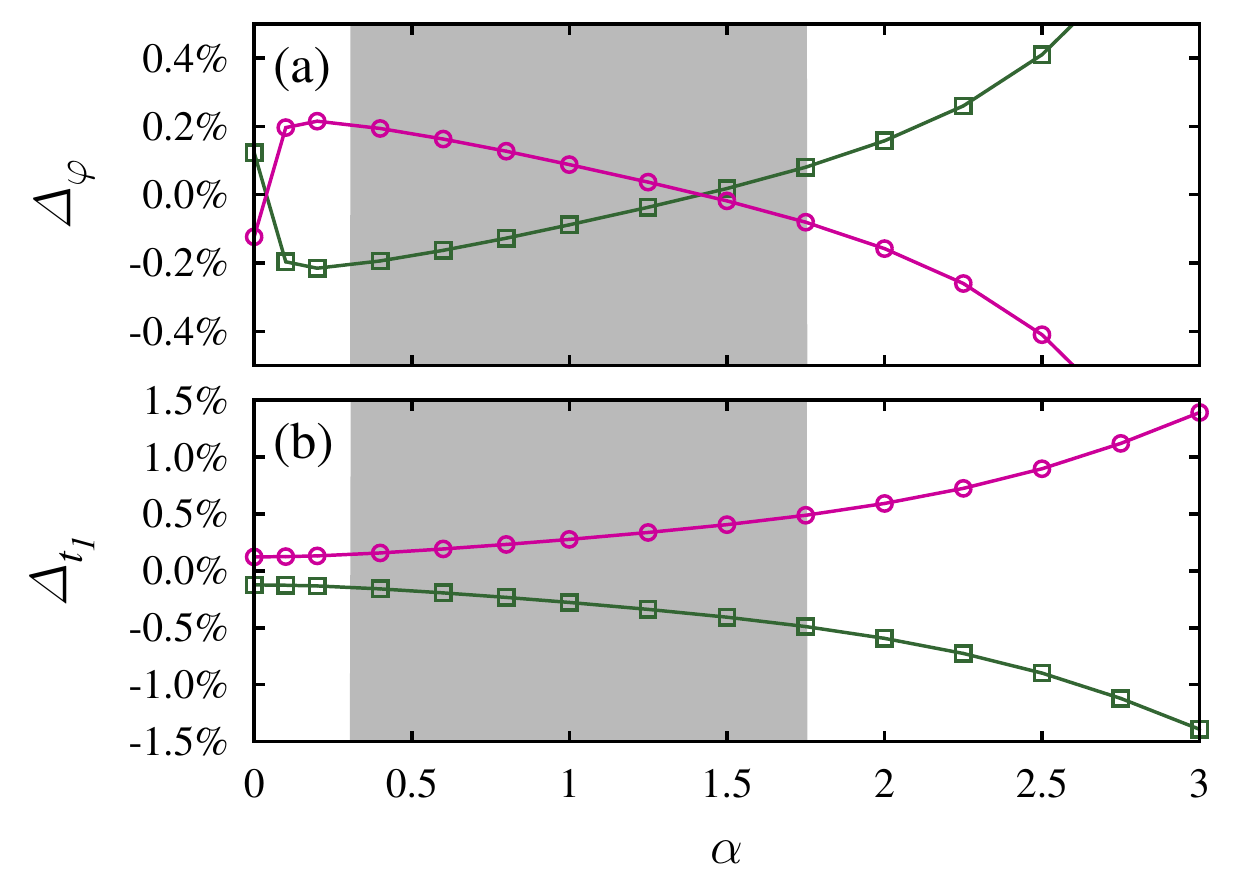}}
\caption{(Color online) Relative deviations from the average values of (a) 
the phase, $\Delta_{\varphi}$, and (b) the magnitude of the next-to-nearest tunneling coefficient, 
$\Delta_{t_{1}}$, for $\chi_{A}=0.001$, $s=5$ $E_{R}$.
These quantities have been calculated \textit{ab-initio} by using the MLWFs approach \cite{ibanez-azpiroz2015}.}
\label{fig:sps}
\end{figure}

Finally, let us comment on the approximations employed in deriving the Haldane model. We remind that the Haldane model relies on the following approximations: $|t_{1A}|=|t_{1B}|\equiv|t_{1}|$, $\varphi_A=-\varphi_B\equiv\varphi$, that in principle are not consistent with the broken degeneracy between sites $A$ and $B$ in the presence of parity breaking. Their accuracy can be checked by using the \textit{ab initio} values of those terms \cite{ibanez-azpiroz2015}.
This is done in Fig. \ref{fig:sps}, where we compare the relative deviations from the average values of the phase, $\Delta_{\varphi}\equiv1-\varphi_{A,B}/\varphi$, and of the magnitude of the next-to-nearest tunneling coefficient, 
$\Delta_{t_{1}}\equiv1-|t_{1A,B}|/|t_{1}|$, for $\chi_{A}=0.001$, $s=5$. 
This figure demonstrates that the maximum relative deviation in both cases is
below $\sim 1\%$. 
It can be verified that this holds for all values of $s$ and $\chi_{A}$ considered here, 
thus justifying the assumptions of the Haldane model in the whole range of parameters.

\subsubsection{Breakdown of the Peierls substitution}
\label{sec:peierls}

Remarkably, the behavior of the tunneling coefficients as a function of the amplitude $\alpha$ of the vector potential, shown e.g. in Figure \ref{fig:phi-alpha}a, is dramatically different from that dictated by the so-called \textit{Peierls substitution}, a widely employed approximation for describing tight-binding electrons in the presence of a slowly varying external vector field. We recall that in the tight-binding formulation the Peierls substitution\footnote{The \textit{Peierls substitution}, named after the original work by R. Peierls \cite{peierls1933}, was originally formulated within the semiclassical approximation as a modification of the dispersion relation, $E(\bm{k})\to E(-i\hbar\bm{\nabla} -(e/c)\bm{A})$ \cite{luttinger1951,hofstadter1976,alexandrov1991a}.} consists in replacing the tunneling coefficients $t_{ij}$ with $t_{ij}\exp\{ie\int_{i}^{j}\bm{A}d\bm{r}\}$ \cite{shao2008,bernevig2013}, with the integral to be evaluated along the straight path connecting sites $i$ and $j$ \cite{haldane1988,boykin2001}. 
Its formal demonstration requires the hypothesis of a same-site, same-orbital interaction with the vector field, $\langle w_{\bm{j}\nu}|\bm{A}(\bm{r})|w_{\bm{j'}\nu'}\rangle=\bm{A}(\bm{R}_{j\nu})\langle w_{\bm{j}\nu}|w_{\bm{j'}\nu'}\rangle$ \cite{boykin2001}, corresponding to a vector potential varying on a length scale that is much larger than the lattice spacing. Though this condition is explicitly violated in the Haldane model -- the vector field $\bm{A}(\bm{r})$ has the same periodicity of the underlying lattice -- this point is often underrated in the literature, see e.g. \cite{haldane1988,shao2008}. In fact, the explicit failure of the Peierls substitution has been reported only recently \cite{ibanez-azpiroz2014} (see also \cite{alexandrov1991,alexandrov1991a} for what concerns the semiclassical approach).

Actually, Figure \ref{fig:phi-alpha}a shows that the value of the phase obtained from the Peierls substitution, $\varphi_{P}\equiv\int_{\bm{r}_{A}}^{\bm{r}_{A}-\bm{a}_{1}}\bm{A}\cdot d\bm{r}=({2\pi}/{\sqrt{3}})\alpha$ \cite{shao2008}, differs by more than one order of magnitude from the actual values, even in the regime of low vector potential amplitude ($\alpha<0.5$), where the calculated phase is also linear.
Moreover, $\varphi_{P}$ does not account for the dependence on the amplitude $s$ of the scalar potential, that is appreciable even in the full tight-binding regime \cite{ibanez-azpiroz2014}, nor the fact that when one moves away from the linear regime, both tunneling coefficients $t_{0}$ and $t_{1}$ are strongly suppressed, and that the phase $\varphi$ deviates from the linear behavior. This originates from the usual implicit assumption that the basis of localized orbitals is not affected by the vector potential (see e.g. \cite{boykin2001}), whereas in fact  the presence of the vector potential may significantly affect both the Bloch eigenfunctions $\psi_{m\bm{k}}$ \cite{kohn1959} and the gauge transformation $U_{\nu m}$ entering eq. (\ref{eq:mlwfs}) \cite{alexandrov1991a}. Remarkably, the fact that the phase $\varphi$ is limited by a maximal value implies that $\varphi$ can only access a restricted range of values, so that only a limited portion of the nominal phase diagram can be physically accessed \cite{ibanez-azpiroz2015}.

\subsubsection{Topological phase diagram}
\label{sec:topo}

\begin{figure}[H]
\centerline{\includegraphics[width=0.9\columnwidth]{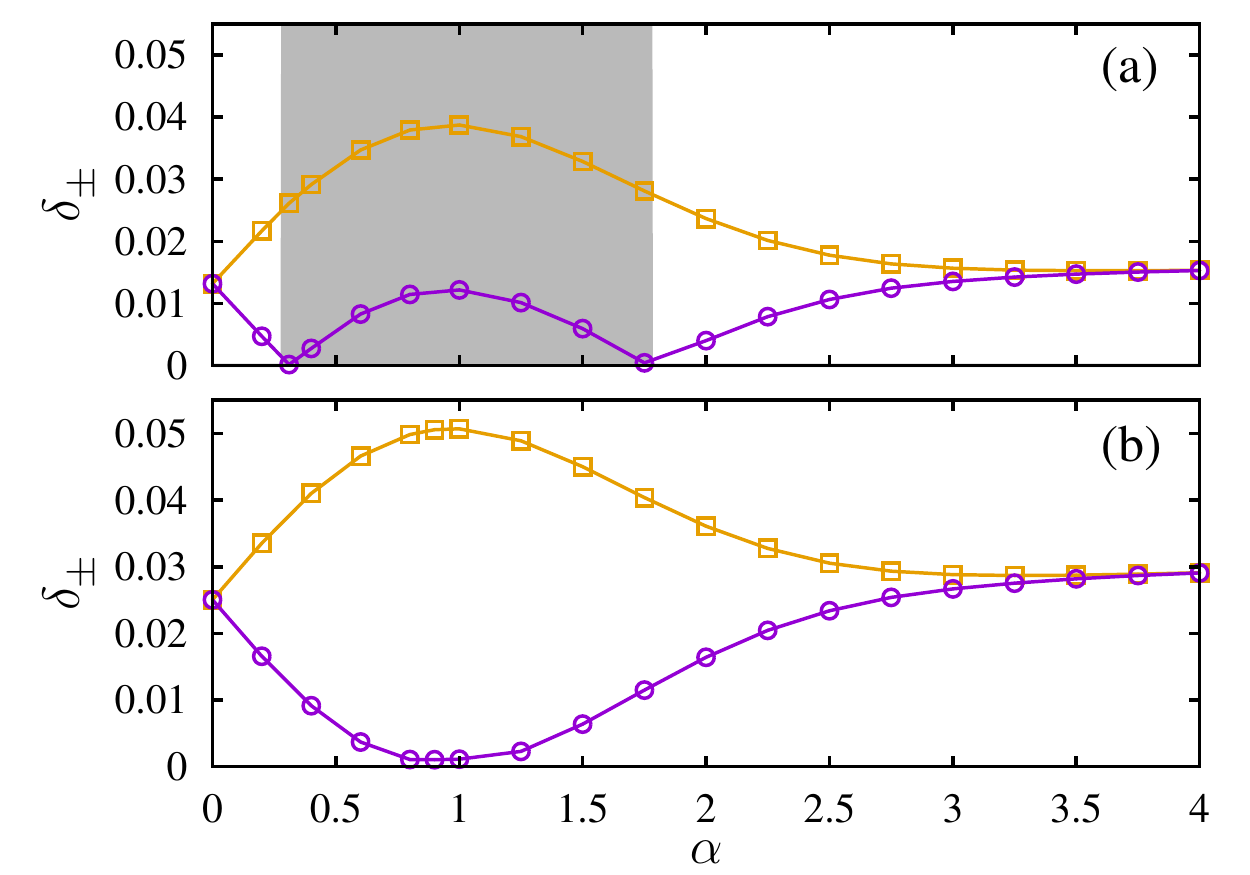}}
\caption{(Color online) behavior of the gaps $\delta_{+}$ (black, squares)
and $\delta_{-}$ (red, circles) as a function of $\alpha$, for $s=5$ and
$\chi_{A}=2\cdot10^{-3}$ (a), and $1.9\cdot10^{-3}$ (b).
The latter corresponds to the maximal value of $|\chi_{A}|$ for which the system can be in the topological insulating phase. The grey shaded area identifies the topological insulator phase ($C=1$), whereas the white background corresponds to a normal insulating state ($C=0$).}
\label{fig:gaps-Chern}
\end{figure}

The Haldane model is characterized by different insulating phases, 
associated to the values of the Chern number (or topological index) \cite{thonhauser2006}
\begin{equation}
\label{eq:chern}
C=\frac{i}{2\pi}\int_{BZ}\!\!\!\!d\bm{k}
\sum_{\nu}^{occ}
\langle{\partial_{\bm{k}}u_{\nu\bm{k}}|\times|\partial_{\bm{k}}u_{\nu\bm{k}}}\rangle,
\end{equation}
where $u_{\nu\bm{k}}(\bm{r})=e^{-i\bm{k}\cdot\bm{r}}\psi_{\nu\bm{k}}(\bm{r})$
are the periodic part of the Bloch eigenfunctions, the sum over $\nu$ being restricted only to occupied bands\footnote{The Haldane model consists just of a valence and a conduction band, 
so that only the lowest band enters the sum over occupied states. }. 
$C$ takes only integer values, and it is non vanishing for a topological insulator.
The structure of the phase diagram is intimately connected to the presence of the gaps at the Dirac points.
When only time-reversal symmetry is broken ($\alpha\neq0$, $\chi_{A}=0$), the gap at the Dirac points is always finite, and the system is in a topological insulating state ($C\neq 0$). On the other hand, if a gap is opened solely by inversion symmetry breaking, the state of the system is topologically trivial ($C=0$).
When both symmetries are broken, the behavior of the system depends on the relative strength of the inversion and time-reversal symmetry breaking. In particular, when $\chi_{A}$ is relatively small, the gap $\delta_{-}$\footnote{The role of $\delta_{+}$ and $\delta_{-}$ is exchanged for $\alpha<0$.} vanishes for two different values of $\alpha$, as shown in Figures \ref{fig:gaps-Chern}a. In between these two values (grey shaded area in the Figures) the state of the system corresponds to a topological insulator ($C=1$); this phase shrinks and eventually disappears by increasing $\chi_{A}$, see Figure \ref{fig:gaps-Chern}b.

\begin{figure}[H]
\centerline{\includegraphics[width=0.9\columnwidth]{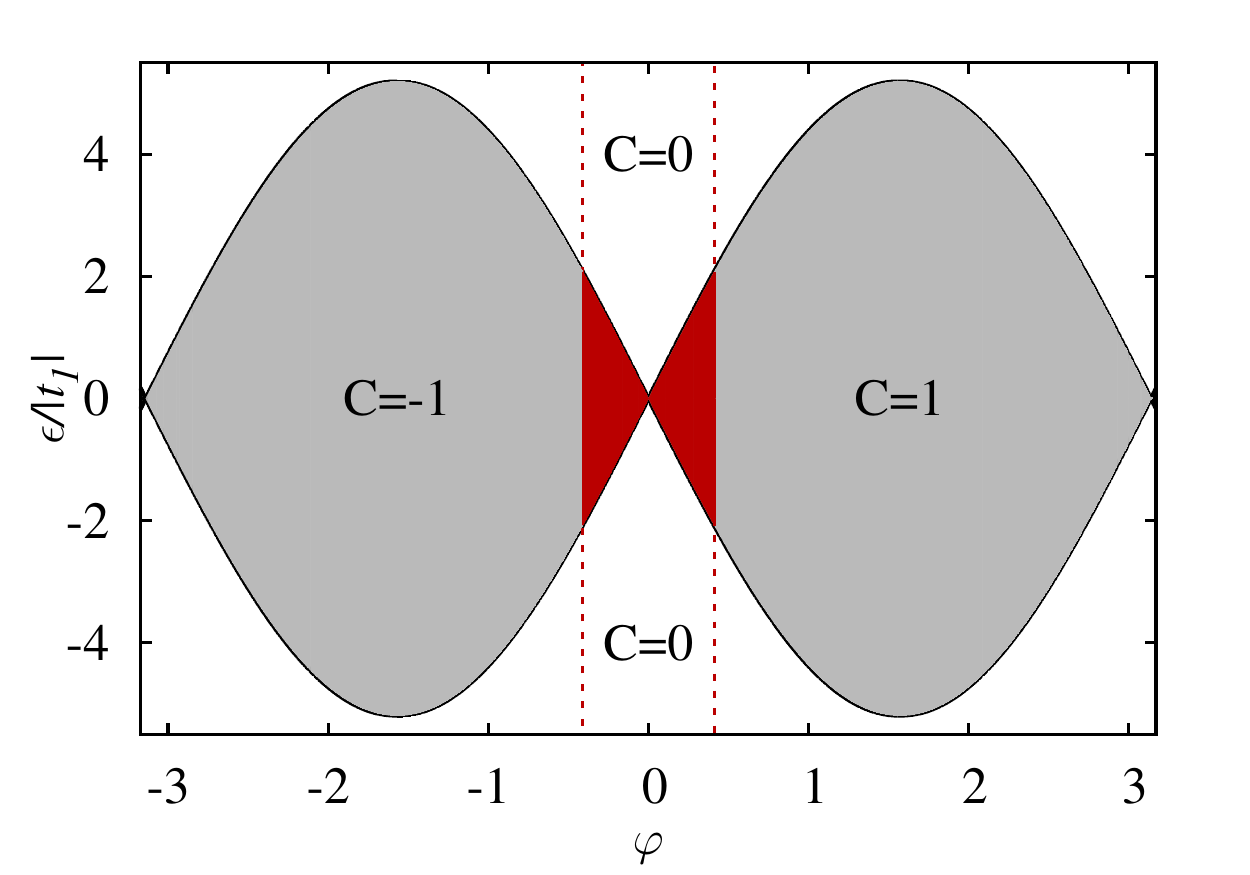}}
\caption{(Color online) Nominal phase diagram of the Haldane model as a function of $\varphi$ and $\epsilon/|t_{1}|$.
The solid (black) line denotes the analytical boundary $\epsilon/|t_{1}|=3\sqrt{3}\sin\varphi$ between the topological insulating phases ($C=\pm1$, colored areas) and the normal insulator phase ($C=0$, white areas). 
Actually, only the region in between the two vertical red dashed lines - corresponding to the maximum value of $\varphi$ shown in Figure \ref{fig:phi-alpha} - is physically accessible (here $s=5$; for higher values that region shrinks even further). }
\label{fig:C-phi-M-t1}
\end{figure}

\begin{figure}[H]
\centerline{\includegraphics[width=0.9\columnwidth]{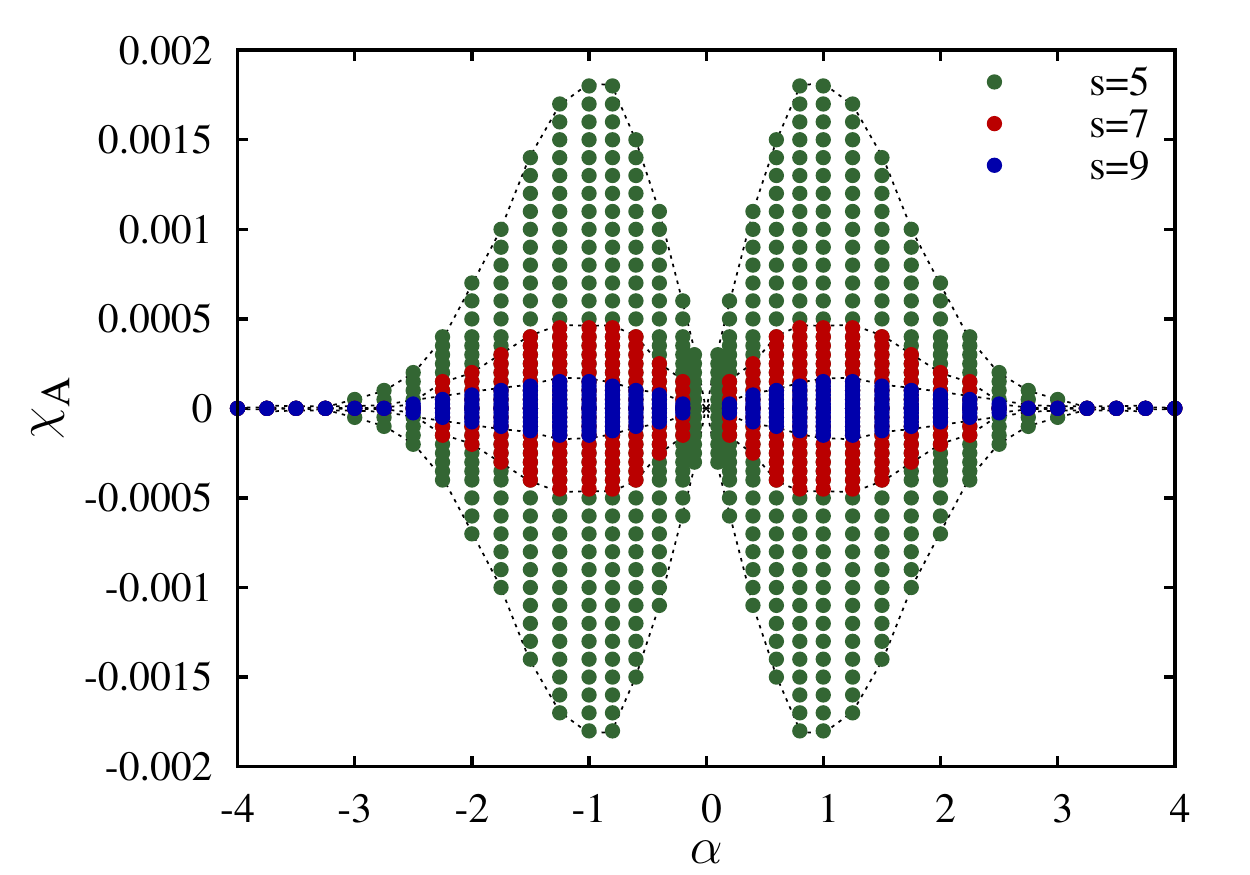}}
\caption{(Color online) Topological phase diagram of 
the continuous Hamiltonian in eq. (\ref{eq:hamiltonian}), as a function of $\alpha$ and $\chi_{A}$, 
for three different values of the scalar potential amplitude $s$. 
The non-trivial topological state is indicated by the colored dots.
The black dashed lines represent a guide to the eye for the phase boundaries for each value of $s$.}
\label{fig:C-chi-alpha}
\end{figure}

The phase diagram is traditionally drawn as a function of $\varphi$ and $\epsilon/|t_{1}|$~\cite{haldane1988,shao2008}, with the boundary between normal and topological insulator phases corresponding to the vanishing of the gap at one of the two inequivalent Dirac points in eq. (\ref{eq:gaphaldane}), namely $\epsilon/{|t_1|}=\pm 3\sqrt{3}\sin\varphi$. 
The original formulation of the model is obtained by means of the Peierls substitution~\cite{haldane1988,shao2008}, so that the whole phase diagram is accessible. However, since the possible values of $\varphi$ are actually limited to a finite range that depends on $s$, only a finite portion of the nominal phase diagram can be accessed, see Figure \ref{fig:C-phi-M-t1}. 
The topological phase diagram can also be drawn in terms of the physical parameters that
characterize the underlying continuous Hamiltonian. This is shown in Figure \ref{fig:C-chi-alpha}, 
where we plot the phase diagram in the 
$\alpha-\chi_A$ plane, for three different values of $s$. Remarkably, the topological insulating phase shrinks substantially by increasing $s$ (that is, as the system 
becomes more and more tight-binding). 

\subsection{A stretched honeycomb}
\label{sect:stretched}

Another interesting setup for exploring the topological transition at the merging of Dirac points is represented by the tunable honeycomb lattice of Tarruell \textit{et al.} \cite{tarruell2012}. In this case the system can be described in terms of a minimal tight-binding model defined on a \textit{square} lattice, characterized by three tunneling coefficients
$t_{0}$, $t_{1}$ and $t_{2}$, or by means of an universal tight-binding Hamiltonian that provides a low energy effective description in the vicinity of the Dirac points \cite{montambaux2009,montambaux2009b,lim2012,uehlinger2013a}.
Obviously, one can also employ the ab-initio approach discussed in the preceding sections, that allows to take into account the full geometry of the system. This has been considered in Ref. \cite{ibanez-azpiroz2013a}, and will be reviewed in the following. 

\begin{figure}[H]
\centerline{\includegraphics[width=0.7\columnwidth]{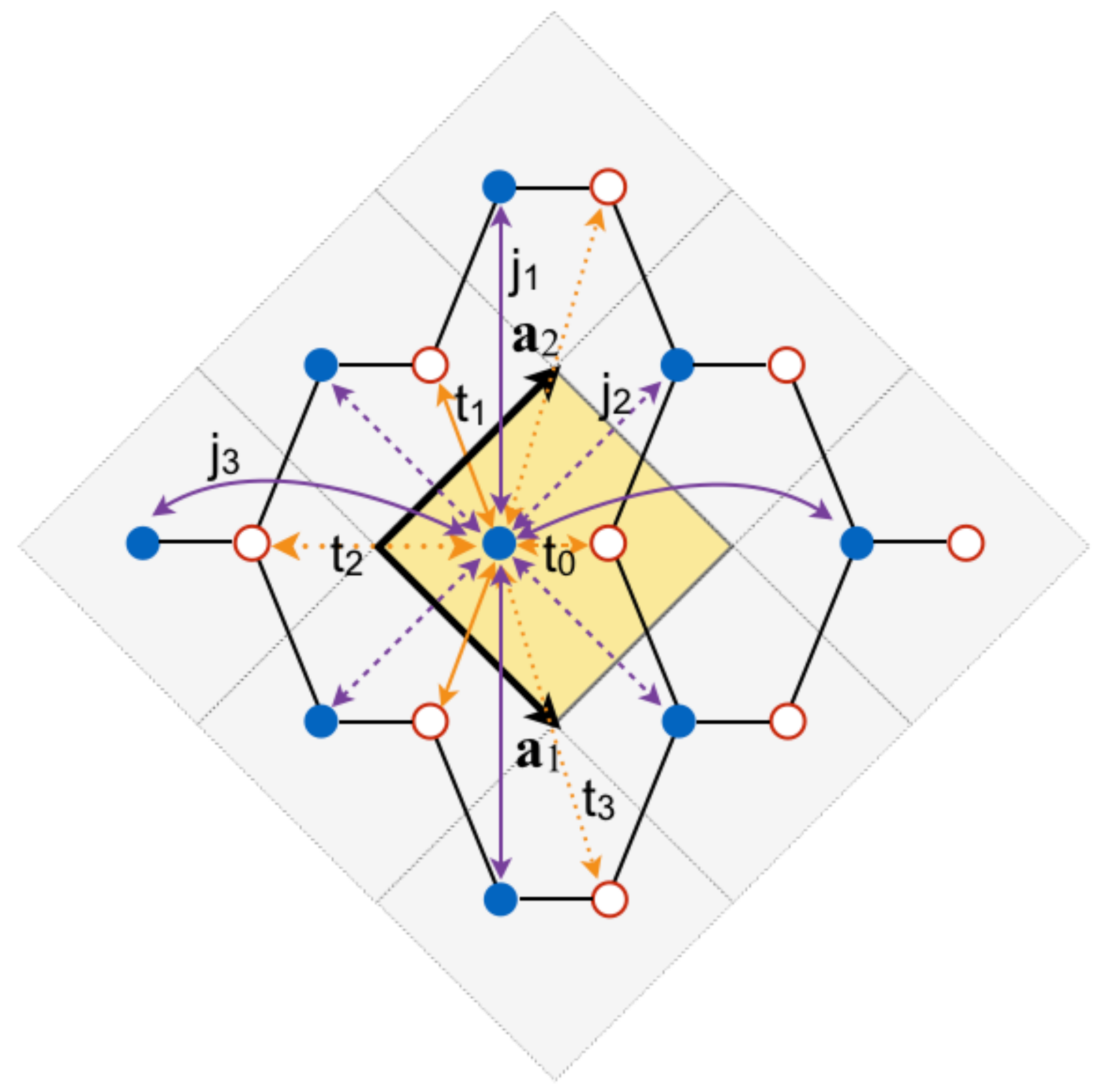}}
\caption{(Color online) 
Bravais lattice for the stretched honeycomb configuration of the potential in eq. (\ref{eq:tarruellpot}). Filled and empty circles refers to minima of type $A$ and $B$, respectively. The elementary cell is highlighted in yellow. The various diagonal and off-diagonal tunneling coefficients considered here are indicated for the site of type A in the central cell.}
\label{fig:bravais}
\end{figure}

\begin{figure}[H]
\centerline{\includegraphics[width=\columnwidth]{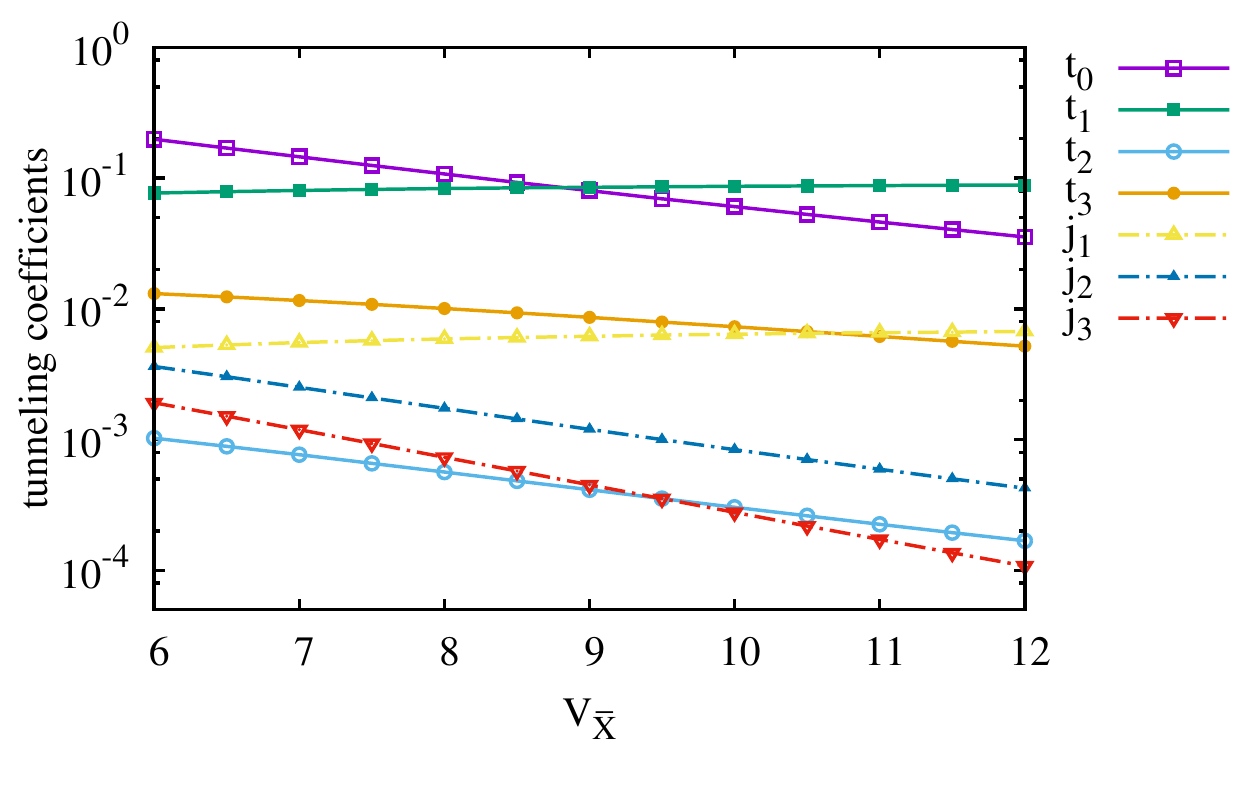}}
\caption{(Color online) Plot of the various tunneling coefficients as function of $V_{\overline{X}}$. 
}
\label{fig:sym-tunnel-tarruel-str-hon}
\end{figure}

Let us start with the \textit{tunable honeycomb} potential employed in Ref. \cite{tarruell2012}
\begin{align}
\label{eq:tarruellpot}
&V(x,y)=-V_{\overline{X}}\cos^2(k_{L}x+\theta/2)-V_{X}\cos^2(k_{L}x)\\
&\quad-V_{Y}\cos^2(k_{L}y)-2\alpha\sqrt{V_{X}V_{Y}}\cos(k_{L}x)\cos(k_{L}y)\cos(\varphi),
\nonumber
\end{align}
where, by varying the laser intensities 
$V_{\overline{X}},V_{X}$ and $V_{Y}$, several structures
can be realized by continuous deformations, including square, triangular, chequerboard, dimer, honeycomb 
and 1D chain geometries~\cite{tarruell2012}.
Here we consider a set of parameters that guarantee a proper tight-binding regime, with $V_{X}=0.56$, $V_{Y}=3.6$, and
 $V_{\overline{X}}$ variable in the range $[6,12]$ \cite{ibanez-azpiroz2013a}. 
The corresponding Bravais lattice -- see Figure \ref{fig:bravais} -- is generated by the two basis vectors
$\bm{a}_{1,2}=\pi(\bm{e}_{x}\mp\bm{e}_{y})/k_{L}$, with the reciprocal vectors being 
$\bm{b}_{1,2}=k_{L}(\bm{e}_{x}\mp\bm{e}_{y})$. 
Here we shall consider an \textit{extended} tight-binding model including all the tunnelings coefficients indicated in Figure \ref{fig:bravais}. Notice that the ordering of the tunneling coefficients does not necessarily correspond to the hierarchy of their magnitudes. The latter depends on the regime of the potential parameters considered, see e.g. Figure \ref{fig:sym-tunnel-tarruel-str-hon}.

In this case, the functions $z(\bm{k})$ and $f^{\nu}(\bm{k})$ are given by (see eqs. (\ref{eq:epsnu}),(\ref{eq:zetanu}))
\begin{align}
z(\bm{k})&= -\left[t_{0}+2t_{1}\cos(\pi k_{y})e^{-i\pi k_{x}}+t_{2}e^{-2i\pi k_{x}}
+2t_{3}\cos(2\pi k_{y})\right],
\label{eq:zetatb}
\\
f^{\nu}(\bm{k})&=2\left[j^{\nu}_{1}\cos\left(2\pi k_{y}\right)+2j^{\nu}_{2}\cos\left(\pi k_{y}\right)\cos
\left(\pi k_{x}\right)
\right.
\nonumber\\
&
\qquad\left.
+j^{\nu}_{3}\cos\left(2\pi k_{x}\right)\right].
\label{eq:effenu1}
\end{align}
The behavior of the tunneling coefficients as a function of $V_{\overline{X}}$ is shown in Figure \ref{fig:sym-tunnel-tarruel-str-hon}.

\begin{figure}[H]
\centerline{\includegraphics[width=0.95\columnwidth]{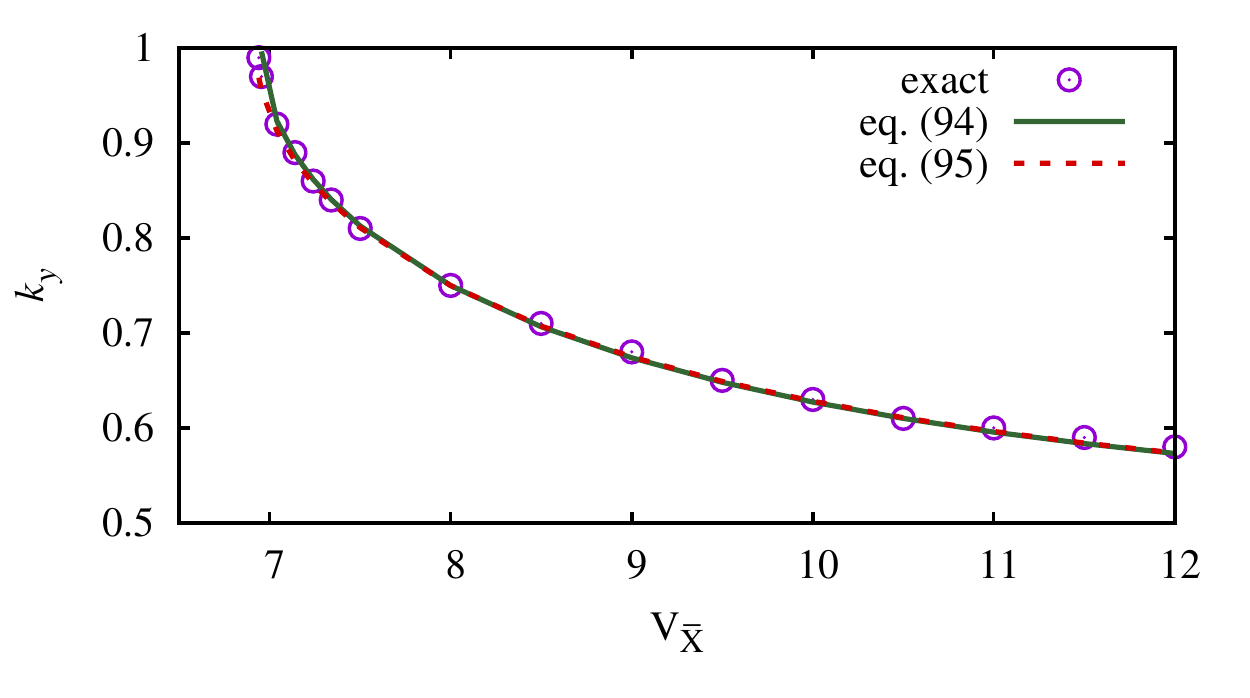}}
\caption{(Color online) Position of the Dirac points along the $k_{y}$-axis as a function of $V_{\overline{X}}$ for the parameter regime of Tarruell \textit{et al.} \cite{tarruell2012}.
The exact positions (circles) extracted from the Bloch spectrum are compared with the predictions of eqs. (\ref{eq:diracpointloc}) (solid line) and (\ref{eq:diracmont}) (dashed dotted line), that are almost indistinguishable. Only around the merging point, at $V_{\overline{X}}\simeq6.94$, it is preferable to use the complete expression (\ref{eq:diracpointloc}) instead of eq. (\ref{eq:diracmont}).}
\label{fig:dirac}
\end{figure}

\begin{figure}[H]
\centerline{\includegraphics[width=0.6\columnwidth]{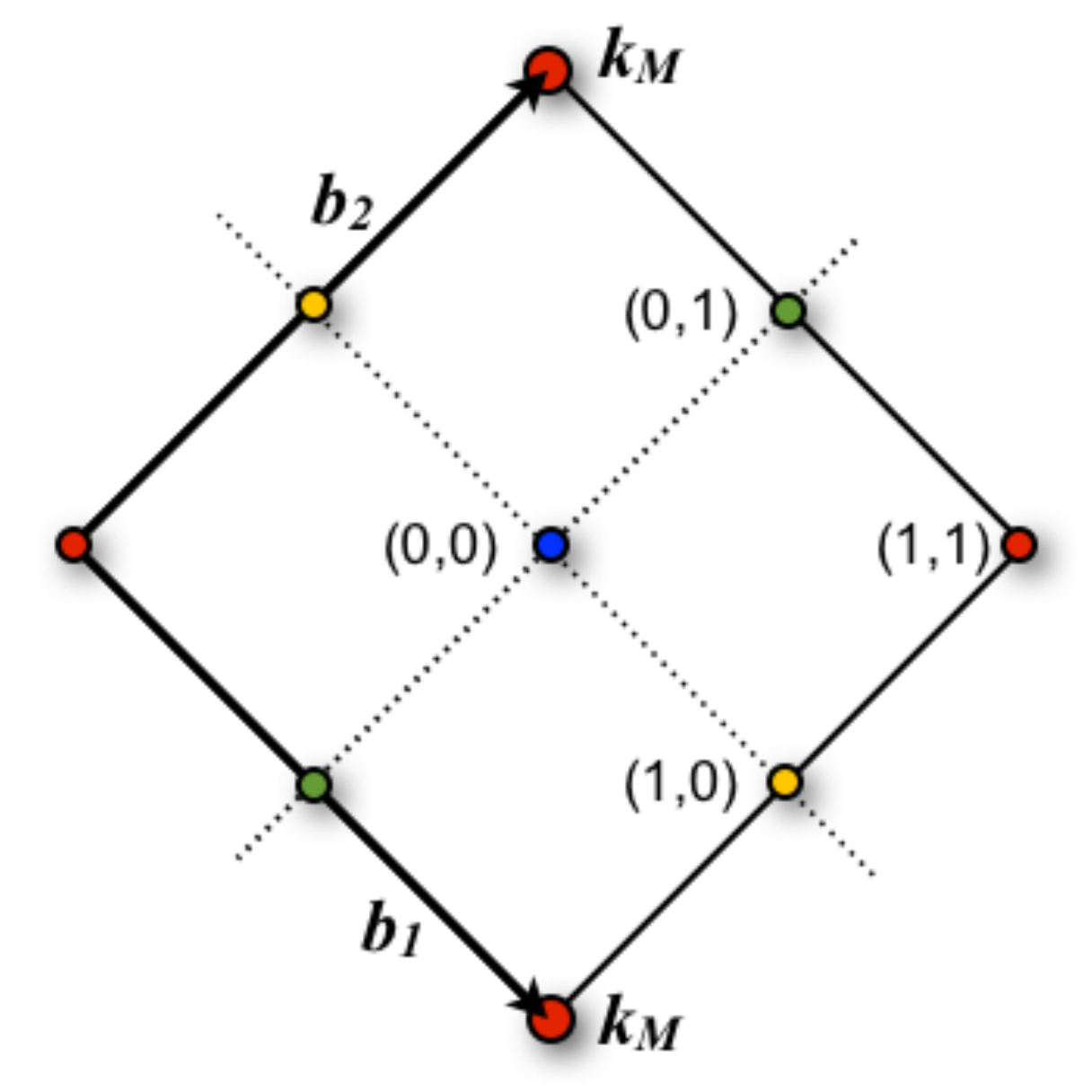}}
\caption{(Color online) Unit cell in reciprocal space (first Brillouin zone). The location of all possible locations of the merging of the Dirac points are indicated by color dots. Equivalent points (connected by a reciprocal space vector $\bm{G}$)
are depicted with the same color. Given the actual values of the tunneling coefficients,
only the points at $\bm{k}_{M}\equiv (0,\pm 1)$ can be realized (larger red dots). 
}
\label{fig:merging}
\end{figure}

\begin{figure}[H]
\centerline{\includegraphics[width=0.9\columnwidth]{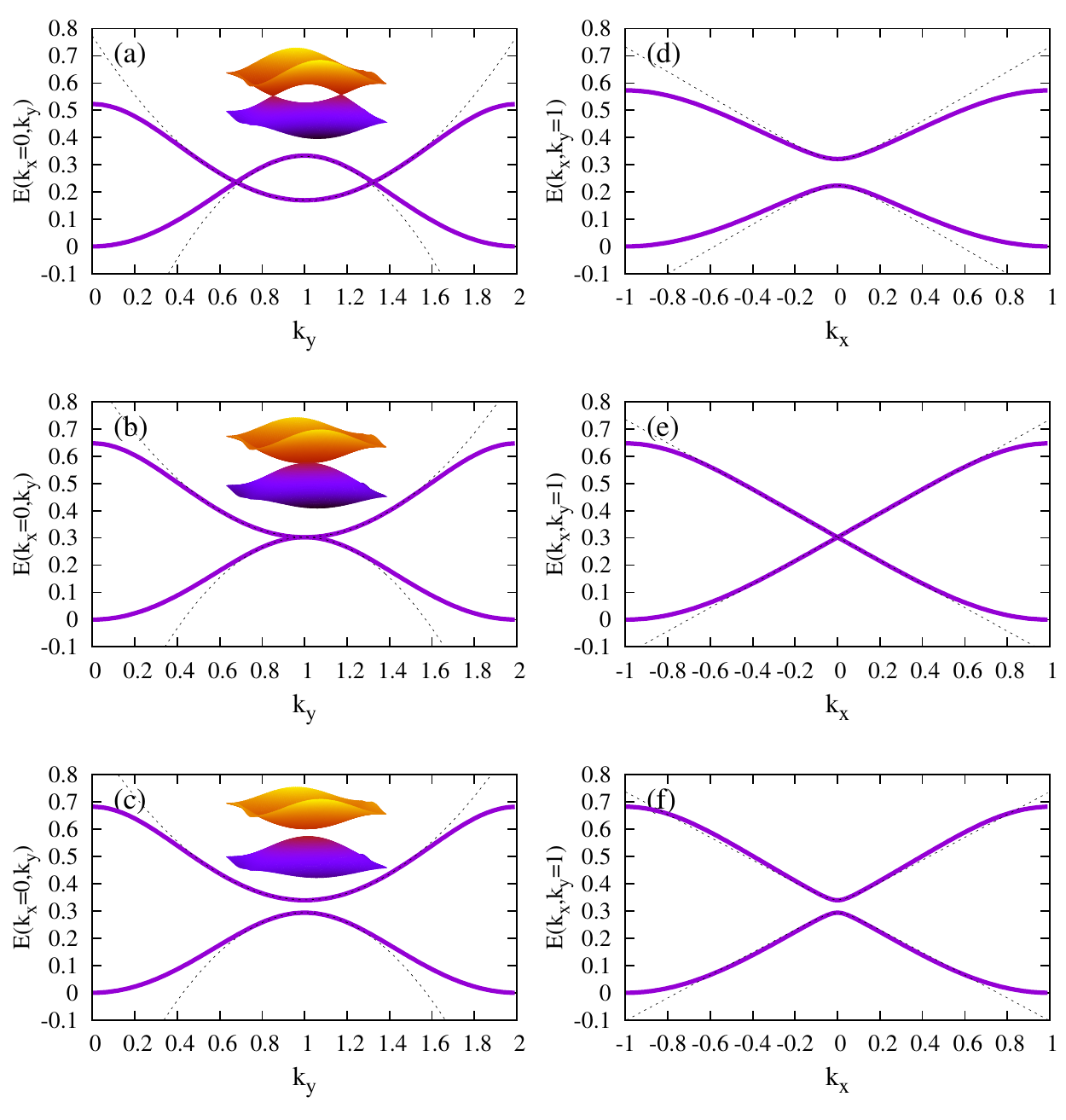}}
\caption{(Color online) Cuts of the energy bands around the merging point $\bm{k}_{M}= (0,1)$, for $V_{X}=0.56 $, $V_{Y}=3.6 $. The exact Bloch bands (red solid lines) are compared to the approximate expressions in eq. (\ref{eq:bandsclosemerging}), as a function of $k_{y}$ (at $k_{x}=0$) (a,b,c), and of $k_{x}$ (at $k_{y}=1$) 
(d,e,f). Each column corresponds to a different value of $V_{\overline{X}}$: (a,d) $V_{\overline{X}}=8 $, (b,e) $V_{\overline{X}}=6.94 $ (merging point), $V_{\overline{X}}=6.54$ (c,f) . Note that the cut along $k_{x}$ in (d) does not cross the Dirac point, as the latter is located at $k_{y}\simeq 0.68$.
}
\label{fig:merging-cut}
\end{figure}

\textit{Dirac points.}
Among all the possible lattice configurations \cite{tarruell2012,ibanez-azpiroz2013a}, let us consider the case with two degenerate minima per unit cell ($\theta=\pi$, $\varphi=0$), that is particularly interesting owing to the presence of massless Dirac points. The corresponding spectrum is $\varepsilon_{\pm}(\bm{k})=f(\bm{k})\pm|z(\bm{k})|$ (see eq. (\ref{eq:degeneratecase})). It is characterized by Dirac points where the two bands are degenerate, with a linear dispersion along at least one direction. These points are defined by $z(\bm{k}_{D})=0$, and their existence and position depend on the geometry of the lattice. In the present case they can be moved inside the Brillouin zone, as shown in \cite{tarruell2012}. In particular, given the actual hierarchy of the tunneling coefficients \cite{ibanez-azpiroz2013a}, the solutions inside the first Brillouin zone correspond to $k_{x}=0$, and $k_{y}$ given by the following expression
\begin{equation}
k_{y}=\pm \frac{1}{\pi}\cos^{-1}\!\!\left[\displaystyle{\frac{-t_{1}+
\sqrt{t_{1}^2+4t_{3}\left(2t_{3}-t_{0}-t_{2}\right)}}{4t_{3}}}\right].
\label{eq:diracpointloc}
\end{equation}
When $t_{3}$ is negligible, this expression reduces to \cite{montambaux2009,montambaux2009b}
\begin{equation}
 k_{y}\simeq \pm \frac{1}{\pi}\cos^{-1}\left[\displaystyle{-\frac{t_{0}+t_{2}}{2t_{1}}}\right].
\label{eq:diracmont}
\end{equation}
The behavior of $k_{y}$ as a function of $V_{\overline{X}}$ is shown in Figure \ref{fig:dirac}, where the predictions of both eqs. (\ref{eq:diracpointloc}) and (\ref{eq:diracmont}), are compared with the exact values extracted from the Bloch spectrum.
As the position of the Dirac points is not fixed, for certain values of the parameters they merge and eventually disappear \cite{montambaux2009,montambaux2009b,gail2012}. This is particularly interesting, as it is associated to a topological phase transition from a semimetallic to an insulating phase. 
The merging occurs when the two solutions of eq. (\ref{eq:diracpointloc}) coincide modulo a reciprocal space vector $\bm{G}=p\bm{b}_{1}+q\bm{b}_{2}$ (with $p, q\in \mathbb{Z}$), namely at $\bm{k_{M}}=\bm{G}/2=(p\bm{b_{1}}+q\bm{b}_{2})/2$ \cite{montambaux2009,montambaux2009b}.
In principle, the current geometry of the lattice would permit four possible inequivalent merging points, namely $(p,q)=(0,0), (0,1), (1,0), (1,1)$, see Figure \ref{fig:merging}. However, considering the actual values of the tunneling coefficients, the only possible solutions inside the first Brillouin are at $\bm{k}_{M}\equiv (0,\pm 1)$.
For present parameter regimes, the merging occurs at $V_{\overline{X}}\simeq6.94$, see Figure \ref{fig:dirac}.
Following Refs. \cite{montambaux2009,montambaux2009b,lim2012}, one can expand the Hamiltonian density around one of the two merging points\footnote{A similar expansion can be derived around a generic Dirac point.}, by defining $\tilde{\bm{k}}\equiv \bm{k}-\bm{k}_{M}$.
The real and imaginary parts of the off-diagonal component $z(\bm{k})$ contributes with a linear term in $\tilde{k}_{x}$ and a quadratic term in ${\tilde k}_y$, respectively.
The leading terms of the expansion are
\begin{align}
z_{R}(\tilde{\bm{k}})&\simeq
-\left[t_{0}-2t_{1}+t_{2}+2t_{3}\right]+\pi^{2}\left[\left(4t_{3}-t_{1}\right)\tilde{k}_{y}^{2}\right]
\nonumber\\
z_{I}(\tilde{\bm{k}})&\simeq2\pi\left(t_{2}-t_{1}\right)\tilde{k}_{x}.
\end{align}
As for the diagonal term $F(\bm{k})$, it also contributes with a quadratic term in ${\tilde k}_y$, that accounts for the asymmetry between the two bands \cite{ibanez-azpiroz2013a}. Neglecting an irrelevant constant term, one has
\begin{equation}
F(\tilde{\bm{k}})\simeq-2\pi^{2}(2j_{1}-j_{2})\tilde{k}_{y}^2.
\end{equation}
Then, close to the merging point, the
Hamiltonian density can be cast into the following form
\begin{equation}
h_{\nu\nu'}(\tilde{\bm{k}})\simeq \displaystyle{\frac{\tilde{k}_{y}^2}{2\mu}}\otimes I + \left(\Delta+\displaystyle{\frac{\tilde{k}_{y}^2}{2m^{*}}}\right)
\otimes\sigma_x + c\tilde{k}_{x}\otimes\sigma_y
\end{equation}
with
\begin{align}
\Delta&\equiv -\left[t_{0}-2t_{1}+t_{2}+2t_{3}\right]
\label{eq:delta1}\\
\displaystyle{\frac{1}{2m^{*}}}&\equiv\pi^{2}\left(4t_{3}-t_{1}\right)\\
c&\equiv 2\pi\left(t_{1}-t_{2}\right)\\
\displaystyle{\frac{1}{2\mu}}&\equiv -2\pi^{2}(2j_{1}-j_{2}).
\label{eq:mu}
\end{align}
The corresponding dispersion law is
\begin{equation}
 \varepsilon_{\pm}(\tilde{\bm{k}})\simeq\displaystyle{\frac{\tilde{k}_{y}^2}{2\mu}}\pm\sqrt{\left(\Delta+\displaystyle{\frac{\tilde{k}_{y}^2}{2m^{*}}}\right)^{2}
+c^{2}\tilde{k}_{x}^2},
\label{eq:bandsclosemerging}
\end{equation}
that accurately reproduces the exact Bloch spectrum close to the merging point, as shown in Figure \ref{fig:merging-cut}. As in the present regime of parameters $m^{*}$ is always negative, the topological transition between the semi-metallic and insulating phases is driven by the sign of $\Delta$. When $\Delta$ is positive the system describes a semi-metal, characterized by the presence of Dirac points as in panels (a),(d). For $\Delta=0$ two Dirac points belonging to adjacent Brillouin zones eventually merge (panels (b),(e)). Finally, when $\Delta$ is negative a gap opens at the merging point, see panels (c),(f).

Notice that a gap can be opened also by breaking the parity symmetry, when the angle $\theta$ is tuned away from $\pi$ \cite{tarruell2012}. In fact, in this case the two minima in the unit cell  -- and so the diagonal terms $\epsilon_{\nu}$ and $j^{\nu}$ ($\nu=A,B$) -- are no longer degenerate, and a finite Dirac mass is generated. Even in this case, the \textit{extended} tight-binding model discussed here provides an accurate description of the microscopic Hamiltonian \cite{ibanez-azpiroz2013a}. 

\section{Interaction terms}
\label{sec:interactions}

\begin{figure}[H]
\centerline{\includegraphics[width=0.8\columnwidth]{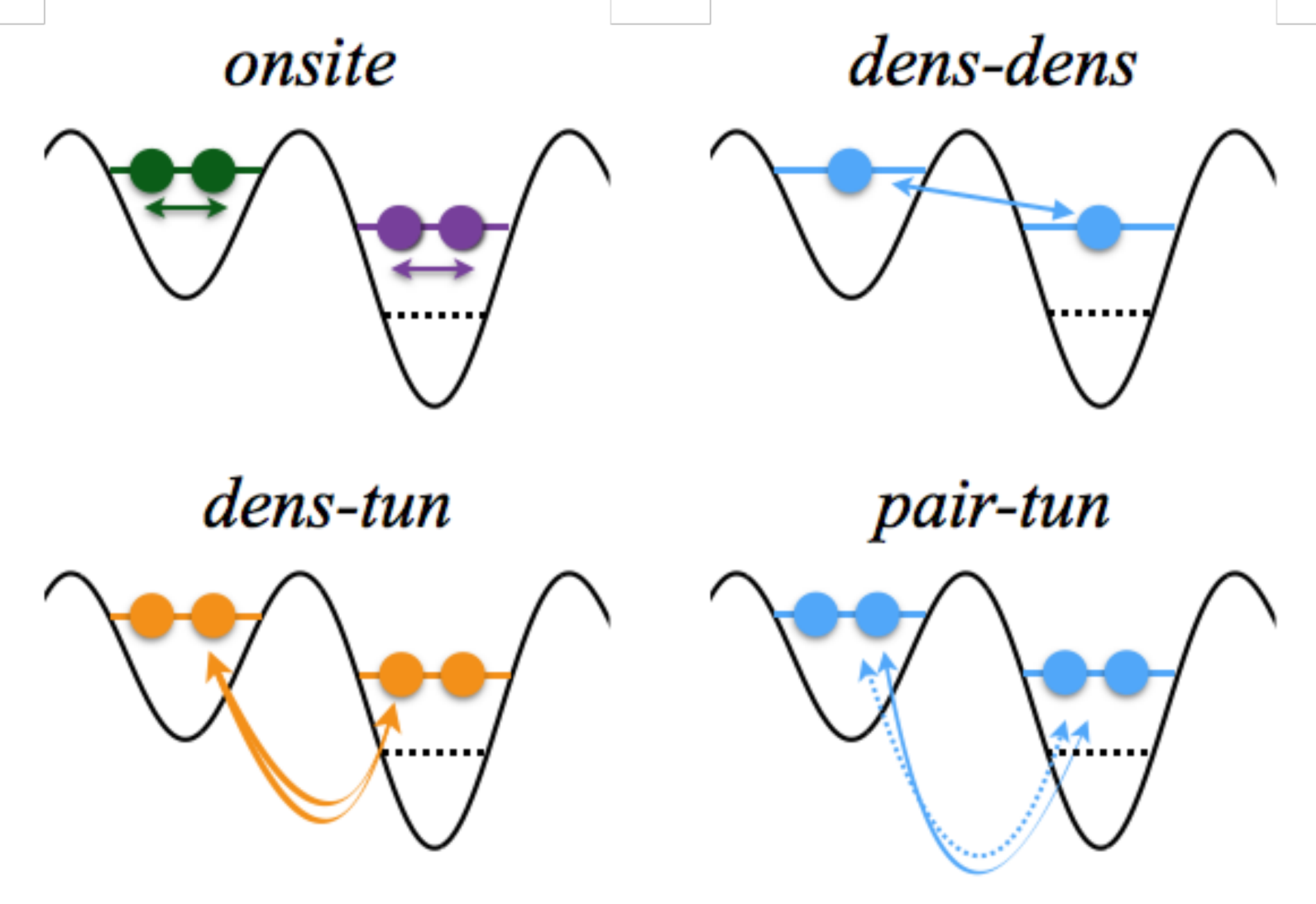}}
\caption{(Color online) A sketch of the various tunneling and interaction terms of the extended Bose-Hubbard model.}
\label{fig:sketch-int}
\end{figure}

In this final section we shall discuss how to deal with the effect of interaction between the particles, by considering the specific case of bosonic particles\footnote{The fermionic case is analogous, one should only make attention to the Pauli exclusion principle.} with contact interaction.  Namely, we consider an Hamiltonian term of the form
\begin{equation}
\hat{\cal{H}}_{int}=\frac{g}{2}\int d^{D}\bm{r}~\left|{\hat{\psi}}^\dagger(\bm{r}){\hat{\psi}}(\bm{r})\right|^{2},
\label{eq:hamint}
\end{equation}
with $g$ being a coupling constant.
Then, it is rather obvious that if interactions are not too strong, the same optimal basis of MLWFs obtained in the free particle limit is the natural choice for expanding the Hamiltonian (\ref{eq:hamint}), though other approaches - aimed at minimizing the contribution of the terms not included in the expansion - can also be found in the literature \cite{luhmann2014}. In general, one has
\begin{align}
\hat{\cal{H}}_{int} &= \frac{g}{2}\sum_{\{\nu_{i}\}=A,B}
\sum_{\{\bm{j}_{i}\}}\hat{a}_{\bm{j}_{1}\nu_{1}}^\dagger\hat{a}_{\bm{j}_{2}\nu_{2}}^\dagger\hat{a}_{\bm{j}_{3}\nu_{3}}\hat{a}_{\bm{j}_{4}\nu_{4}}\cdot
\nonumber\\
&\cdot\int d\bm{r}~ w^{\ast}_{\bm{j}_{1}\nu_{1}}(\bm{r})w^{\ast}_{\bm{j}_{2}\nu_{2}}(\bm{r})
w_{\bm{j}_{3}\nu_{3}}(\bm{r}) w_{\bm{j}_{4}\nu_{4}}(\bm{r}).
\label{eq:Hint-full}
\end{align}
The leading term is represented by the usual Bose-Hubbard on-site interaction \cite{jaksch1998}, namely
\begin{equation}
\hat{H}_{onsite}= \sum_{\nu=A,B}U_{\nu}\sum_{\bm{j}\nu}\hat{n}_{\bm{j}\nu}\left(\hat{n}_{\bm{j}\nu}-1\right),
\label{eq:Hint-onsite}
\end{equation}
with $U_{\alpha}=(g/2)\int\!d\bm{r}~\left| w_{\bm{j}\nu}(\bm{r})\right|^4$. 
In addition, one may also consider the effect of next-to-leading terms that couple neighboring wells inside the elementary cell \cite{luhmann2014,ganczarek2014}, see Figure \ref{fig:sketch-int}. 
They include a density-density interaction term
\begin{equation}
\hat{H}_{dens-dens}=\frac{g}{2} I_{2A2B}\cdot\hat{n}_{\bm{j}A}\hat{n}_{\bm{j}B},
\end{equation}
a density induced tunneling
\begin{equation}
\hat{H}_{dens-tun}= 
\frac{g}{2} I_{1A3B}\hat{a}_{\bm{j}A}^\dagger\hat{n}_{\bm{j}B}\hat{a}_{\bm{j}B} + 
\frac{g}{2} I_{3A1B}\hat{a}_{\bm{j}A}^\dagger\hat{n}_{\bm{j}A}\hat{a}_{\bm{j}B} + (A\leftrightarrow B),
\end{equation}
and the tunneling of pairs
\begin{equation}
\hat{H}_{pair-tun}= \frac{g}{2} I_{2A2B}\cdot\hat{a}_{\bm{j}A}^\dagger\hat{a}_{\bm{j}A}^\dagger\hat{a}_{\bm{j}B}\hat{a}_{\bm{j}B} + h.c..
\end{equation}
Here $I_{iAkB}$ represent a shorthand notation for the superposition integral in eq. (\ref{eq:Hint-full}), namely
\begin{equation}
I_{iAkB}\equiv\int d\bm{r}~ (w_{\bm{j}A}(\bm{r}))^{i} (w_{\bm{j}B}(\bm{r}))^{k}.
\label{eq:w4}
\end{equation}

\begin{figure}[H]
\centering
\includegraphics[width=\columnwidth]{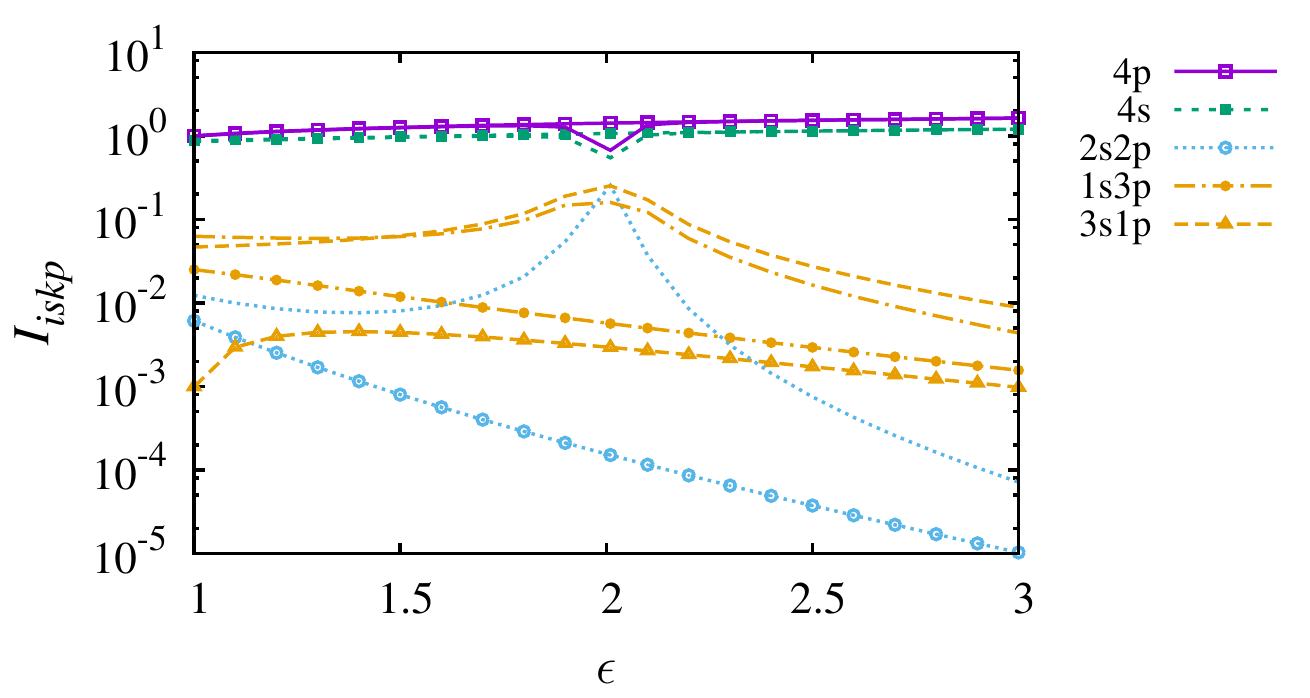}
\caption{(Color online) Plot of the (modulus of) the integrals $I_{iskp}$ characterizing the amplitude of various interaction terms, as a function of $\epsilon$. Lines with symbols correspond to the composite band approach, whereas plain lines refer to the single band case. The color code is the same as that in Figure \ref{fig:sketch-int}.}
\label{fig:integrals}
\end{figure}

As an example, let us consider the one-dimensional case discussed in sect. \ref{sec:oned}.
The relative weight of the various integrals in eq. (\ref{eq:w4}) is reported in Figure~\ref{fig:integrals}, for both the single- and composite-band approaches. This Figure shows that, as far as interactions are concerned, the composite-band model outperforms the single-band one, as the next-to-leading terms are significantly smaller in the former case. Notice also that (outside the resonant region) the values for on-site interaction given by the two approaches are almost indistinguishable. This is expected due to the very similar behavior of the bulk profiles of the Wannier functions.

From Figure~\ref{fig:integrals} it is also evident that, in the range of $\epsilon$ considered here, the $s-p$ density-density interaction is completely negligible with respect to the on-site interaction. Similarly small is the pair tunneling contribution - as the integrals involved are exactly the same for contact interactions. The significance of the density induced tunnelings with respect to the leading order tunneling in $\hat{H}_{0}$ depends on the value of the interaction constant $g$ and the lattice filling factors, and in general when $g$ is small they can be safely neglected. 

Let us conclude this section reminding the reader that the relative role of different terms may be different for long-range interactions like in dipolar condensates, see Ref. \cite{dutta2015} and references therein.

\section{Conclusions}
\label{sec:conclusions}

In this paper we have reviewed a general method for constructing tight-binding models for ultracold atoms in optical lattices, by means of the maximally localized Wannier functions (MLWFs) for composite bands introduced by Marzari and Vanderbilt \cite{marzari1997}. The MLWFs, obtained through a gauge transformation that minimizes their spread, constitute a powerful tool for calculating the tight-binding coefficients \textit{ab initio}, allowing for a direct connection between the continuous microscopic potential and the discrete tight-binding hamiltonian. A general method for extracting the values of the tight-binding parameters from the spectrum, by means of a set of suitable analytical relations,  has also been discussed. Several applications to one and two dimensional systems with two lattice sites per unit cell -- whose minimal description requires a set of two Bloch bands -- have been considered. 
In the one dimensional case, where the gauge transformation can be obtained by solving  a set of ordinary differential equations with periodic boundary conditions, we have considered the case of quasi resonance between the two lowest bands, and that between $s-p$ orbitals that is particularly interesting due to the presence of a Dirac point. The role of the MLWFs in the derivation of the effective Dirac equation close to the Dirac point has also been discussed. 
As for two-dimensional systems, we have considered several applications to regular and stretched honeycomb lattices, which represent a powerful platform for simulating the physics of graphene and for investigating different topological phases.  In particular, we have considered the case of the Haldane model, a paradigmatic model for topological insulators, that can be realized by adding an artificial magnetic field with the same periodicity of the lattice and vanishing flux through the unit cell. The present analysis, based on first principles, has permitted to reveal a number of important results, including the breakdown of the Peierls substitution and the fact that, in general, only a small portion of the nominal phase diagram of the Haldane model can be accessed. 
For the case of stretched honeycomb lattices, whose Dirac points can be moved and merged by tuning the lattice parameters, we have discussed the low-energy expansion around the merging points, that follows naturally from the tight-binding expansion. In all cases, the approach based on the MLWFs allows to accurately parametrize the optimal tight-binding parameters, providing a direct connection with the parameters that are accessed experimentally.

The present formulation allows also to include the effects of inter-particle interactions, that can be accounted for by expanding the corresponding terms on the basis of noninteracting MLWFs, in a perturbative approach. This has been briefly discussed in the last section of this Review, where we have considered the case of bosonic particles in a one dimensional superlattice, as an example. Regarding this point, we mention that only in the limit of strong interactions the single particle basis may not be the optimal one, and other approaches may be considered \cite{johnson2009,mering2011,bissbort2012,luhmann2012,lcacki2013}.

\vspace*{2mm} \Acknowledgements{\bahao 
Most of the material presented in this review is the result of previous collaborations with A. Bergara, A. Eiguren, X. Lopez-Gonzalez, J. Sisti, J. Zakrzewski. We would like to thank them all. 
This work has been supported by the Universidad del Pais Vasco/Euskal Herriko Unibertsitatea under Program No. UFI
11/55, the Ministerio de Economia y Competitividad through Grants No. FIS2012-36673-C03-03,  the Basque Government through Grant No. IT-472-10, the Helmholtz Gemeinschaft Deutscher-Young Investigators Group Program No. VH-NG-
717 (Functional Nanoscale Structure and Probe Simulation Laboratory)
and the Impuls und Vernetzungsfonds der Helmholtz-Gemeinschaft Postdoc Programme.}

\end{multicols}

\end{document}